\documentclass[12pt]{article}
\usepackage{amsmath} 
\usepackage{amssymb}
\usepackage{bm}
\usepackage{bbm}
\usepackage{cite}
\usepackage{color} 
\usepackage{graphicx}
\usepackage{slashed} 
\usepackage{tabularx}
\usepackage{comment} 

\newcommand{\Sec}[1]{Sec.~\ref{#1}}  

\newcommand{\fig}[1]{Fig.~\ref{#1}}

\newcommand{\tab}[1]{Tab.~\ref{#1}}

\newcommand{\eq}[1]{Eq.~(\ref{#1})}

\begin{document}
\title{
\vspace{-1.2truecm}
\Large\bf
Implications of the first AMS-02 measurement for  \\
dark matter annihilation and decay
}
\author{
  Hong-Bo Jin$^a$, \ 
Yue-Liang Wu$^{a,b}$ \ 
 and  Yu-Feng Zhou$^a$\ 
  \\ \\
 \textit{$^a$State Key Laboratory of Theoretical Physics},\\
  \textit{Kavli Institute for Theoretical Physics China}, \\
  \textit{Institute of  Theoretical Physics, Chinese Academy of Sciences}\\ 
\textit{Beijing, 100190, P.R. China}\\
   \textit{$^b$University of Chinese Academy of Sciences},\\
    \textit{Beijing, 100049, P.R. China}\\
}
\date{}
\maketitle
\begin{abstract}
In light of the first  measurement on 
the positron fraction by the AMS-02 experiment,
we perform a detailed  global analysis  on 
the  interpretation of  the latest data of 
PAMELA, Fermi-LAT, and AMS-02 in terms of 
dark matter (DM) annihilation and decay
in various propagation models.
%
The allowed regions for 
the DM particle mass and annihilation cross section or decay life-time 
are obtained for  channels with leptonic final states:
$2e$, $2\mu$, $2\tau$, $4e$, $4\mu$ and $4\tau$. 
We show that 
for the conventional astrophysical background 
the AMS-02 positron fraction data alone favour 
a DM particle mass $\sim 500 (800)$ GeV
if DM particles annihilate dominantly into $2\mu (4\mu)$ final states, 
which  is significantly lower than that  favoured by  
the Fermi-LAT data on the total flux of electrons and positrons. 
The allowed regions by the two experiments do not overlap 
at a high confidence level ($99.99999\%$C.L.).
We consider a number of propagation models with different
halo height $Z_{h}$, diffusion parameters $D_{0}$ and $\delta_{1/2}$, 
and power indices of primary nucleon sources $\gamma_{p1/p2}$.
The normalization and the slope of the electron background are 
also allowed to vary. 
We find that the tension between  the two experiments can 
be only slightly reduced in the model with large $Z_{h}$ and  $D_{0}$.
The consistency of fits is improved for 
annihilation channels with  $2\tau$ and $4\tau$ final states 
which favours TeV scale DM particle with large cross sections above 
$10^{-23} \text{cm}^3\text{s}^{-1}$.
In all the  considered leptonic channels,  
the current data favour the scenario of DM annihilation over  DM decay. 
In the decay scenario, 
the charge asymmetric DM decay is slightly favoured.
\end{abstract}


\section{Introduction}
Compelling evidence from astronomical observations has indicated that
dark matter contributes to
nearly 27\% of the energy density  of the Universe
\cite{Ade:2013xsa
}.
Popular DM candidates such as 
the weakly interacting massive particles (WIMPs) 
are expected to annihilate or decay into 
standard model (SM) final states in the Galactic halo and beyond, 
which may leave imprints in the fluxes of cosmic-ray particles, 
including  electrons, positrons, antiprotons, and cosmic gamma-rays.
The ongoing satellite-borne experiments, 
such as PAMELA, Fermi-LAT, and AMS-02 etc.   
are  searching for such potenital indirect signatures of DM with high precision.

Significant progresses have been made in the recent years. 
For instance,
the PAMELA  collaboration has reported  that 
the ratio of the positron flux to the total flux of electrons and positrons 
rises  with increasing energy in the range 10--100~GeV,
which is not  expected from  conventional astrophysical backgrounds
\cite{Adriani:2008zr, 
1001.3522 
}.
This result was later confirmed by Fermi-LAT,
which further showed  that 
the positron fraction continues to rise in the higher energy range 100--200~GeV
\cite{1109.0521 
}. 
The total flux of electrons and positrons  measured by 
the ballon-borne experiment ATIC and BBP-BETS
also showed an excess in 300--700 GeV with a peak located at  around 600~GeV
\cite{Chang:2008zzr, 
0809.0760  
}.
Although the ATIC/BBP-BETS  ``bump" was not confirmed by 
Fermi-LAT
\cite{Abdo:2009zk,
Ackermann:2010ij 
}  
and  HESS
\cite{Aharonian:2008aa,
Aharonian:2009ah 
}, 
the featureless power-law spectrum measured by Fermi-LAT 
corresponds to a power index $3.08$~\cite{Ackermann:2010ij} 
which is  harder than what expected from 
the convention astrophysical background,
and may also require an exotic source of 
cosmic-ray electrons/positrons.

Recently, 
the Alpha Magnetic Spectrometer (AMS-02) collaboration has released  the first measurement  of 
the positron fraction based on the collected $6.8\times 10^6$ events of electron and positron
with unprecedented accuracy
\cite{PhysRevLett.110.141102
}.
The result shows a steadily increasing of positron fraction from 10 to $\sim250$ GeV,
which is consistent with  the previous measurements by PAMELA and Fermi-LAT. 
The spectrum measured  by AMS-02 is slightly lower than 
PAMELA for electron energy larger than $\sim 40$ GeV,
and the slope of the positron fraction spectrum decreases 
by an order of magnitude from 20 to $\sim250$ GeV. 
Furthermore, the current AMS-02 data show no significant fine structure or 
anisotropy in the positron flux.

The rising spectrum of positron fraction may have astrophysical origins, 
such as from nearby pulsars 
\cite{Hooper:2008kg,
Profumo:2008ms 
}
and supernovae remnants
\cite{Blasi:2009hv,
Ahlers:2009ae 
}
. 
These explanations will be constrained by the anisotropy and the high energy 
behaviour of the positron flux.
Halo DM annihilation or decay can provide an alternative explanation.
Given the precisely measured  shape of the positron fraction spectrum by AMS-02,
more insights on the nature of DM can be obtained.
Although at this stage it is still impossible to distinguish different type of exotic sources,
this precision measurement on the positron fraction spectrum may shed new light on 
the origin of  high energy cosmic-ray positrons.

In this work,
we perform an updated global analysis on  
the DM interpretation of the current measurement of 
cosmic-ray electrons and positrons,
for both annihilation and decay scenarios. 
We calculate the propagation of cosmic-ray particles using 
the numerical package GALPROP.
Typical channels of DM annihilation and decay with 
final satates $2e$, $2\mu$, $2\tau$, $4e$, $4\mu$ and $4\tau$ 
are investigated  in
a number of propagation models with different
halo heights $Z_{h}$, diffusion parameters $D_{0}$, $\delta_{1,2}$ and 
power indices of primary nucleon sources $\gamma_{p1,p2}$, etc..
The normalization and the slope of the electron background are 
also allowed to vary. 
The results show that
for DM annihilating into  $2\mu$  and $4\mu$ final states, 
the allowed parameter regions determined by AMS-02 positron fraction data is  
highly inconsistent with 
that favoured by  Fermi-LAT data on the total flux of electrons and positrons.
We  find that the tension between  the two experiments can
be slightly reduced in the case of large $Z_{h}$ and $D_{0}$.
More consistent fits are obtained  for $\tau$-lepton final states, which favours TeV scale
DM with large cross sections above $10^{-23} \text{ cm}^3\text{s}^{-1}$.
In all the  considered leptonic channels,  
we find that
the current data favour the scenario of DM annihilation over  DM decay. 
For the DM decay scenario, 
both the charge symmetric and asymmetric cases are investigated. 
In the decay scenario, 
the charge asymmetric DM decay is slightly favoured.

This paper is organized as  follows. 
In \Sec{sec:framework}, 
we outline the framework for 
calculating the propagation of the cosmic-ray particles and
the primary sources from DM annihilation and decays.
In \Sec{sec:fitSchemes},
we describe the data selection and 
the strategy of the data fitting in a number of propagation models.
The numerical results are presented in \Sec{sec:results}.
We finally conclude in \Sec{sec:conclusions}.

\section{Sources and propagation of cosmic-ray particles}\label{sec:framework}
%
In the diffusion models of cosmic-ray propagation, 
the Galactic halo within which  the diffusion processes occur is parametrized by 
a cylinder with radius $R\approx 20$ kpc and half-height $Z_{h}$. 
The number densities of cosmic-ray particles are vanishing at the boundary of the halo. 
The processes of energy losses, reacceleration, annihilation,
as well as the secondary sources of cosmic-rays 
are confined  within the Galactic disc. 
%
The diffusion  equation for the cosmic-ray particles is given by
\begin{align}\label{eq:28}
  \frac{\partial \psi}{\partial t} =&
  \nabla (D_{xx}\nabla \psi -\mathbf{V}_{c} \psi)
  +\frac{\partial}{\partial p}p^{2} D_{pp}\frac{\partial}{\partial p} \frac{1}{p^{2}}\psi
  -\frac{\partial}{\partial p} \left[ \dot{p} \psi -\frac{p}{3}(\nabla\cdot \mathbf{V}_{c})\psi \right]
  \nonumber \\
  & -\frac{1}{\tau_{f}}\psi
  -\frac{1}{\tau_{r}}\psi
  +q(\mathbf{r},p)  ,
\end{align}
where $\psi(\mathbf{r},p,t)$ is  the number density per unit of total particle momentum, 
which is related to the phase space density $f(\mathbf{r},p, t)$ as
$\psi(\mathbf{r},p,t)=4\pi p^{2}f(\mathbf{r},p,t) $. 
For steady-state diffusion, it is assumed that  $\partial  \psi/\partial t=0$.
The spatial diffusion coefficient $D_{xx}$ is parametrized as
\begin{align}
\label{eq:29}
D_{xx}=\beta D_{0} \left( \frac{\rho}{\rho_{0}} \right)^{\delta_{1,2}}  ,
\end{align}
%
where $\rho=p/(Ze)$ is the rigidity of the cosmic-ray particle, 
and 
$\delta_{1(2)}$ is the power spectral index 
when $\rho$ is below (above) a reference rigidity $\rho_{0}$.  
The parameter $D_{0}$ is a normalization constant, 
and $\beta=v/c$ is the velocity of the cosmic-ray particle
with $c$ the speed of light.
The values of $D_{0}$ and $\delta_{1,2}$ are determined by 
the  ratio between secondary and primary cosmic-ray species 
such as the ratio of Boron to Carbon (B/C) 
and that of isotopes $^{10}\text{Be}/^{9}\text{Be}$, etc..
The convection term is related to the drift of cosmic-ray particles from 
the Galactic disc due to the Galactic wind.  
The direction of the wind is usually assumed to be 
along the $z$-direction  perpendicular to the galactic disc, 
and 
is a constant $\mathbf{V}_{C}=[2\theta(z)-1] V_{c}$.

The  diffusion in momentum space is described by 
the reacceleration parameter $D_{pp}$ 
which is related to the  Alfv$\grave{\mbox{e}}$n speed $V_{a}$ of 
the disturbances in the hydrodynamical plasma as follows
%
\begin{align}
D_{pp}=
\frac{4V_{a}^{2} p^{2}}
{3D_{xx}\delta_{i}
\left(4-\delta_{i}^{2}\right)
\left(4-\delta_{i}\right)},
\end{align}
where $\delta_{i}=\delta_{1}$ or $\delta_{2}$, depending on the rigidity.
The momentum loss rate is denoted  by $\dot{p}$,
and $\tau_{f}$, $\tau_{r}$  are the time scales for fragmentation and radioactive
decay,  respectively.
%
%
%
High energy electrons/positrons loss  energy due to 
the processes like inverse Compton scattering and synchrotron radiation.
The typical propagation length is around a few kpc for electron energy 
around 100 GeV.
In the calculation of energy loss rate, 
we take the interstellar magnetic field to be 
\begin{align}
 B(r,z)=B_{0} 
 \exp\left(-\frac{r-r_{\odot}}{r_{0}}\right)
 \exp\left(-\frac{|z|}{z_{0}} \right)  ,
 \end{align}
with 
$B_{0}=5\times 10^{-10}$~Tesla,
$r_{0}=10$ kpc,
and 
$z_{0}=2$ kpc.

%
%
The injection spectrum of a primary cosmic-ray particle 
$A$ $(A=e^{-}, p,\dots)$ 
is assumed to have a broken power low behaviour 
$dq_{A}(p)/dp\approx\rho^{-\gamma_{A}}$, with $\gamma_{A}=\gamma_{A1}(\gamma_{A2})$ for 
the rigidity $\rho_{A}$ below (above) a reference rigidity $\rho_{As}$.
%
Secondary  cosmic-ray particles are treated as 
decay products of charged pions and kaons  created in 
collisions of primary cosmic-ray particles with interstelar  gas. 
%

%
%

The   flux of the cosmic-ray particle is related to its density function as 
\begin{align}\label{eq:38}
\Phi= \frac{v}{4\pi} \psi(p) .
\end{align}
For high energy electrons/poistrons $v\approx c$. 
At the top of the atmosphere (TOA) of the Earth, 
the fluxes of cosmic-rays  are affected  by solar winds 
and the helioshperic magnetic field. 
This effect is taken into account using 
the force-field approximation \cite{Gleeson:1968zza}.
The electron/positron flux at the top of the atmosphere of the Earth $\Phi^{\text{TOA}}_{e^{\pm}}$ 
which is measured by the experiments is related to the interstellar flux as follows
\begin{align}
\label{eq:45}
\Phi^{\text{TOA}}_{e}(T_{\text{TOA}})=\left(\frac{2m_{e} T_{\text{TOA}}+T_{\text{TOA}}^{2}}{2m_{e} T+T^{2}}\right)\Phi_{e}(T) ,
\end{align}
where $T_{\text{TOA}}=T-\phi_{F}$ is the kinetic energy of electrons/positrons 
at the top of the atmosphere of the Earth.  
We take $\phi_{F}=0.55$ GV in numerical analysis.
As we are interested in electrons/positrons fluxes at energies above $\sim 20$ GeV,
the effect  of solar modulation is less significant.

In this work, 
we solve the diffusion equation and calculate 
the energy losses for electrons by ionization, Coulomb interactions, bremsstrahlung, inverse Compton, 
and synchrotron using the publicly available  numerical code  GALPROP v54
\cite{astro-ph/9807150,astro-ph/0106567,astro-ph/0101068,astro-ph/0210480,astro-ph/0510335}
which makes use of  realistic astronomical information on 
the distribution of interstellar gas and other data as input, 
and consider various kinds of data including primary and secondary nuclei,
electrons and positrons,
$\gamma$-rays, synchrotron  radiation, etc. 
in a self-consistent way. 
Other approaches based on simplified assumptions on 
the Galactic gas distribution 
which  allows  for fast  analytic solutions can be found in Refs.
\cite{astro-ph/0103150,
astro-ph/0212111,
astro-ph/0306207,
1001.0551,
Cirelli:2010xx
}.
In solving the propagation equation, 
we use a very fine grid in kinetic energy  space with 
a logarithmic scale factor $1.02$. 
%
The calculations of the electron propagation in this work 
are cross-checked by 
comparing the reults with that from the GALPROP webrun
\cite{Vladimirov:2010aq
}.

The primary source term from the annihilation of Majorana DM particles has the form
\begin{align}\label{eq:ann-source}
q_{e}(\mathbf{r},p)=\frac{\rho(\mathbf{r})^2}{2 m_{\chi}^2}\langle \sigma v \rangle 
\sum_X \eta_X \frac{dN^{(X)}_e}{dp} ,
\end{align}
where $\langle \sigma v \rangle$ is 
the velocity-averaged DM annihilation cross section multiplied by relative velocity
(referred to as cross section), and 
$\rho(\mathbf{r})$ is the DM energy spatial distribution function. 
$dN^{(X)}_e/dp$ is the injection spectrum  from
DM particles annihilating into $e^{\pm}$ via all possible 
intermediate states $X$ with  
$\eta_X$ the corresponding branching fractions. 

In the case of DM  decay, 
if the DM particle is not its own antiparticle,  
its decay into charged leptons can be charge asymmetric, 
for instance,
$\chi\to e^+ +Y^-$ while $\bar\chi\to e^- + Y^+$, 
where $Y^\pm$ can be the SM gauge boson $W^\pm$ 
or charged Higgs boson $H^\pm$  or other charged particle in 
new physics models. 
In the generic case where 
the energy density of the relic DM particles is also asymmetric, 
i.e., 
$\rho_\chi(\mathbf{r}) \neq \rho_{\bar\chi}(\mathbf{r})$, 
the source term can be written as
\begin{align}\label{eq:decay-source}
q_{e^\pm}(\mathbf{r},p)=\frac{\rho(\mathbf{r})}{2\tau m_{\chi}} (1\pm \epsilon)  
\sum_X \eta_X \frac{dN^{(X)}_e}{dp} ,
\end{align}
where $\rho(\mathbf{r}) \equiv\rho_\chi(\mathbf{r}) + \rho_{\bar\chi}(\mathbf{r})$,
$\epsilon \equiv (\rho_\chi(\mathbf{r}) - \rho_{\bar\chi}(\mathbf{r}))/(\rho_\chi(\mathbf{r}) + \rho_{\bar\chi}(\mathbf{r}))$,
and $\tau$ is the life-time of the DM particle.  
The case where $\epsilon=1\ (-1)$ corresponds 
to the DM particle decaying into $e^+\ (e^-)$ only,
and 
$\epsilon=0$ corresponds to the DM decay equally into $e^+$ and $e^-$.
The  charged leptons from the decay of $Y^{\pm}$ are not considered,
which corresponds to the case with maximal  charge asymmetry. 
The phenomenology of charge asymmetric decay has been investigated  previously in 
Refs.~\cite{
Feldstein:2010xe,
Chang:2011xn 
}.

The injection spectra $dN^{(X)}_e/dp$ from DM annihilation and decay are calculated using 
the numerical package PYTHIA v8.175
\cite{
Sjostrand:2007gs 
}. 
For the decay scenario, we assume that  $X^\pm$ are much lighter than the DM particle, 
and neglect its mass effect in the kinematics of DM decay.
%


The fluxes of cosmic-ray electrons and positrons from DM annihilation depend only
weakly on the DM halo profile.   
%
%
In this work, we shall take the  Einasto profile 
\cite{Einasto:2009zd 
}
\begin{align}
\rho(r)=\rho_\odot \exp
\left[
-\left( \frac{2}{\alpha_E}\right)
\left(\frac{r^{\alpha_E}-r_\odot^{\alpha_E}}{r_s^{\alpha_E}} \right)
\right] ,
\end{align}
with $\alpha_E\approx 0.17$ and $r_s\approx 20$ kpc. 
The local DM energy density is fixed at $\rho_\odot=0.43 \text{ GeV}\text{ cm}^{-3}$
\cite{Salucci:2010qr
}.


\section{Data selection and fitting schemes}\label{sec:fitSchemes}
In order to avoid  uncertainties caused by solar modulation, 
we consider the latest cosmic-ray data from satellite-borne experiments with 
electron energy above 20 GeV which include
4 data points from PAMELA  in 2010
\cite{1001.3522 
}, 
10 data points  from Fermi-LAT data in 2011 
\cite{1109.0521 
},
and 31 data points  from AMS-02
\cite{PhysRevLett.110.141102
}
for the positron fraction. 
For the total flux of electrons and positrons 
we consider the updated data of 
Fermi-LAT in 2010 \cite{Ackermann:2010ij 
}
which contain  28 data points.
%
We also include the data  of  electron flux measured by 
PAMELA (18 data points)~\cite{Adriani:2011xv}
and very recently by AMS-02 (32  data points)~\cite{AMS02electron}.
They are important for constraining the primary electron background.
Thus in total 123 data points are included in the global fits. 
We do not include the data of positron flux 
and the total flux of electrons and positrons
recently reported by by AMS-02,
as they are less accurate in comparison with the AMS-02 result
of positron fraction and the electron flux. 
Some previous global fits to the earlier data can be found in Refs.
\cite{
Cirelli:2008pk,
Bergstrom:2009fa,
Cirelli:2009dv
}.
Note that 
the electron spectrum of the Fermi-LAT 2010 data is 
smoother than the  one previously reported in 2009 
\cite{Abdo:2009zk},
which results in visible modifications to the best-fit parameters 
such as the DM particle mass and 
annihilation cross section or decay lift-time.

 In this work, 
 the DM particle annihilating and decay into two-body and four-body 
 charged leptonic final states $e$, $\mu$, and $\tau$ are investigated.
The relevant quantities  are determined 
through $\chi^2$-fits to the data.  The expression of $\chi^2$ is given by
\begin{align}
\chi^2=\sum_i \frac{(f_i^{\text{th}}-f_i^{\text{exp}})^2}{\sigma_i^2},
\end{align} 
where $f_i^{\text{th}}$ are the theoretical predictions. 
%
$f_i^{\text{exp}}$ and $\sigma_i$ are the central values  and errors of experimental data, respectively. 
The index $i$ runs over all the available data points.

The outcome of the global fits depends on 
the choice of the parameters appearing in 
the propagation equation \eq{eq:28}.
The uncertainties related to  these parameters 
need to be discussed separately.
The height of the propagation halo $Z_{h}$ and  
the diffusion parameters such as 
$D_{0}$ and $\delta_{1,2}$ affect 
both the cosmic-ray backgrounds and 
the DM-induced cosmic-ray fluxes,
while the primary injection indices 
$\gamma_{e1,e2}$ and $\gamma_{p1,p2}$ 
affect the backgrounds of electrons and positrons.
%
%
%
%
%
We first consider two benchmark propagation models which are extensively studied in the literature
\begin{itemize}
\item {\bf Model A},  
the so-called conventional  diffusive reacceleration model \cite{astro-ph/0101068,astro-ph/0510335}
which is commonly adopted by the current experimental collaborations 
such as  PAMELA \cite{Adriani:2008zq,Adriani:2010rc,Adriani:2011xv}  
and Fermi-LAT \cite{Ackermann:2010ij,FermiLAT:2012aa}
as a benchmark model for 
the astrophysical backgrounds and 
the propagation of cosmic antiparticles from DM annihilation/decay. 
The location of the observed peak in the spectrum of B/C at about 1 GeV is 
well reproduced in this model. 
The propagation parameters in this model are determined from 
fitting  the ratio of the secondary to primary nuclei such as B/C,
the flux of primary such as Carbon, 
and the Galactic distribution of cosmic-ray sources are determined from 
the EGRET gamma-ray data.
In this model,
the break in the diffusion coefficient is $\rho_{0}=4$ GV with 
$\delta_{1}=\delta_{2}=0.34$.  
The break of the primary electron source and proton sources
are $\rho_{e}=4$ GV, and $\rho_{p}=9$ GV, respectively.
%
%
The Alfv$\grave{\mbox{e}}$n velocity is set to $V_{a}=36.0 \text{ km s}^{-1}$.
The power indices $\delta_{1,2}$, $\gamma_{e1,e2}$, $\gamma_{p1,p2}$
as well as other parameters in this model are listed in \tab{PowerIndexInjection}.
%
%
\item {\bf Model B}, 
the parameter set determined from 
a comprehensive global Byesian analysis to  the  data of 
B/C, ${}^{10}\text{Be}/{}^{9}\text{Be}$,
Carbon and Oxegen, etc.,
using nested sampling  Markov Chain Monte Carlo method~\cite{1011.0037}.
The gamma-ray data of Fermi-LAT are used to determine 
the distribution of cosmic-ray sources. 
In this model,
the break in the diffusion coefficient is the same $\rho_{0}=4$ GV.
The break of the primary electron source and proton sources
are $\rho_{e}=4$ GV, and $\rho_{p}=10$ GV, respectively.
%
The Alfv$\grave{\mbox{e}}$n speed is $V_{a}=39.2 \text{ km s}^{-1}$.
Other parameters in the model are listed in \tab{PowerIndexInjection}.
\end{itemize}
In  Ref.~\cite{1011.0037}, 
the uncertainties as well as the correlations of 
the  propagation parameters of Model B were carefully studied,
which facilitates the  investigation on the  uncertainties  
induced  by each propagation parameter separately.
The allowed ranges for these parameters at $95\%$ C.L. are given by~\cite{1011.0037}
\begin{align}\label{limits}
Z_{h}&=(3.2-8.6)\text{ kpc}, 
D_{0}=(5.45-11.2)\times10^{28}\text{ cm}^{2}\text{s}^{-1}, 
\delta_{2}=0.26-0.35, 
\nonumber\\
\gamma_{p1}&=1.84-2.00,
\gamma_{p2}=2.29-2.47,
V_{a}=(34.2-42.7) \text{ km s}^{-1}.
\end{align}
Some of the parameters such as $\delta_{1}$, $\gamma_{p1}$ and $V_{a}$ only affect
the predicted fluxes at low energies.
The parameters which are most relevant to  the electron and positron fluxes at high 
energies above 20 GeV are $Z_{h}$, $D_{0}$, $\delta_{2}$ and  $\gamma_{p2}$.
To see how the fit results change within 
the uncertainties of the parameters,
we consider several  limiting cases 
in each case 
one of teh parameters 
is set at its upper or lower limit given in \eq{limits}.
The differences in the fit results can be regarded as an estimation of 
the uncertainties from  that propagation parameter.
Thus besides Model A and B, 
we further consider the following six  propagation models:
\begin{itemize}
%
\item {\bf Model C1 (C2)},
the halo half-height $Z_{h}$ is taken to its lower (upper)
limit $Z_{h}=3.2\ (8.6)$ kpc.  
Since it has been shown that $Z_{h}$ and $D_{0}$ are positively correlated~\cite{1011.0037},
we must  take the value of $D_{0}$ to be 
$5.45\ (11.2)\times10^{28}\text{ cm}^{2}\text{s}^{-1}$ accordingly.
The rest of parameters in this model are fixed at their best-fit values as that
in Model B. 
\item {\bf Model D1 (D2)},
the power index $\delta_{2}$  is taken to is lower (upper) limit $\delta_{2}=0.26\ (0.35)$, 
and the relation $\delta_{1}=\delta_{2}$ is still assumed.
The rest of parameters are the  same  as that in  Model B. 
\item {\bf Model E1 (E2)},
the power index of the injection spectrum of proton $\gamma_{p2}$ 
which is related to the source of secondary positrons is 
taken to its lower and upper limit $\gamma_{p2}=2.29\ (2.47)$.
The rest of parameters are the  same  as that in Model B. 
\end{itemize}
%
%
%
%
%
The corresponding parameters in all the eight propagation models 
from Model~A to Model~E2 are  summerized in \tab{PowerIndexInjection}.
\begin{table}[tb]
\begin{center}
\begin{tabular}{llllll}
\hline\hline
Model& 

$z_h(\text{kpc})$ & $D_0$& $\delta_{2}$ &$\gamma_{e1}/\gamma_{e2}$&
$\gamma_{p1}/\gamma_{p2}$ 
\\ \hline
A 		&4.0		&5.75		&0.34	&1.6/2.5	&1.82/2.36\\
B 		&3.9		&6.59		&0.30	&1.6/2.5	&1.91/2.42\\
C1(C2) &3.2(8.6)	&5.45(11.2)	&0.30 &1.6/2.5	&1.91/2.42\\
D1(D2) &3.9		&6.59      & 0.26(0.35)&1.6/2.5	&1.91/2.42\\
E1(E2) 	&3.9		&6.59		&0.30		&1.6/2.5&1.91/2.29(2.47)\\
\hline\hline
\end{tabular}
\end{center}
\caption{
Parameters of eight propagation models from Model A to Model E2.
The diffusion coefficient $D_0$ is in units of  $10^{28}\text{ cm}^{2}\text{s}^{-1}$.}
\label{PowerIndexInjection}
\end{table}

%
In all the considered models 
the power indices of primary electron $\gamma_{e1}/\gamma_{e2}$ are 
fixed at $1.6/2.5$ which are determined from fitting the early cosmic-ray electron data.
In order to take into account the uncertainties in the primary electrons,
we multiply a scaling  factor  $\kappa$ and 
an energy-dependent factor $(E/\text{GeV})^\delta$ to
the  primary electron flux $\Phi^{\text{bg}}_{e^{-}}$ after propagation.
%
%
The expressions for the positron fraction and 
the  total flux of electron and positron are modified as follows
\begin{align}\label{eq:fluxdef}
\Phi_{\text{tot}} &=
\left(
 \kappa (E/\text{GeV})^{\delta} \Phi_{e^-}^{\text{bg}}+\Phi_{e^+}^{\text{bg}}
 \right)
 +(\Phi_{e^-}^{\text{DM}}+\Phi_{e^+}^{\text{DM}}) ,
\nonumber\\
R_{e^+} &=(\Phi_{e^+}^{\text{DM}}+	\Phi_{e^+}^{\text{bg}})/\Phi_{\text{tot}} ,
\end{align} 
where $\Phi_{e^\pm}^{\text{DM}}$ and $\Phi_{e^\pm}^{\text{bg}}$ are 
the fluxes from DM annihilation/decay and 
background calculated from the GALPROP code, respectively.  
%
%
Both the values of $\kappa$ and $\delta$ are 
treated as free parameters to be determined from the data.
Thus in the case of DM annihilation,  we have in total four free parameters:
$ m_{\chi}$, $\langle \sigma v \rangle$, $\kappa$,  and  $\delta$
to be determined by the experimental data in eight different propagation 
models from Model A to Model E2.
Note that under the approximation
$\Phi^{\text{bg}}_{e^{+}}\ll \Phi^{\text{DM}}_{e^{+}}\ll \Phi^{\text{bg}}_{e^{-}}$,
which is often valid  at high energies,
the positron fraction can be rewritten as
\begin{align}\label{eq:Rapprox}
R_{e^{+}}\approx \frac{\Phi^{\text{DM}}_{e^{+}}}{\kappa (E/\text{GeV})^{\delta} \Phi_{e^-}^{\text{bg}}}.
\end{align}
Since $\Phi^{\text{DM}}_{e^{+}}$ is proportional to $\langle \sigma v\rangle$, 
it is expected that 
in this limit there will be a degeneracy in determining $\langle \sigma v\rangle$ and $\kappa$,
which means that 
if we only consider the data of positron fraction
$\langle \sigma v \rangle$ will be sensitive to $\kappa$.
This degeneracy can be removed by including the measurements of  electron fluxes by PAMELA~\cite{Adriani:2011xv}
and recently by AMS-02~\cite{AMS02electron}.

In comparing  the experimental data with the theoretical predictions, 
we take into account the effect of finite energy resolution of the detectors,
namely, the predicted fluxes are convoluted according to the energy resolution
for each experimental detector. 
The energy resolution of PAMELA is nearly a constant $\sim 5\%$ above 10 GeV
\cite{Picozza:2006nm 
}.
The energy resolution of Ferm-LAT is  $\sim 6\%$ at 7 GeV and $\sim 15\%$ at 1 TeV,
and we take the numerical values from Ref.
\cite{Ackermann:2010ij 
} in the calculation. 
For the resolution of AMS-02 detector, we use the  parametrization
$\sigma(E)/E=[(0.104/\sqrt{E/\text{GeV}})^2+(0.014)^2]^{1/2}$
which reaches
$\sim 1.4\%$ at high energies
\cite{PhysRevLett.110.141102
}.

\section{Results}\label{sec:results}

\subsection{Fits with annihilating dark matter}
\begin{figure}[thb]
\begin{center}
\includegraphics[width=0.40\textwidth]{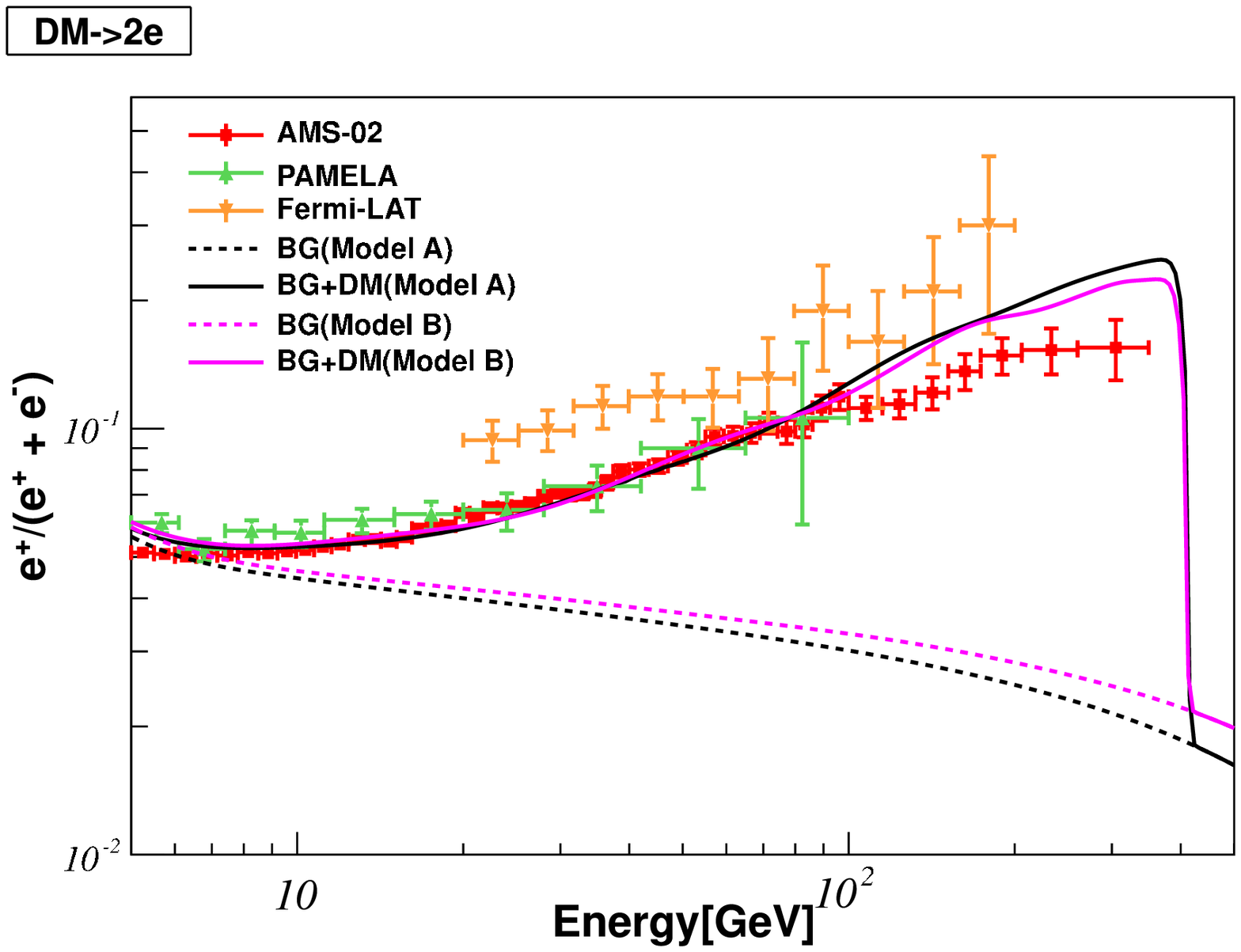}\includegraphics[width=0.40\textwidth]{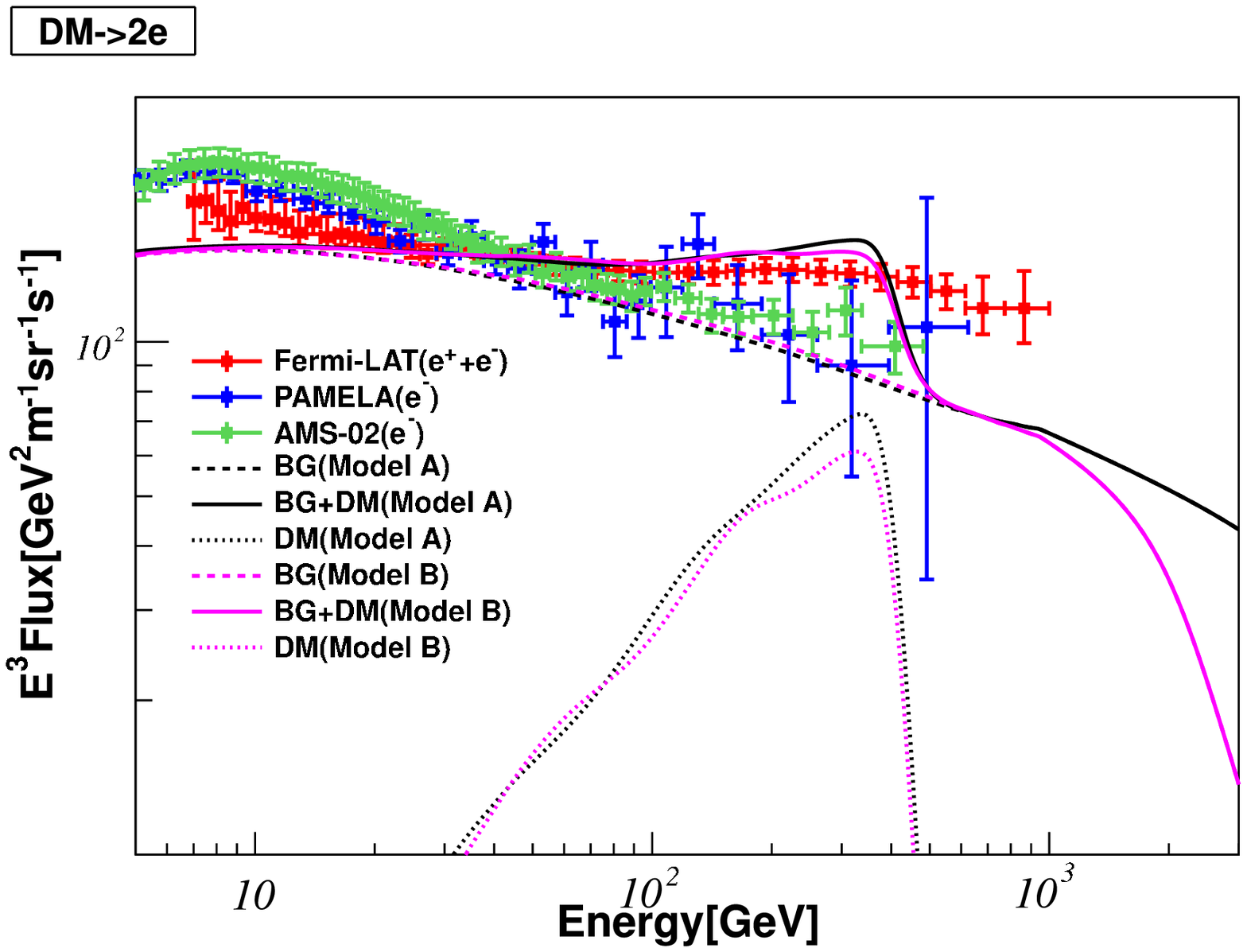}
\\
\includegraphics[width=0.40\textwidth]{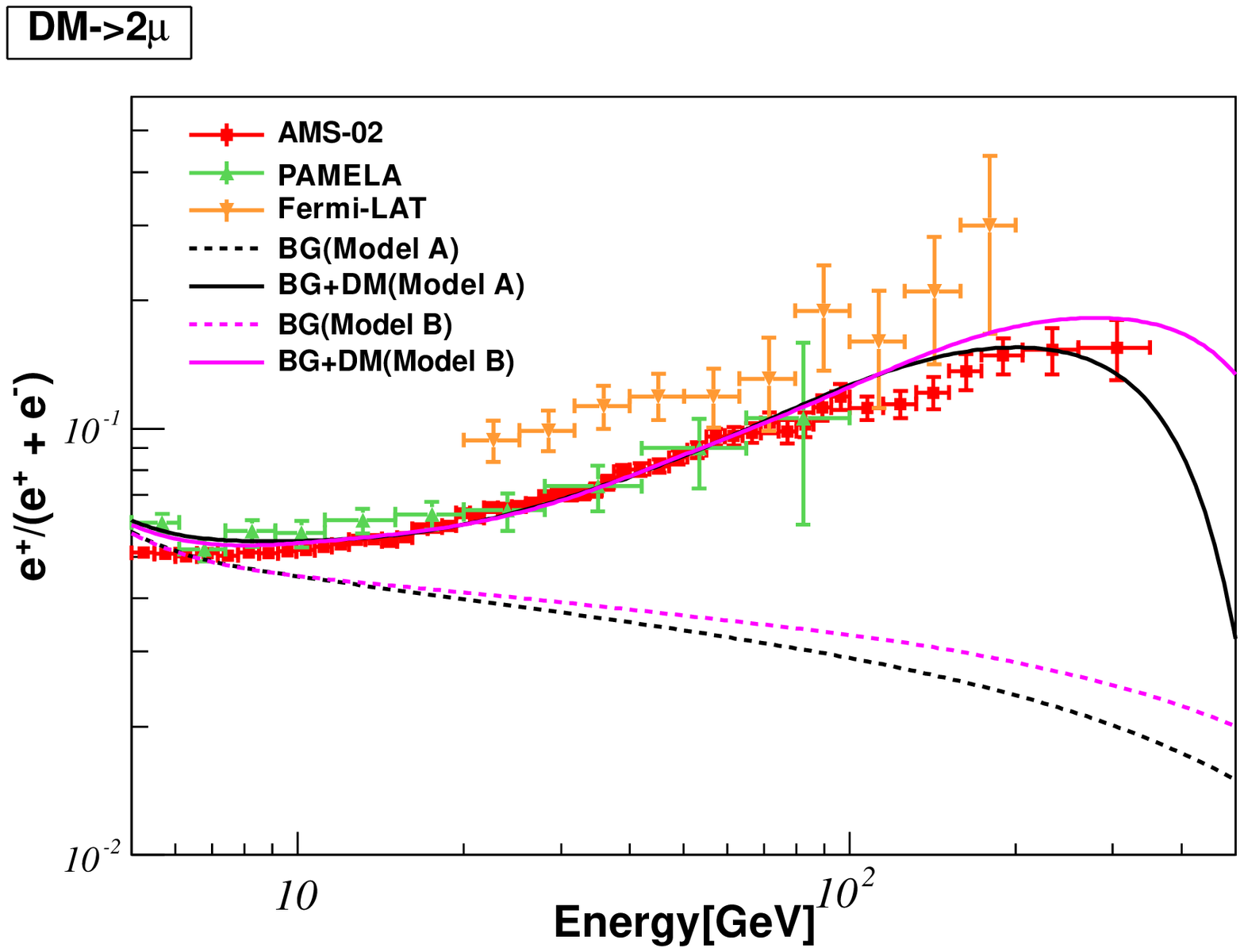}\includegraphics[width=0.40\textwidth]{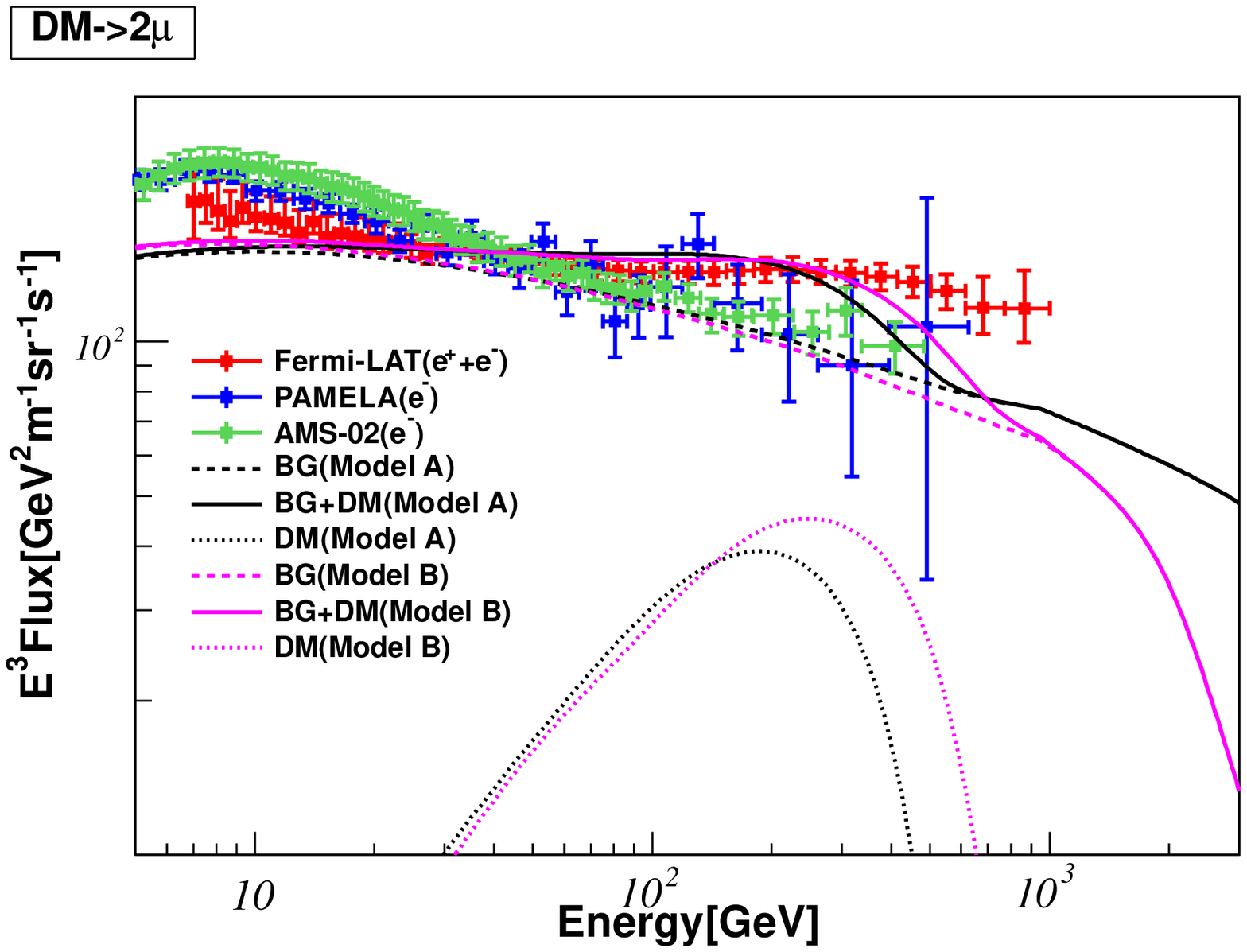}
\\
\includegraphics[width=0.40\textwidth]{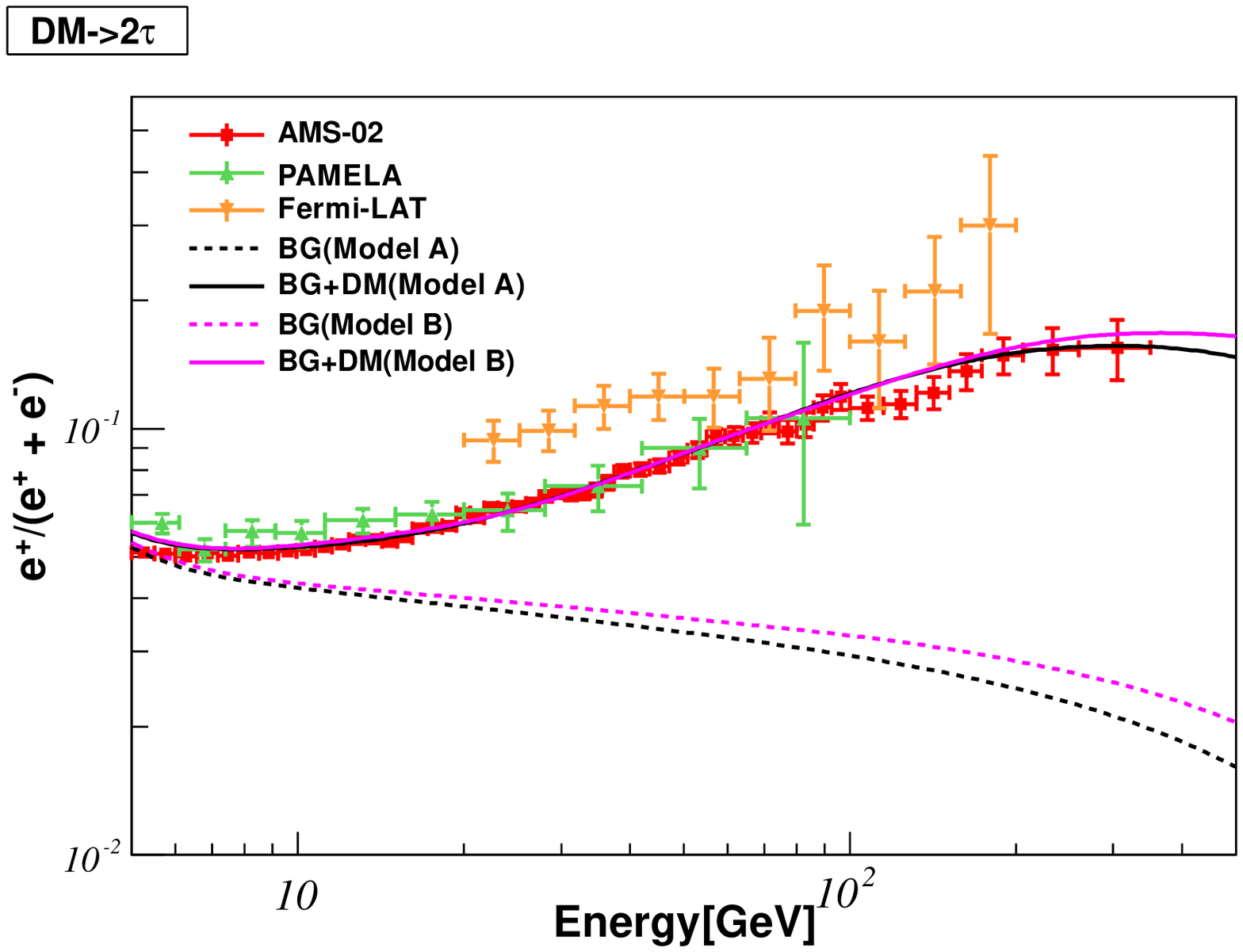}\includegraphics[width=0.40\textwidth]{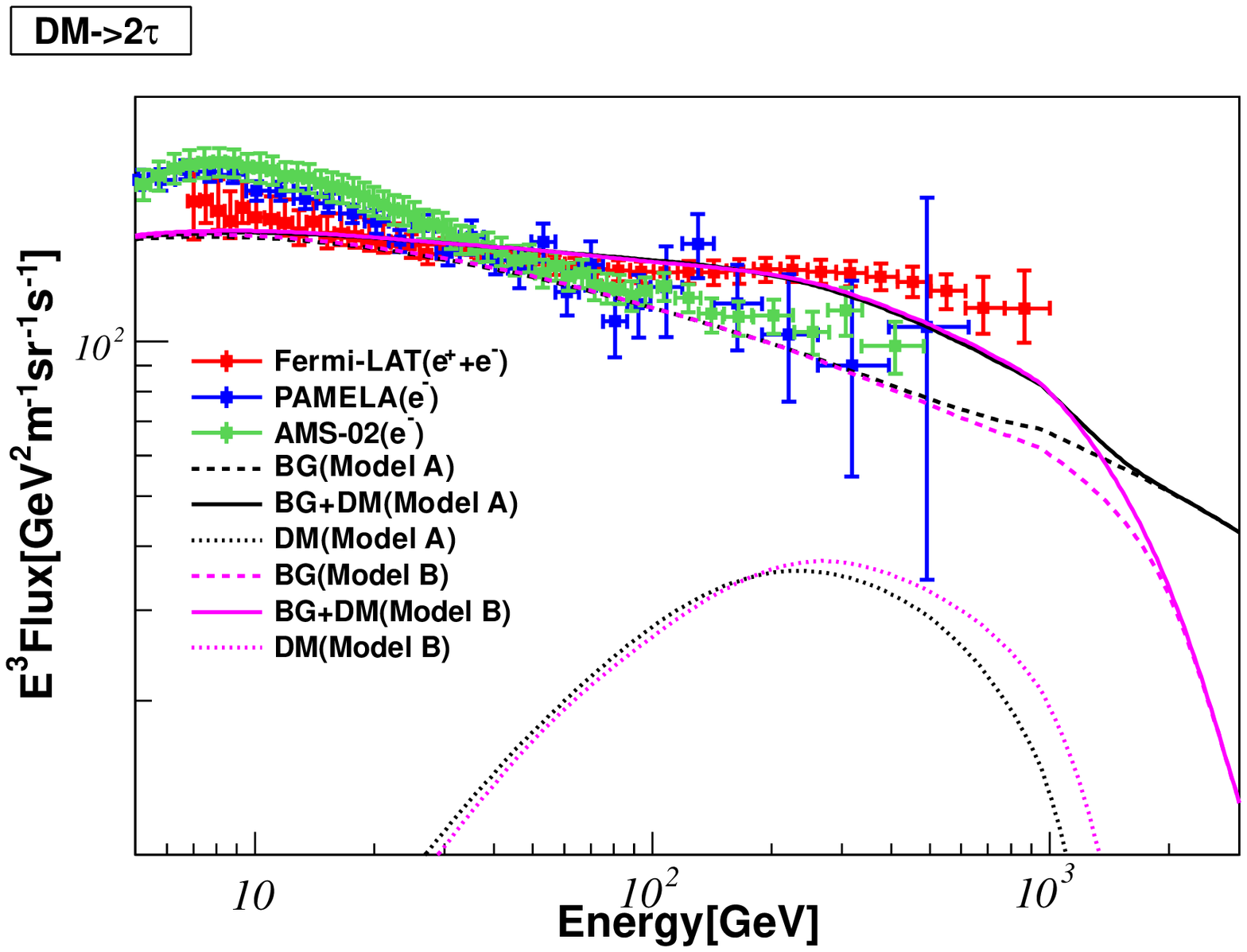}
\caption{
Predicted positron fraction (left) and 
the total flux of electrons and positrons (right) from 
DM annihilating to $2e$, 2$\mu$ and 2$\tau$ final state
according to the best-fit parameters shown in \tab{tab:anni} in Model A and B.
The data  of positron fraction from PAMELA
\cite{1001.3522 }
AMS-02~\cite{PhysRevLett.110.141102}  
and 
Fermi-LAT2012~\cite{1109.0521} 
the total flux from Fermi-LAT~\cite{Ackermann:2010ij}
the electrons flux from PAMELA~\cite{Adriani:2011xv} 
and AMS-02~\cite{AMS02electron}
are also shown.
}
\label{fig:uncertaintiesFlux2FinalAnni}
\end{center}
\end{figure}

\begin{figure}[thb]
\begin{center}
{\includegraphics[width=0.40\textwidth]{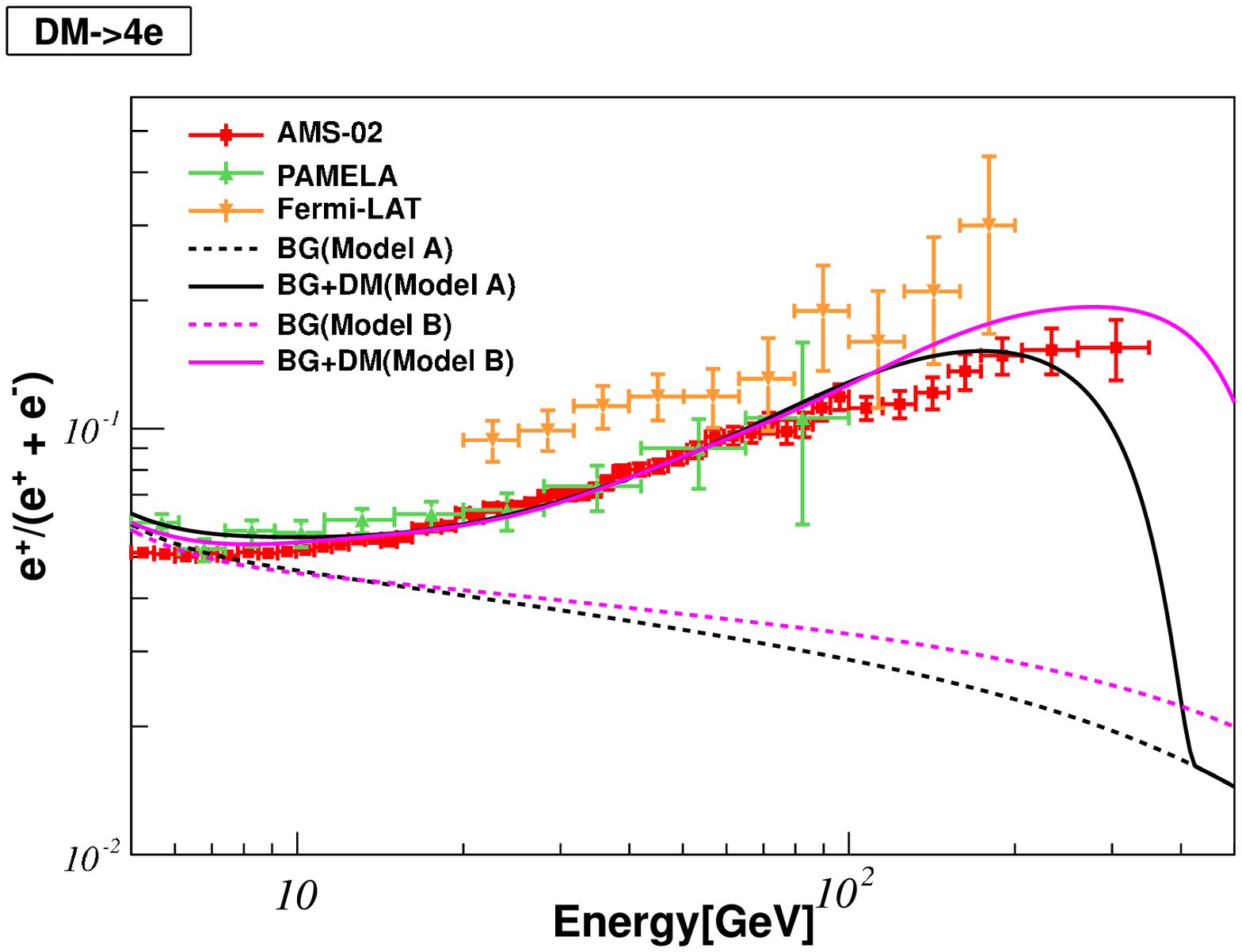}}{\includegraphics[width=0.40\textwidth]{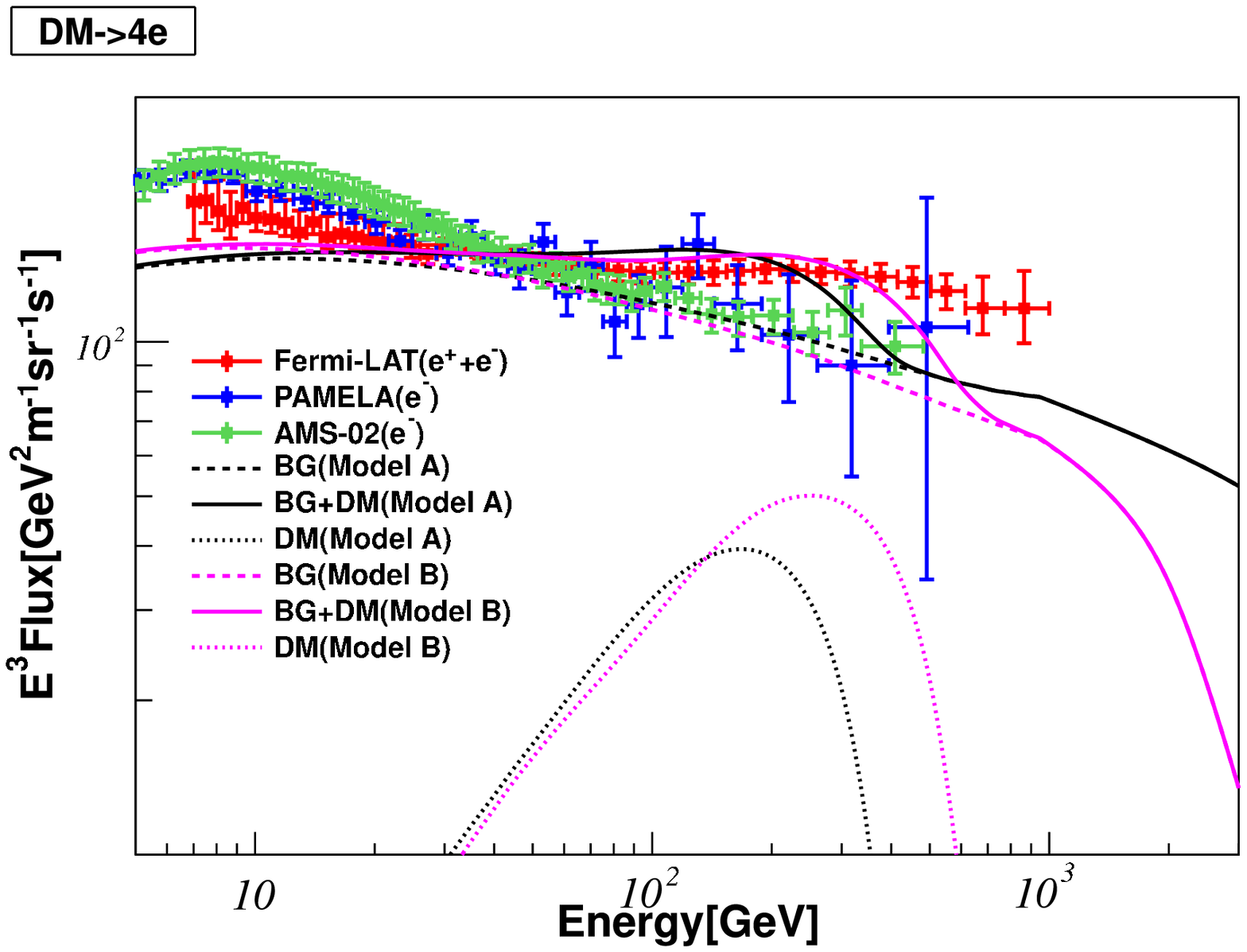}}
\\
{\includegraphics[width=0.40\textwidth]{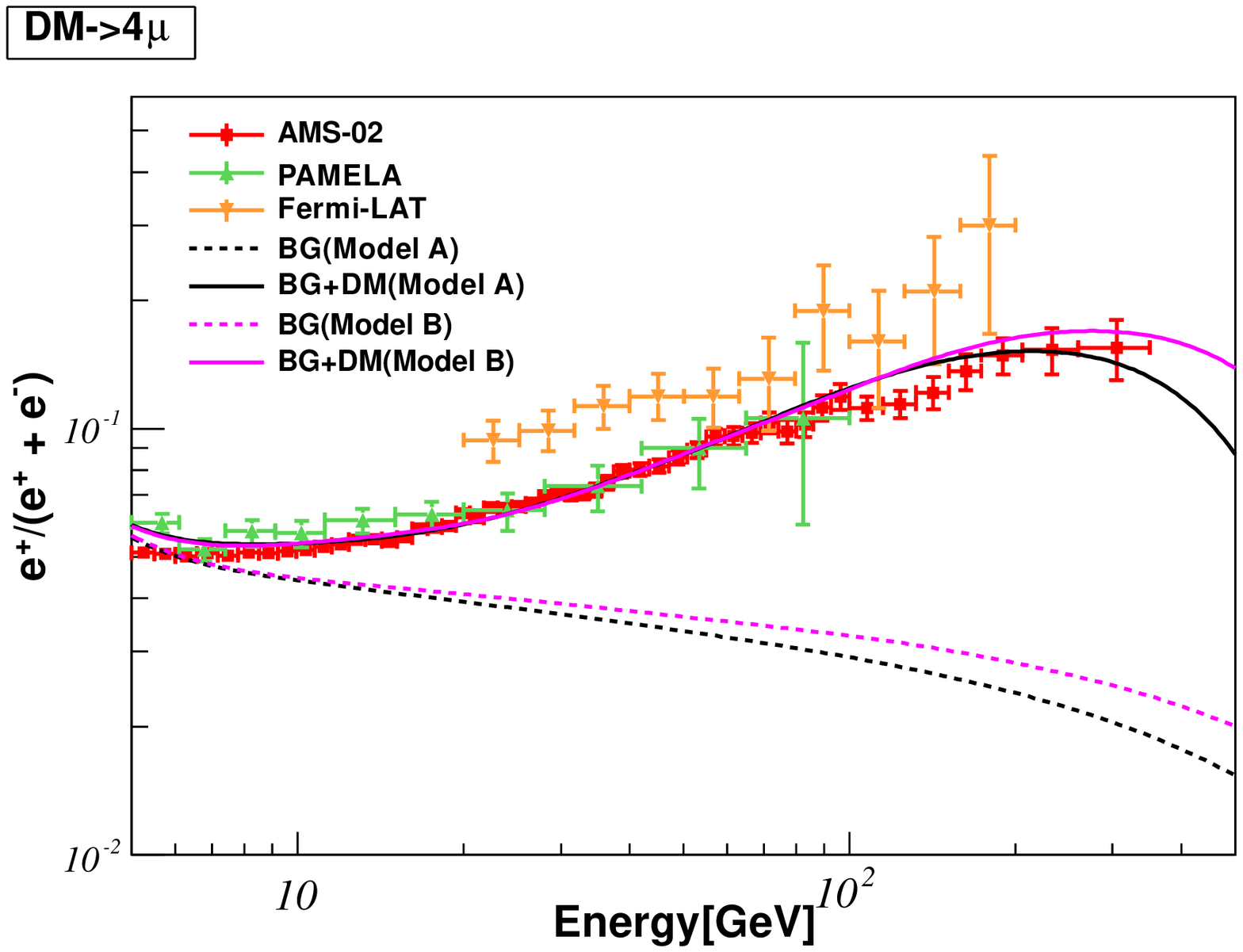}}{\includegraphics[width=0.40\textwidth]{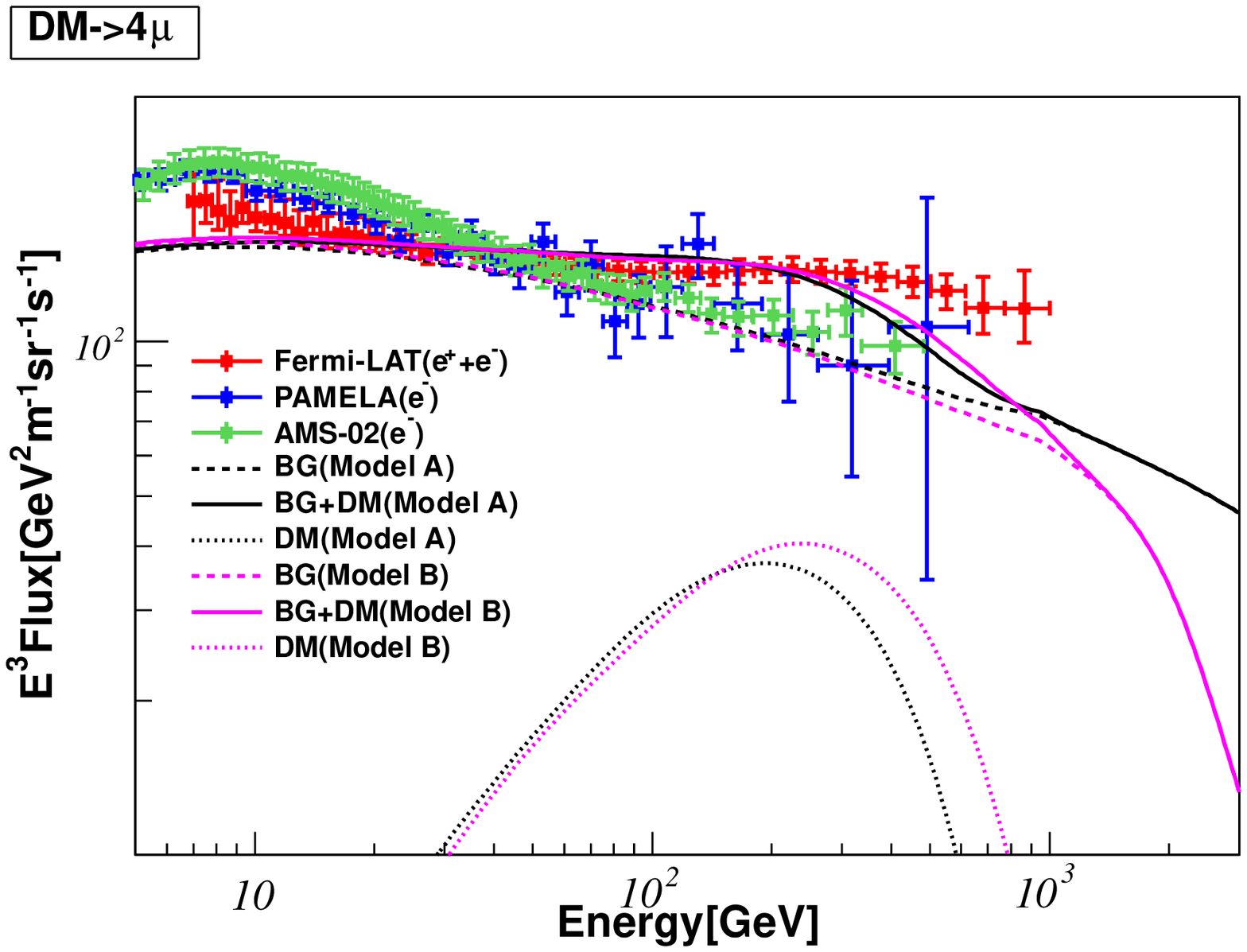}}
\\
{\includegraphics[width=0.40\textwidth]{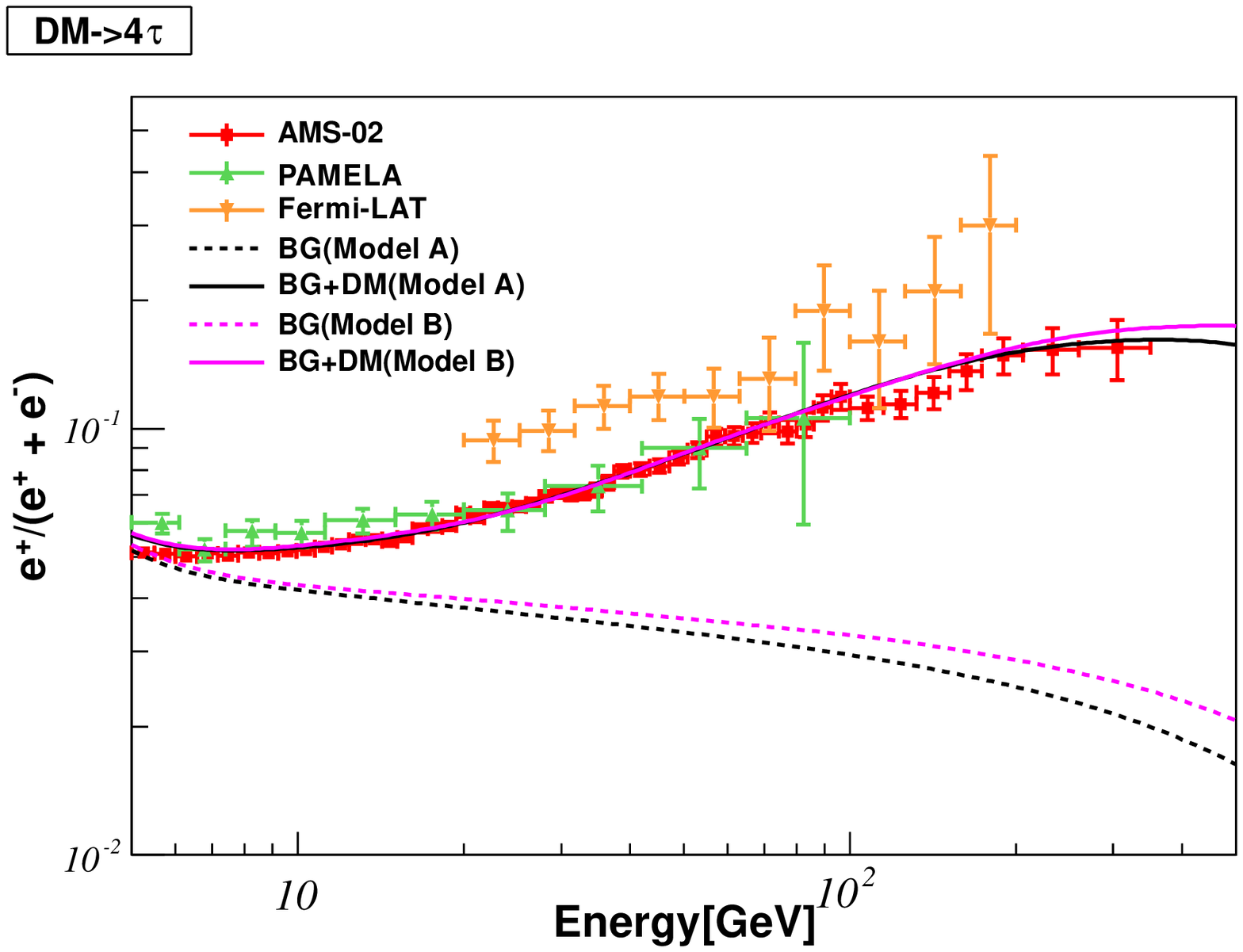}}{\includegraphics[width=0.40\textwidth]{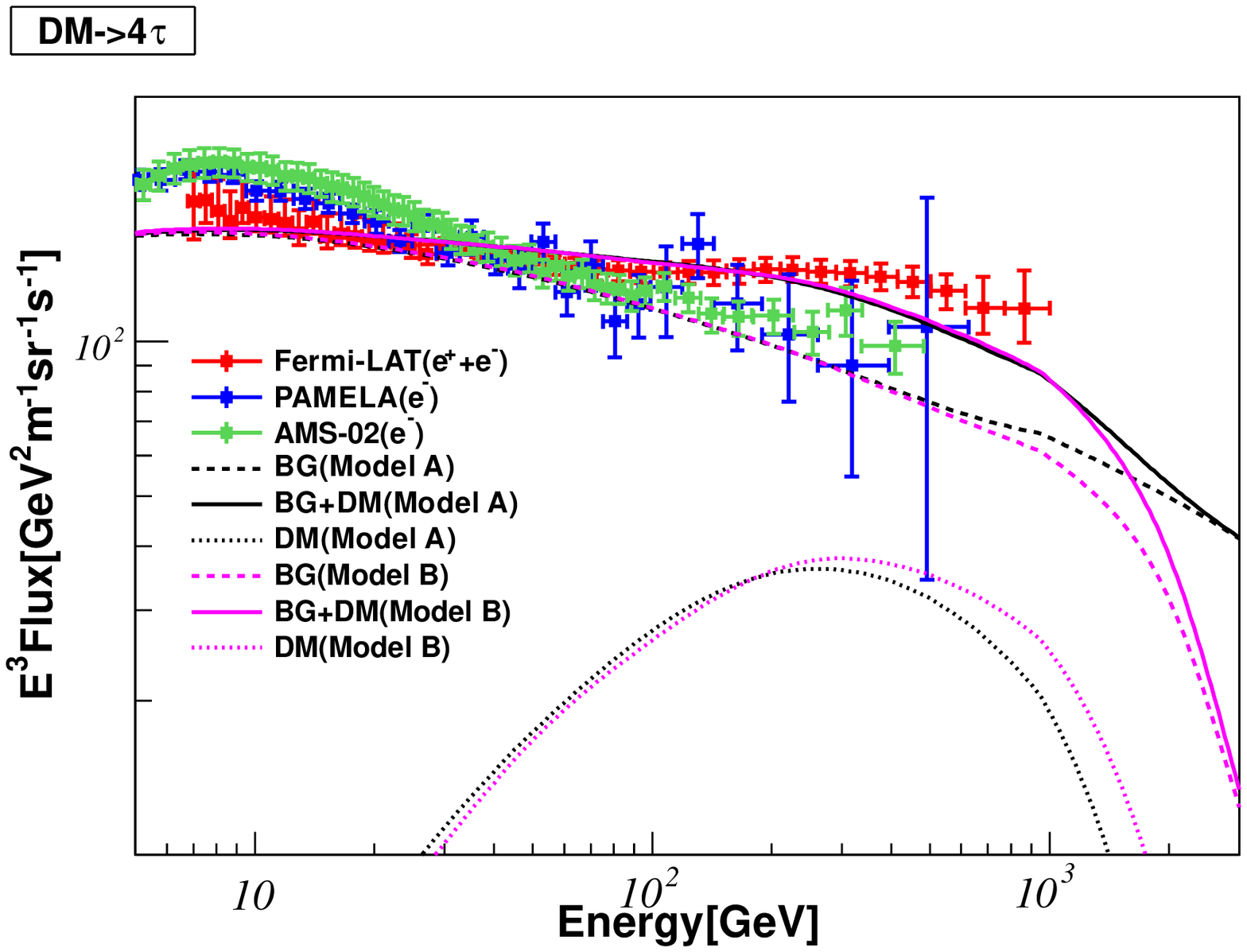}}
\caption{
The same as \fig{fig:uncertaintiesFlux2FinalAnni}, 
but for DM annihilating into $4e$, 4$\mu$ and 4$\tau$ final states.
}
\label{fig:uncertaintiesFlux4FinalAnni}
\end{center}
\end{figure}

\begin{figure}[htb]\begin{center}
\includegraphics[width=0.4\textwidth]{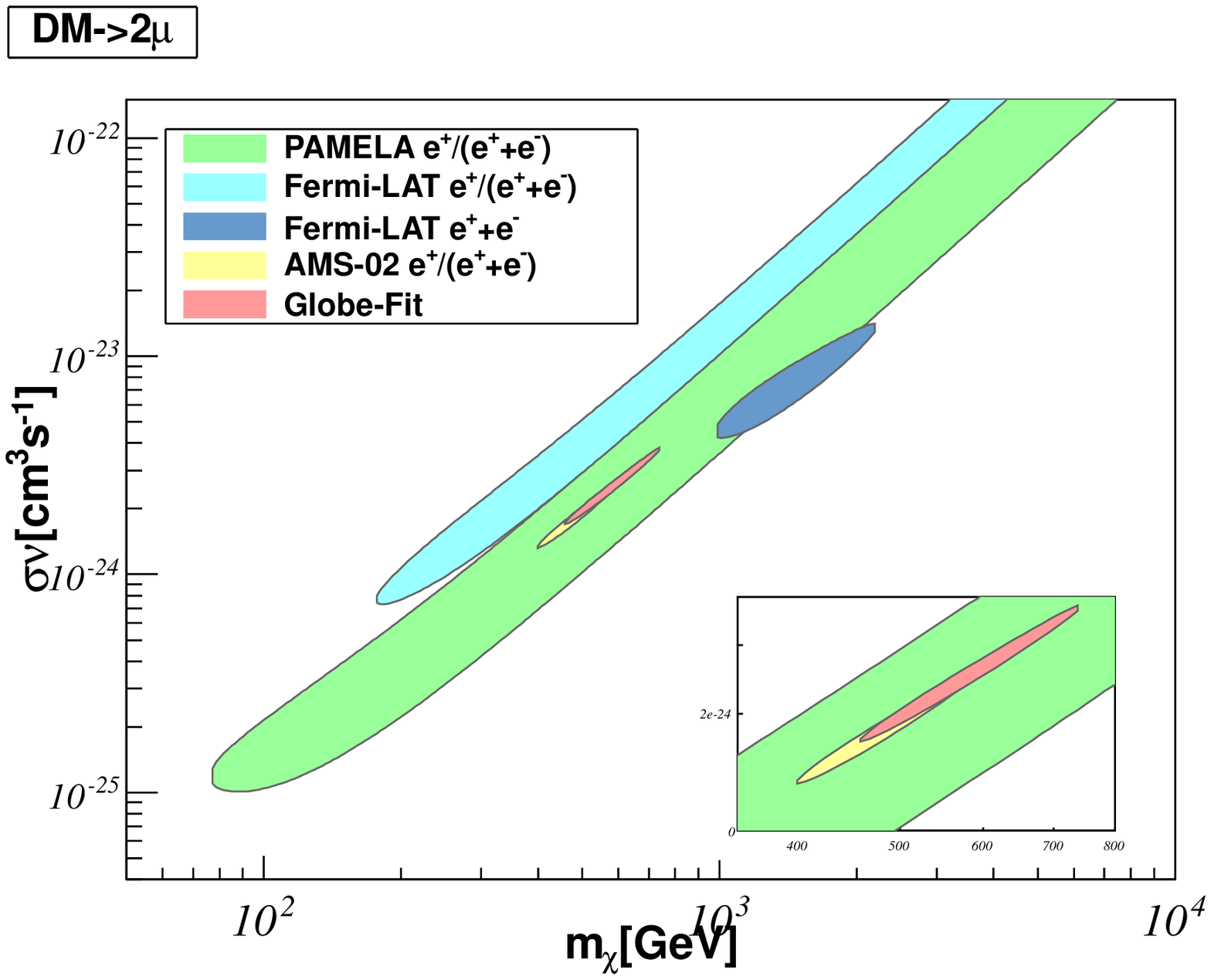}\includegraphics[width=0.4\textwidth]{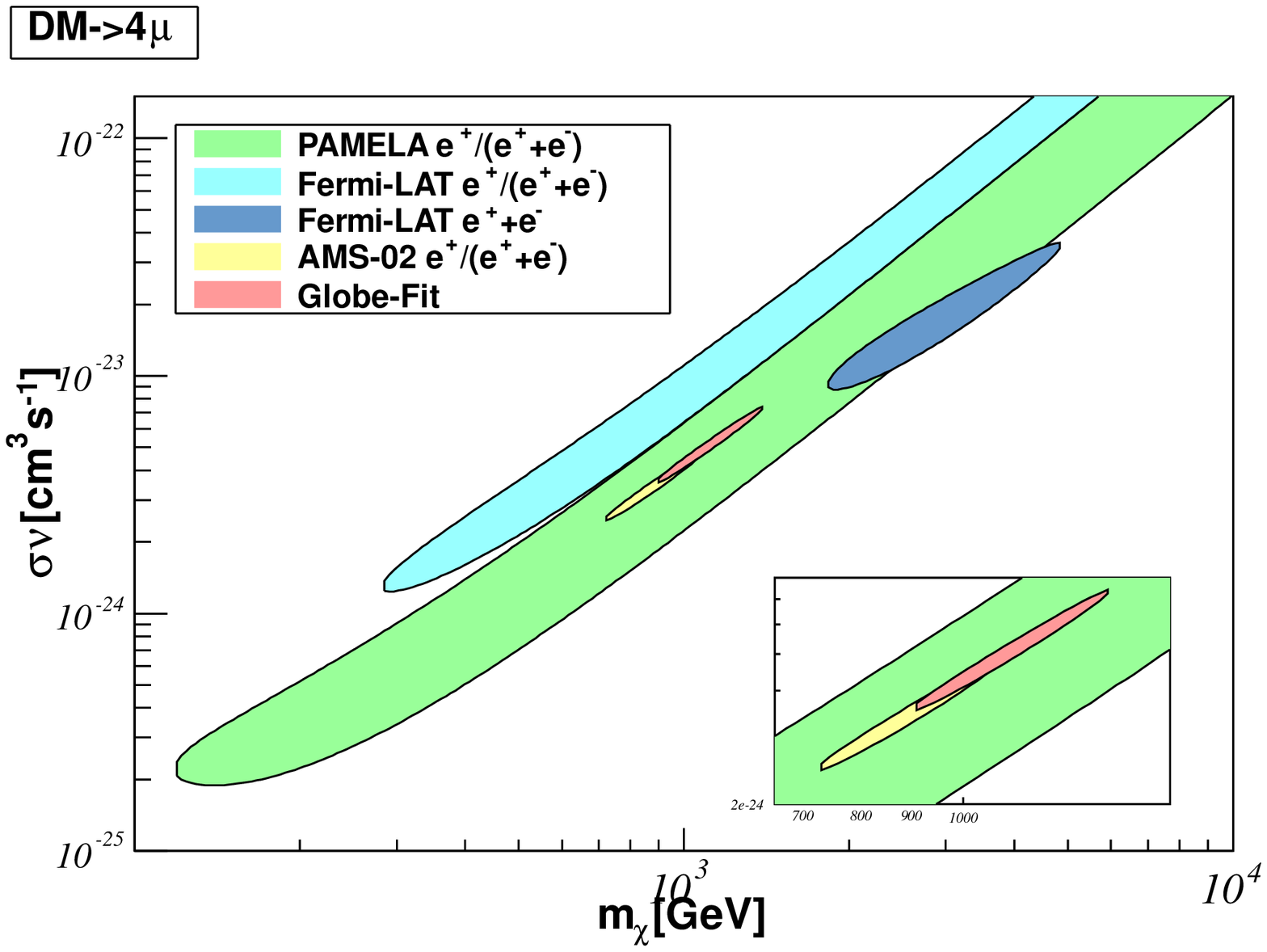}
\\
\includegraphics[width=0.4\textwidth]{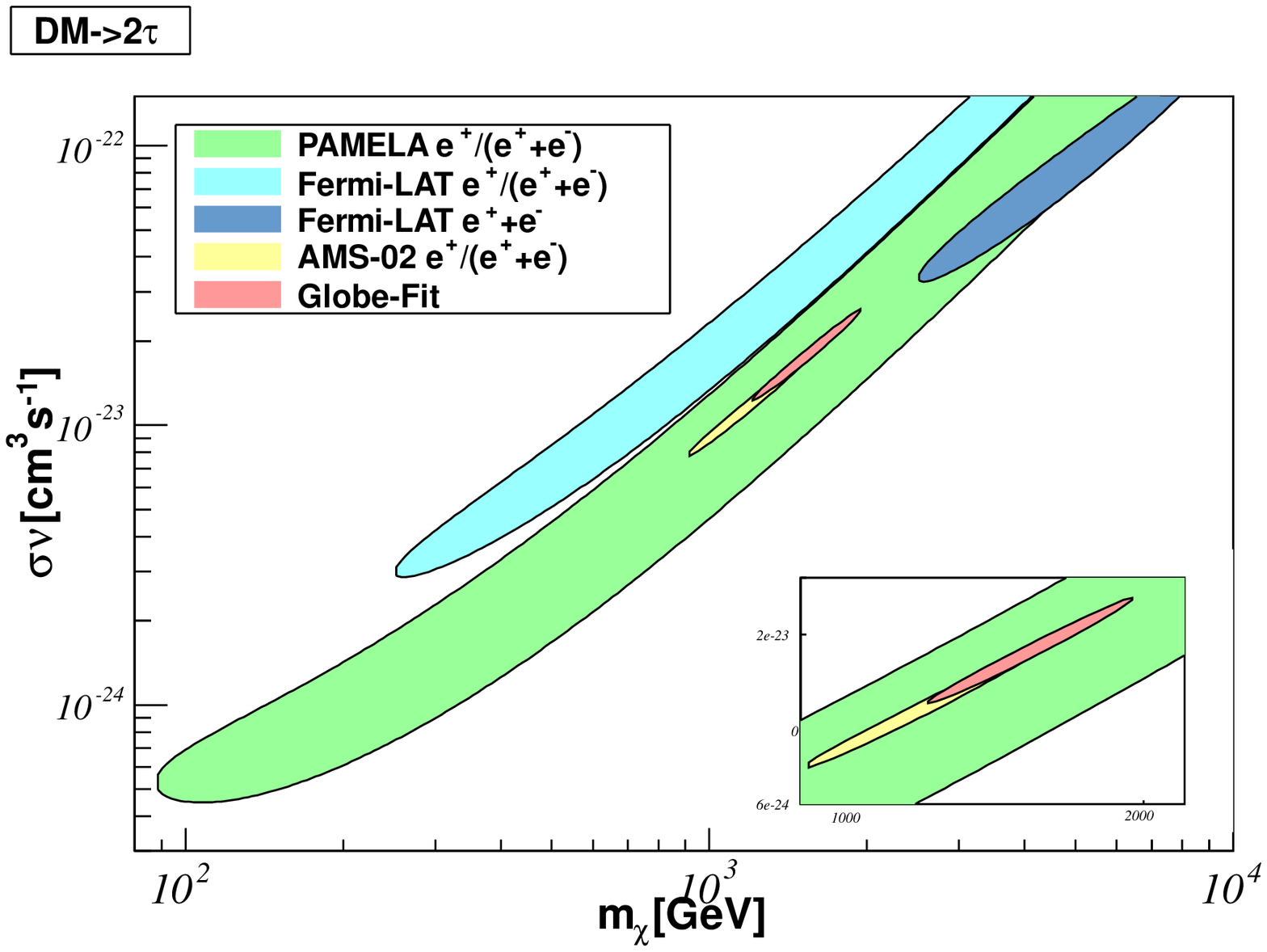}\includegraphics[width=0.4\textwidth]{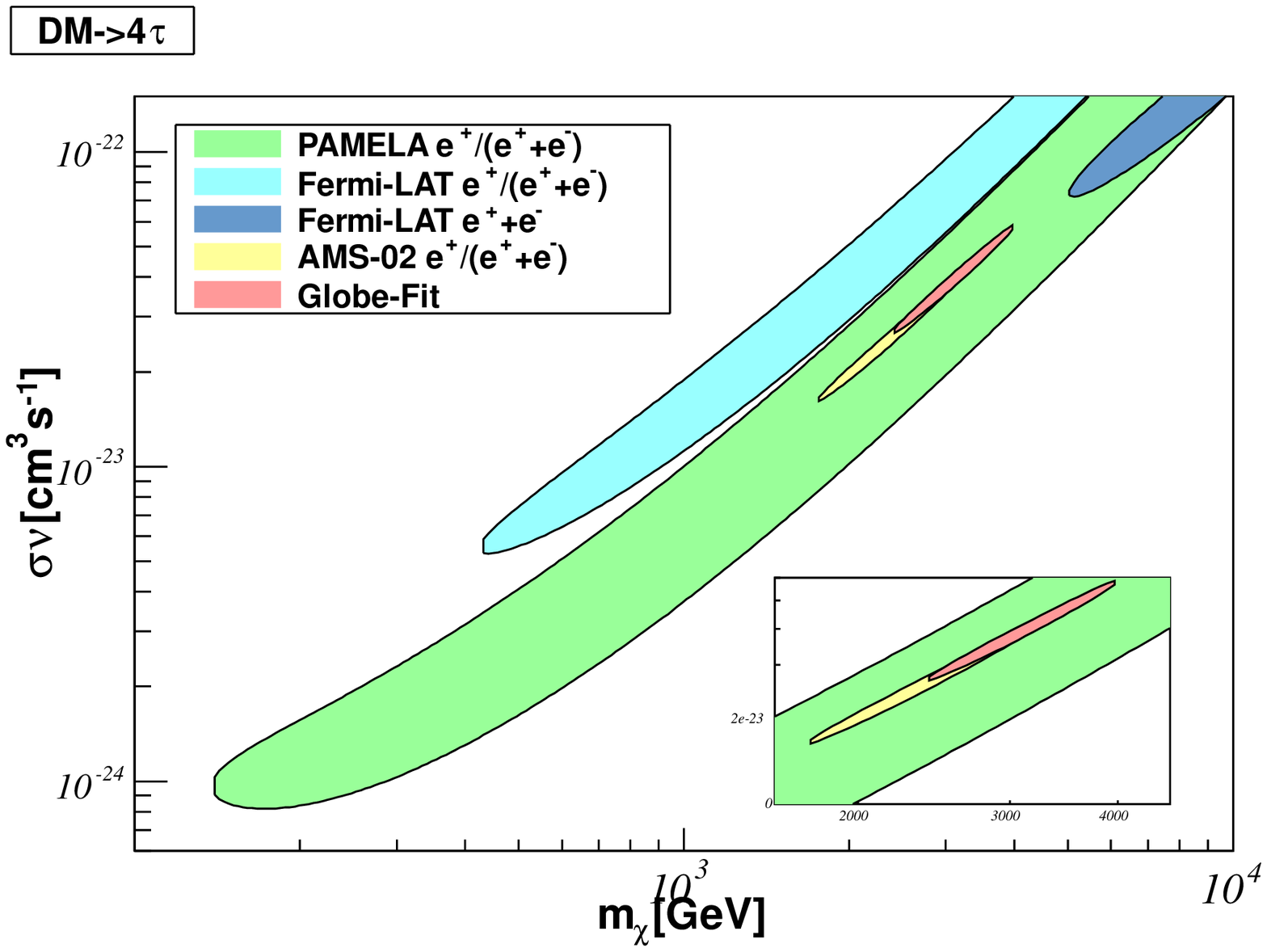}
\caption{ 
Allowed regions in ($m_\chi,\ \langle \sigma v\rangle$)  plane at $99\%$ C.L. 
for DM annihilating into  $2\mu$, $2\tau$, $4\mu$ and $4\tau$ final states 
from the Global fits.
The regions allowed by  data of
PAMELA 
\cite{1001.3522 
}, 
Fermi-LAT \cite{
1109.0521, 
Ackermann:2010ij 
} 
and 
AMS-02 \cite{
PhysRevLett.110.141102
}
are  shown for a comparison.
The  AMS-02 favoured regions  are also displayed in the insets.
See text for explanations. 
}\label{fig:ann}
\end{center}\end{figure}

\begin{figure}[thb]
\begin{center}
\includegraphics[width=0.45\textwidth]{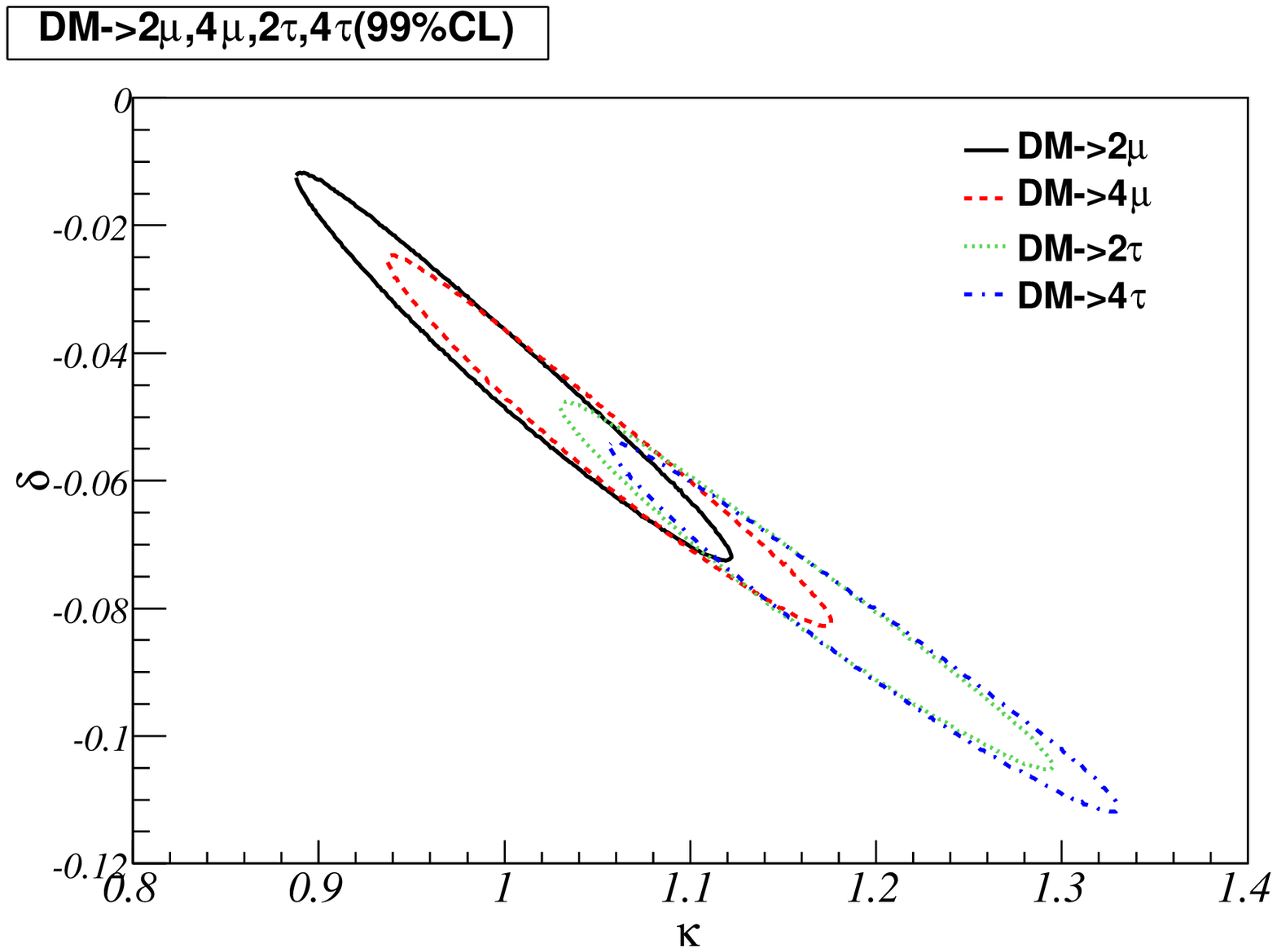}
\includegraphics[width=0.45\textwidth]{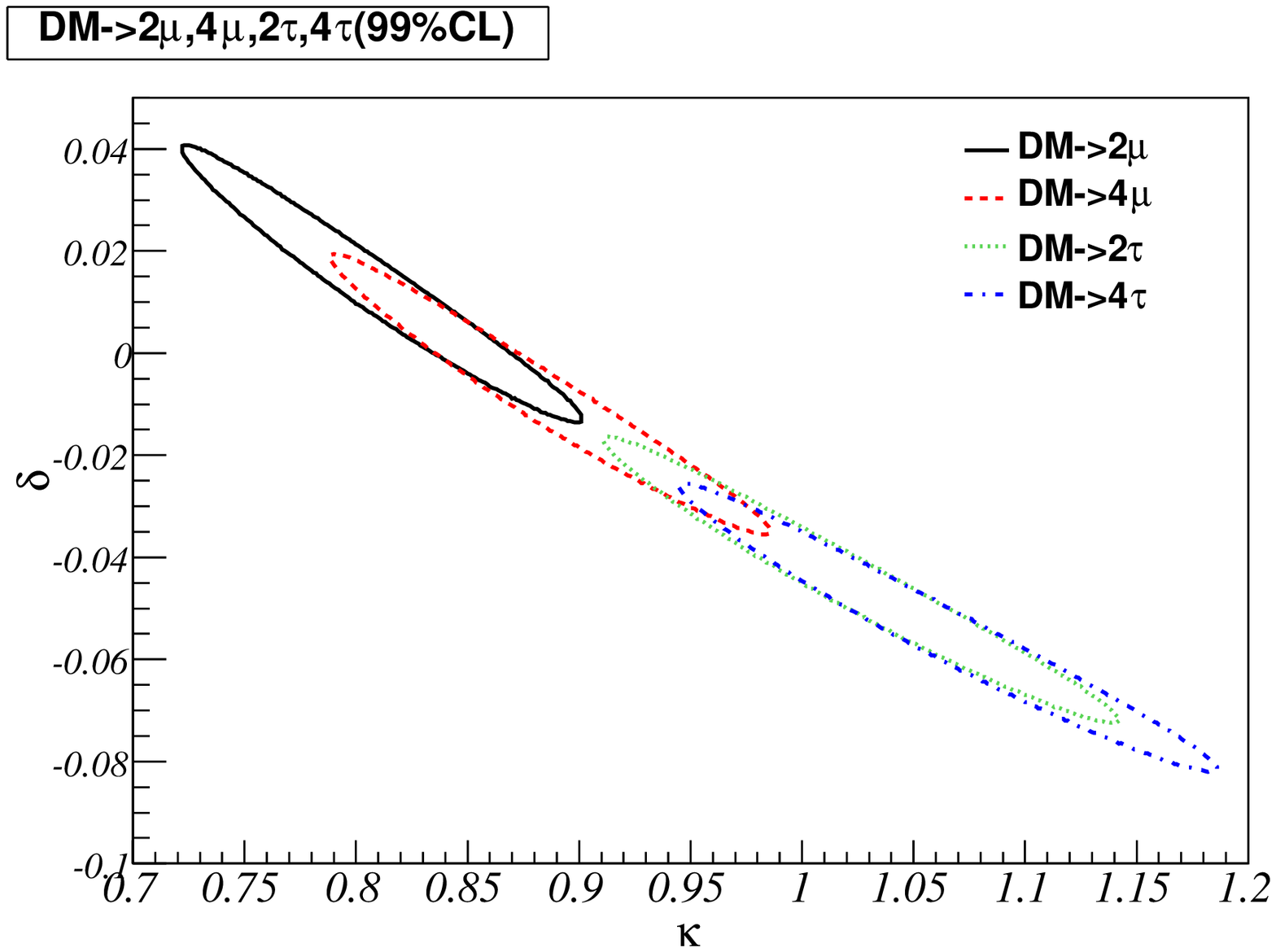}
\end{center}
\caption{(Left) Contours of allowed regions in ($\kappa$, $\delta$) plane at $99\%$ C.L.,
from global fits for DM annihilating into  $2\mu$,  $2\tau$, $4\mu$ and $4\tau$   
final states; (Right) The same as left, but for the case of DM decay.
 } 
 \label{fig:delta_anni}
\end{figure}

We first consider DM annihilation into  charged leptons in Model A and B.
The  best-fit parameters and the corresponding $\chi^{2}/\text{d.o.f}$ value
for each annihilation channel  are given in \tab{tab:anni}.
%
%
%
\begin{table}[tb]
\begin{center}
\begin{tabular}{cccccc}
  \hline\hline
Channel & $m_\chi$(GeV) & 
$\langle \sigma v \rangle $  
& $\kappa$ & 
$\delta(\times 10^{-2})$ &$\chi^2_{\text{tot}}/$d.o.f\\
\hline
$2e$ 	&407.1  & 67.8   &1.064  &-6.43      &450.56/119\\
	   	&     404.9&  55.9&   1.079&  -7.72&      403.40/119\\ 
\hline
$2\mu$	  &570.0  &244   &0.997  &-4.12      &343.25/119\\
              	& 793.8&  387 &  1.136&  -8.71&      299.60/119\\
\hline
$2\tau$ 	&1534.3         &1780  &1.154  &-7.62     &219.67/119\\
			& 1860.1& 2230 &  1.234&  -10.4&      210.78/119\\
\hline
4e      	&423.5  & 59.0  &0.924  &-2.25      &415.21/119\\
		&664.2&  115&   1.106&  -8.22&      355.25/119\\
\hline
$4\mu$  	&1095.7         &497   &1.049  &-5.32     &290.18/119\\
			& 1409.7& 690 &  1.158&  -9.01&      262.22/119\\
\hline
$4\tau$ &3068.4         &3860          &1.186  &-8.26      &205.72/119\\
			&3794.3& 4980 & 1.260&  -10.9&      199.29/119\\
  \hline\hline
\end{tabular}
\end{center}
\caption{
Best-fit values of parameters $m_{\chi}$, $\langle \sigma v \rangle$, $\kappa$ and $\delta$,
as well as the $\chi^{2}/\text{d.o.f}$ for DM particles annihilating into 
$2e$, $2\mu$, $2\tau$, $4e$, $4\mu$ and $4\tau$ final states.
For each final states, the values in the first (second) row corresponds to the results in Model A (B).
The cross section $\langle \sigma v \rangle $ is in units of $10^{-26} \text{ cm}^{3}\text{s}^{-1}$.
}\label{tab:anni}
\end{table}
The predicted spectra of the positron fraction and the total flux of 
electrons and positrons corresponding to the best-fit parameters 
are shown in \fig{fig:uncertaintiesFlux2FinalAnni} and 
\fig{fig:uncertaintiesFlux4FinalAnni}, respectively.
%
In general,  
the qualities of the fits are not good for 
final states with electrons and muons. 
Among all the channels, 
only the $2\tau$ and $4\tau$ channels have $\chi^2/\text{d.o.f} < 2$.
For $2e$ final states, 
the large $\chi^{2}/\text{d.o.f}=3.67\ (3.28)$ in Model A (B) indicates 
a high inconsistency between the theoretical expectation and the experimental data.
The best-fit values are   $m_\chi\approx 407\ (405)$ GeV and 
$\langle \sigma v\rangle \approx 6.8 \ (5.6)\times 10^{-25}\text{ cm}^3\text{s}^{-1}$
in Model A (B).
It is known that the spectra of DM annihilation into $2e$ and $4e$ are 
too sharp to fit the measured relatively smooth fluxes,
which can be seen  clearly in \fig{fig:uncertaintiesFlux2FinalAnni}
and \fig{fig:uncertaintiesFlux4FinalAnni}. 
Thus in the remainder of this section we should focus on DM annihilation/decay into  $\mu$ and $\tau$ final states.
The contours for the allowed regions from the global fit for parameters  ($m_\chi,\langle \sigma v\rangle$)
and $(\kappa,\delta)$ at  $99\%$ C.L.  corresponding to 
$\Delta \chi^2=9.21$ for two variables are shown in \fig{fig:ann} and \fig{fig:delta_anni}, 
respectively. 
%
%
%
%
For all the final states, 
the favoured DM annihilation cross sections are  
larger than the typical thermal WIMP annihilation cross section 
$\langle \sigma v \rangle_{0}\approx 3\times 10^{-26} \text{ cm}^{3}\text{s}^{-1}$
by 2--3 orders of magnitude, 
especially for $2\tau$ and $4\tau$ cases.
The figures show that 
the values of $\langle \sigma v \rangle$ roughly scales with DM particle mass as $m_{\chi}^{2}$,
which is due to the term $\langle \sigma v \rangle/m_{\chi}^{2}$ in the source term in \eq{eq:ann-source}.
The the allowed values of  $\kappa$ and $\delta$ are mostly determined by the electron data 
from AMS-02~\cite{AMS02electron} and  PAMELA~\cite{Adriani:2011xv}. 
%
As shown in \fig{fig:delta_anni}, 
there exists a negative correlation between $\kappa$ and $\delta$, 
as they appear as a combination $\kappa (E/E_{0})^{\delta}$. 
For a given value of  $E$, increasing the value of $\kappa$ leads to a decrease of $\delta$. 

\newpage
%

In \fig{fig:ann}, we also show the  regions allowed by each single measurement 
such as  AMS-02, PEMALA and Fermi-LAT.
As already discussed in the previous section,
the  measurement of positron fraction alone can hardly constrain the value of $\kappa$ and $\delta$.
%
%
We take the values of $\kappa$ and $\delta$ as inputs, 
with central values and uncertainties  determined from each global fit.
The correlations between $\kappa$ and $\delta$ are 
taken into account using the covariance matrix calculated  from 
the global fit using the MINUIT package,
which are consistent with \fig{fig:delta_anni}.
In all the figures, 
there is a visible difference between the AMS-02 favoured region and 
that from the global fit, which is due to the stronger $\kappa$-dependence of 
the positron fraction than that of the total flux of electrons and positrons.
 
%
%

For $2\mu$ final states, 
the favoured  DM particle mass  and the  cross section from the global fit are:
$m_\chi\approx 570\ (794)$ GeV 
and 
$ \langle \sigma v\rangle \approx 2.4\ (3.9)\times 10^{-24}\text{ cm}^3\text{s}^{-1}$
for model A (B).
Previous fits using PAMELA and Fermi-LAT data showed that 
both the positron fraction and the total flux of electrons and positrons can be 
well fitted for $2\mu$ and $4\mu$ channels 
~\cite{
Cirelli:2008pk,
Bergstrom:2009fa,
Cirelli:2009dv
}.
This conclusion is changed after AMS-02.
Unlike the PAMELA positron fraction data which 
can only determine the annihilation cross section as a function of DM particle mass,
the high precision AMS-02 data actually constrain 
both the  mass  and  annihilation cross section in  a relatively small region.
The first AMS-02 data feature a steady decreasing slope of the positron fraction.
From 20 GeV to $\sim 250$ GeV, 
the slope of the  positron fraction  decreases by an order of magnitude.
Such a trend disfavours 
a  TeV scale DM particle preferred  by the Fermi-LAT data
as for such a heavy DM particle 
the  predicted slope of the spectrum will not drop in 
the low energy region $E\lesssim 300$ GeV.
On the other hand, 
as can be seen in \fig{fig:uncertaintiesFlux2FinalAnni}, 
a low mass DM cannot account for the excesses
in Fermi-LAT data.
%

%
%
%
%
%
%

\begin{figure}[htb]
\begin{center}
\includegraphics[width=0.49\textwidth]{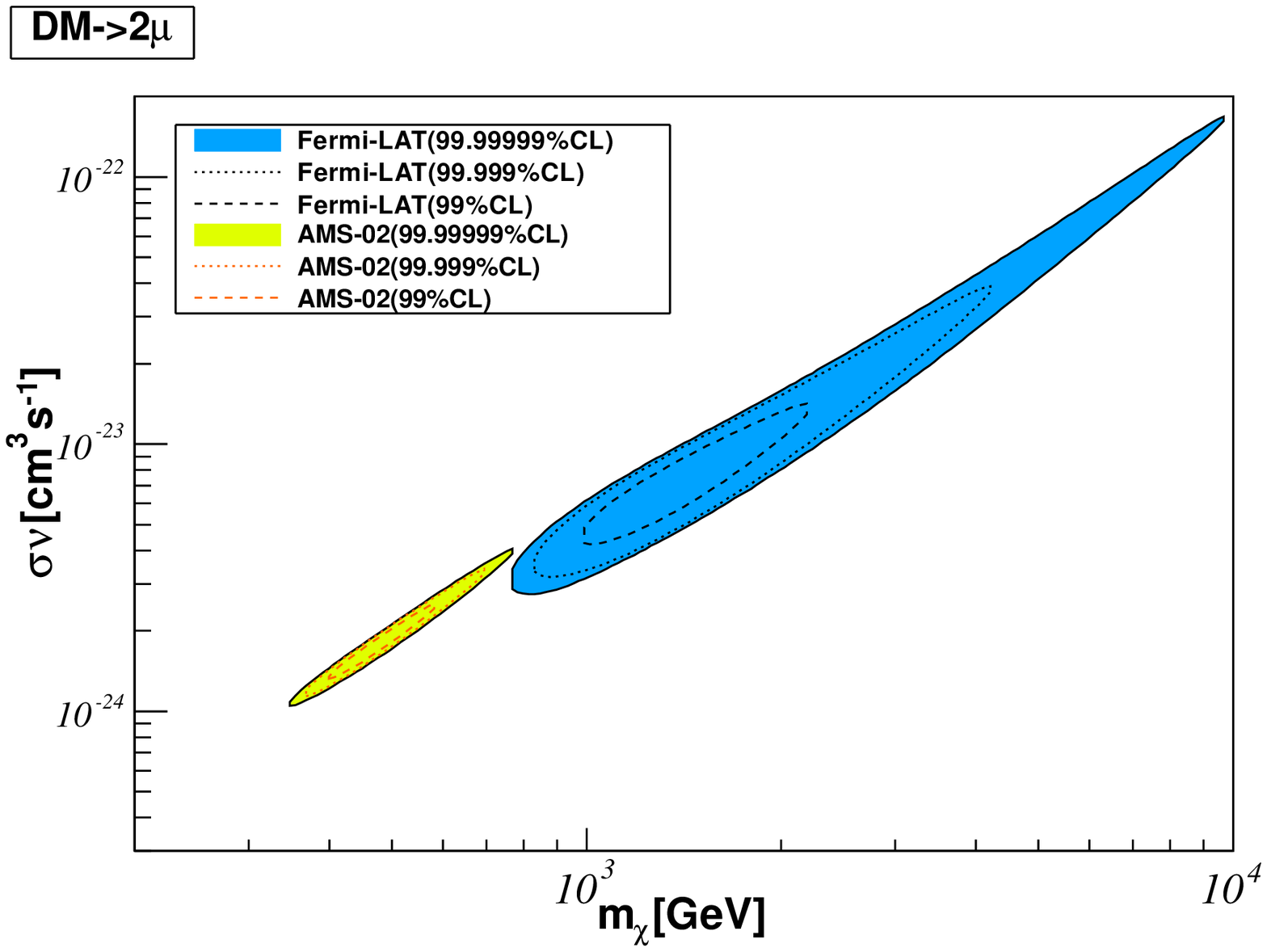}\includegraphics[width=0.49\textwidth]{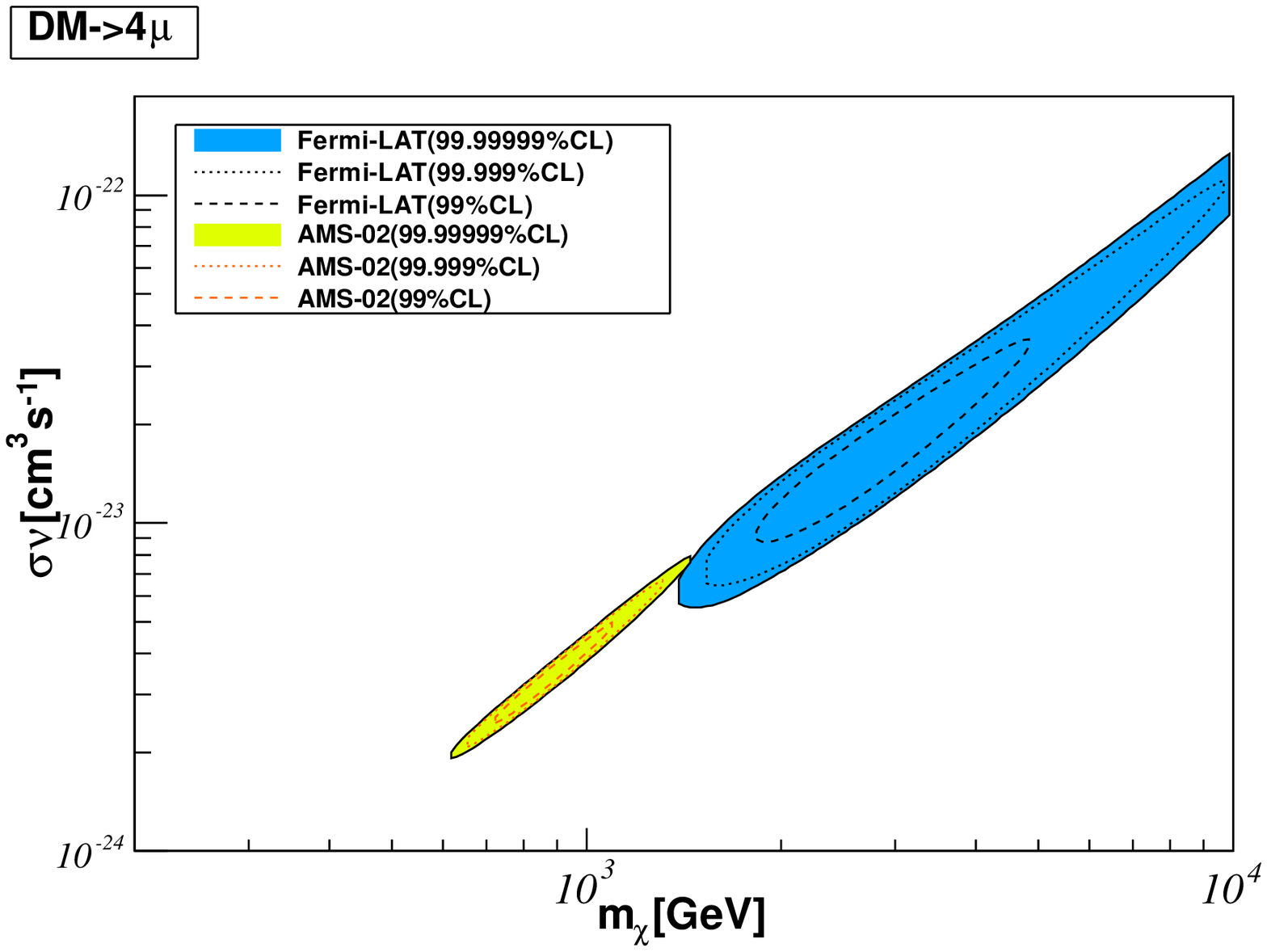}
\\
\includegraphics[width=0.49\textwidth]{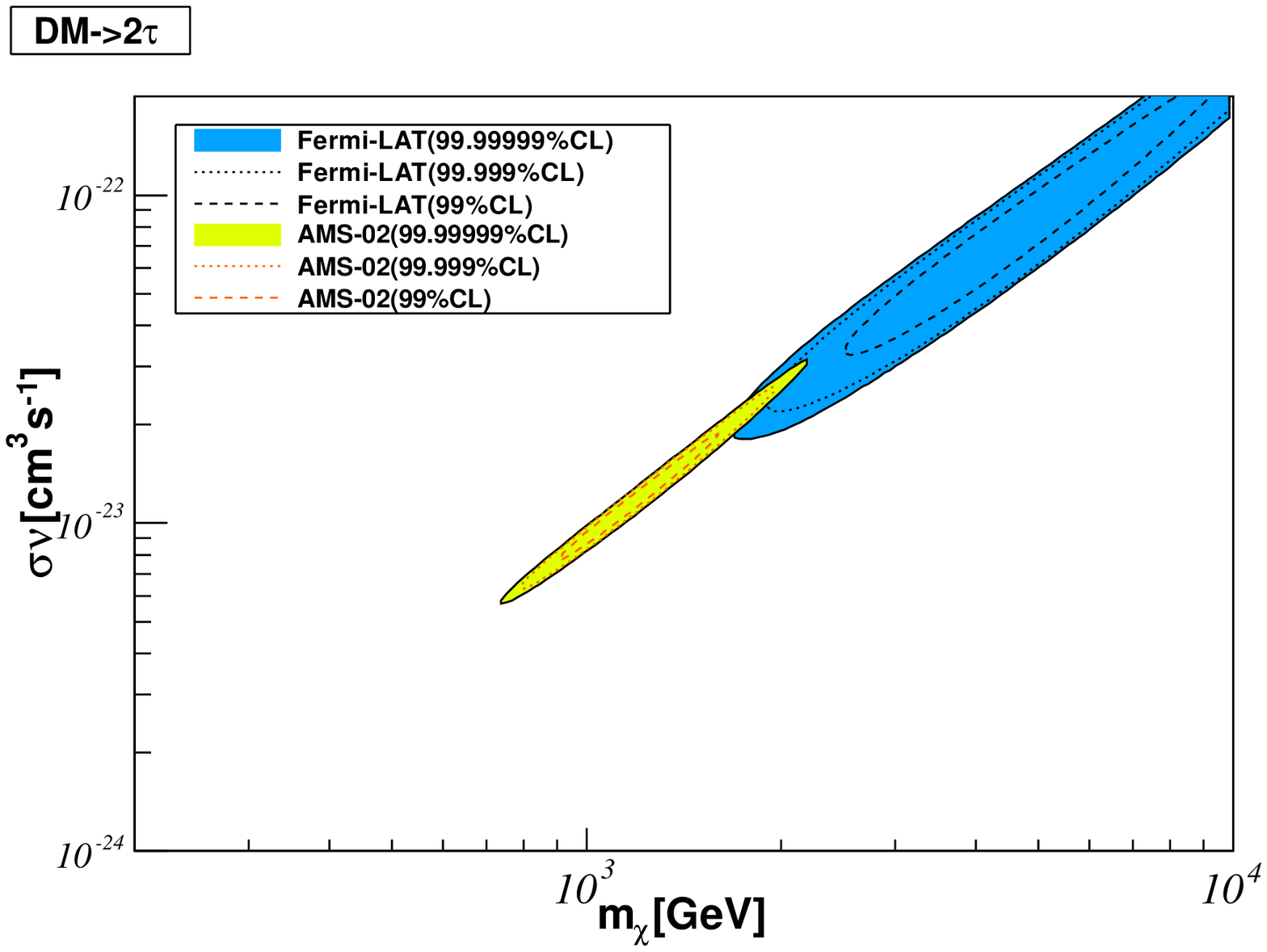}\includegraphics[width=0.49\textwidth]{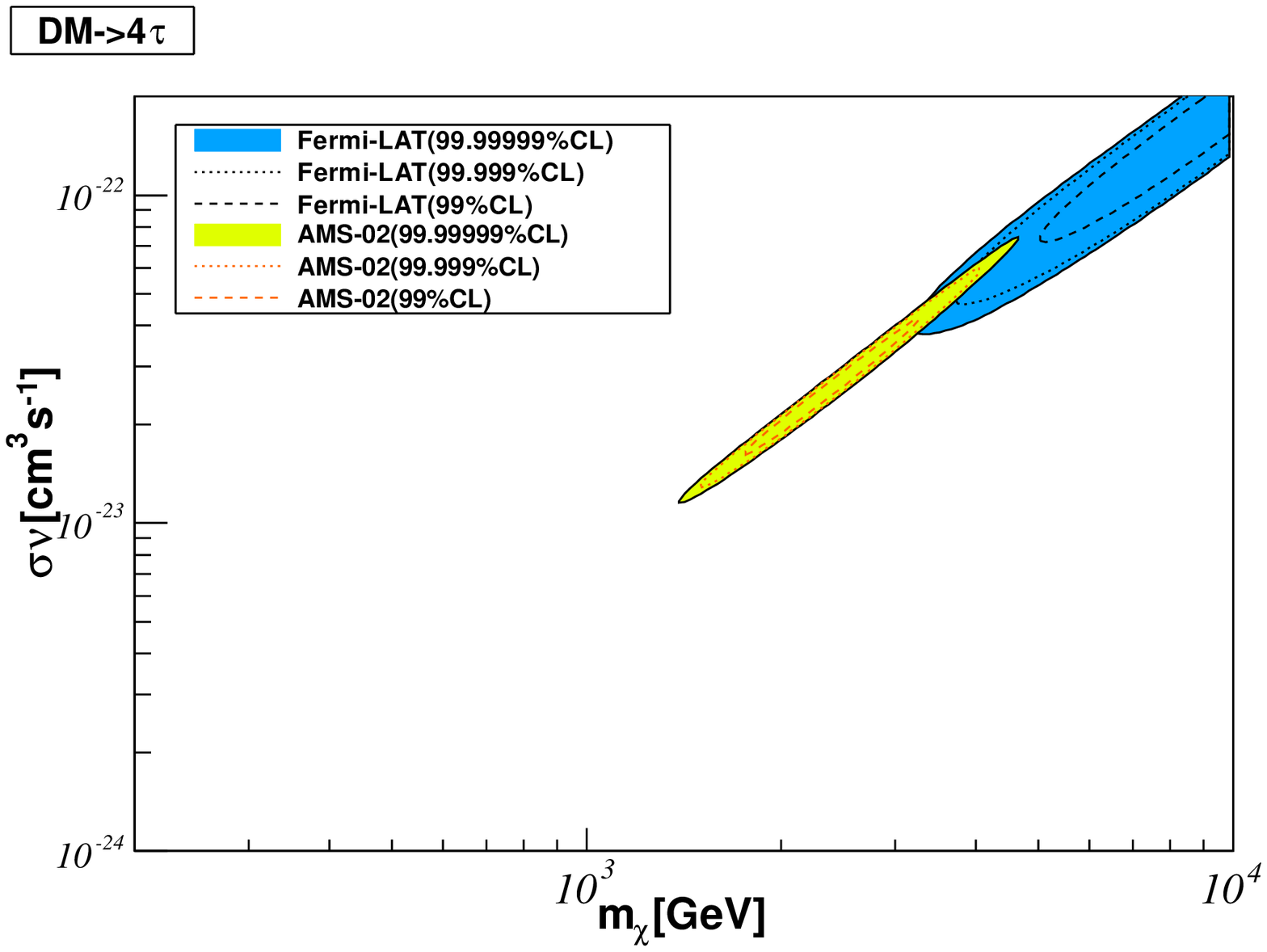}
\caption{Allowed regions by the data of AMS-02 for 
the positron fraction
\cite{
PhysRevLett.110.141102
}
and the data of Fermi-LAT for the total flux of electrons and positrons
\cite{
Ackermann:2010ij 
}
in ($m_\chi,\langle \sigma v\rangle$)  plane at 
$99\%$, $99.999\%$ and $99.99999\%$ C.L. for the case of DM annihilating
into $2\mu$, $2\tau$, $4\mu$, $4\tau$ final states.
 }

\label{fig:uncertaintiesContour99CL}
\end{center}\end{figure}

\begin{figure}
\begin{center}
{\includegraphics[width=0.33\textwidth]{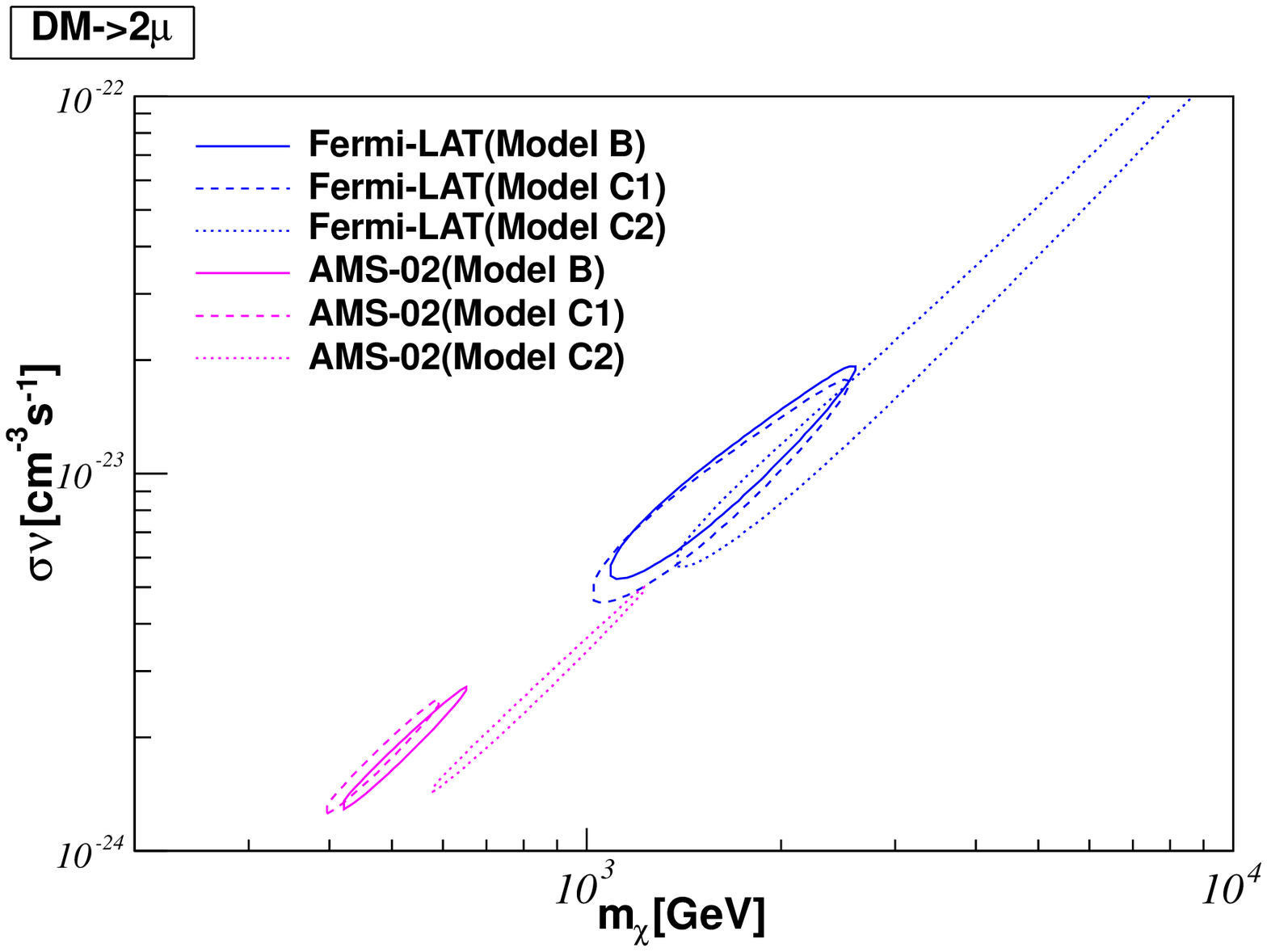}}{\includegraphics[width=0.33\textwidth]{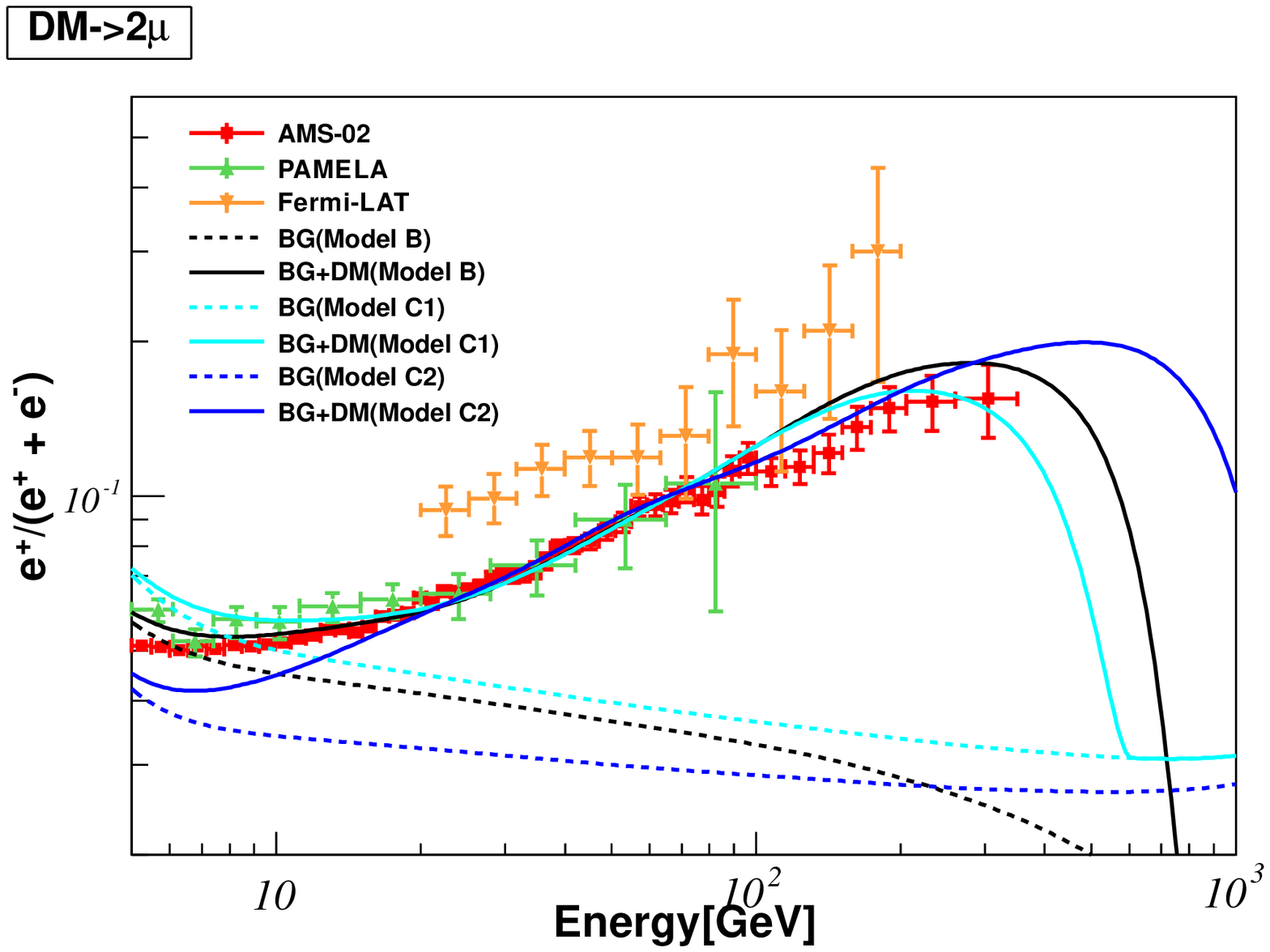}}{\includegraphics[width=0.33\textwidth]{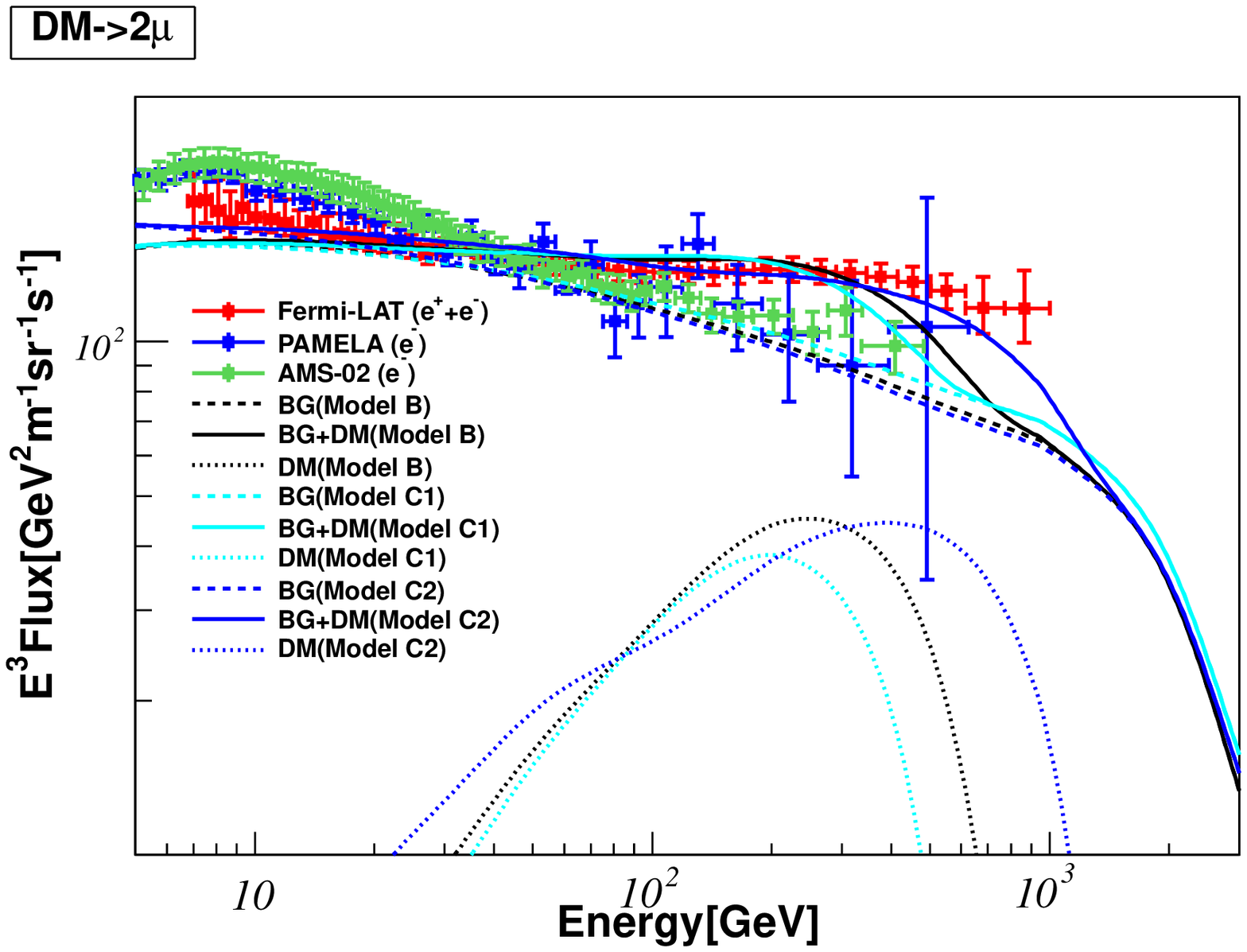}}
\\
{\includegraphics[width=0.33\textwidth]{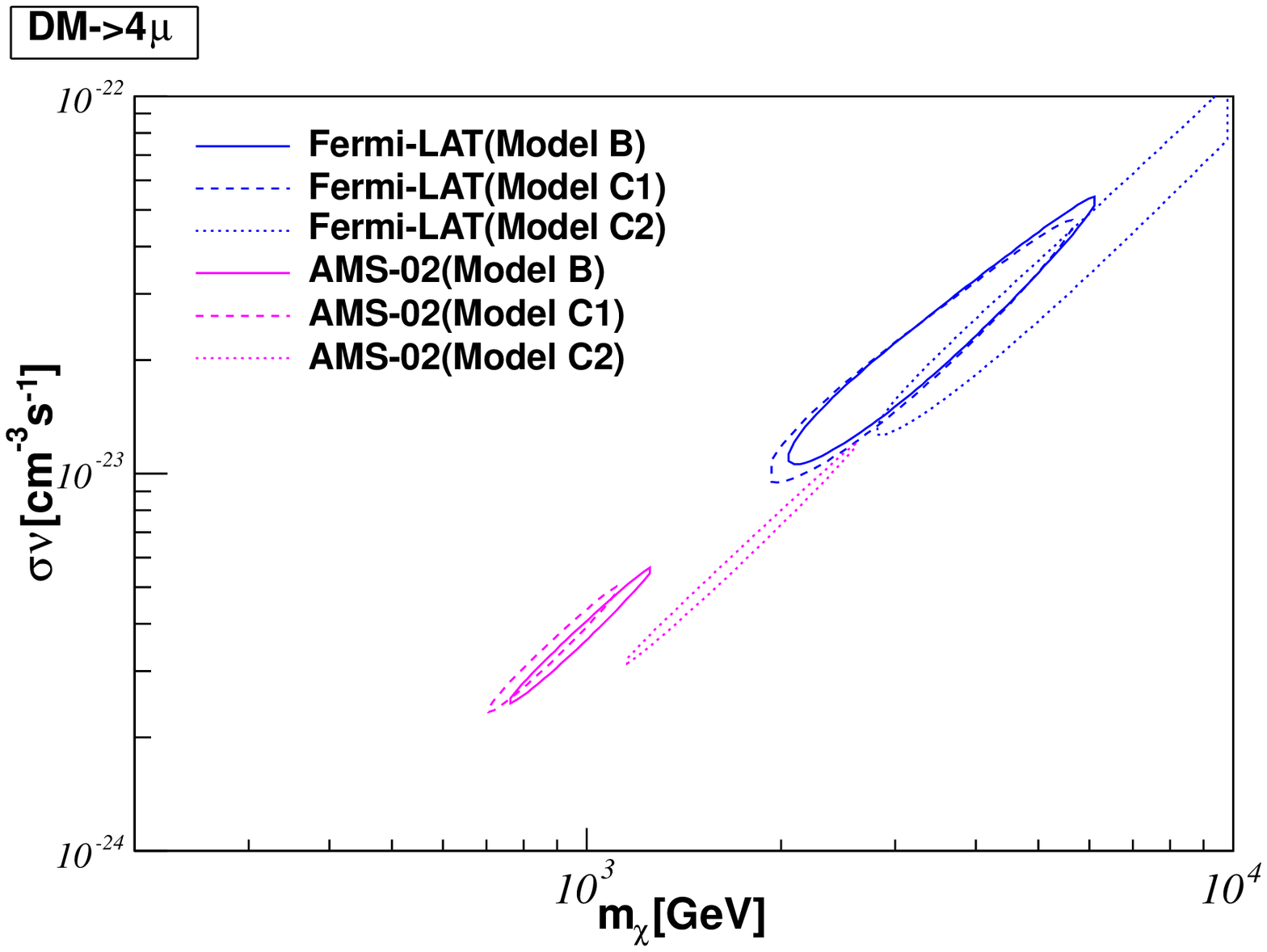}}{\includegraphics[width=0.33\textwidth]{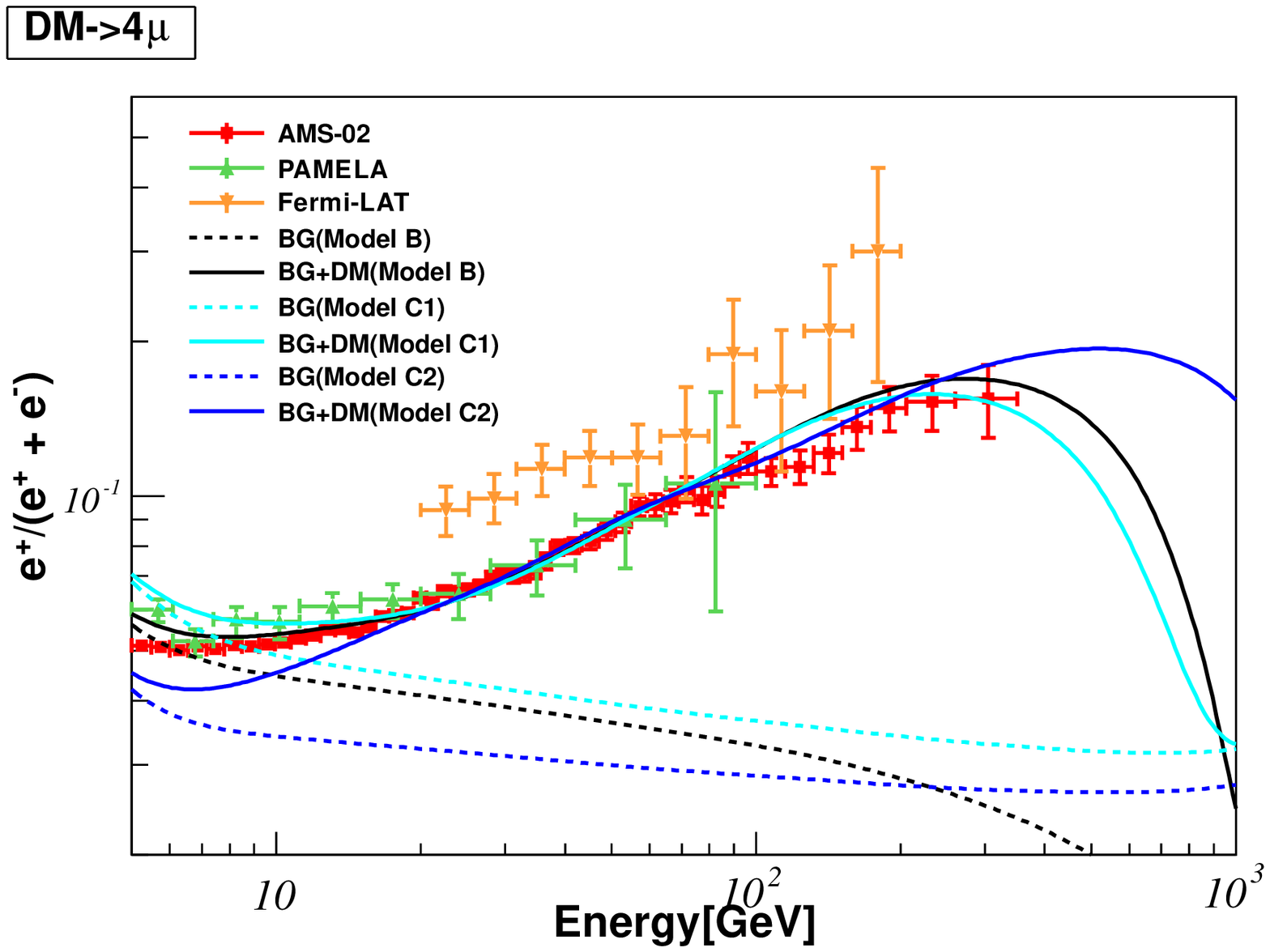}}{\includegraphics[width=0.33\textwidth]{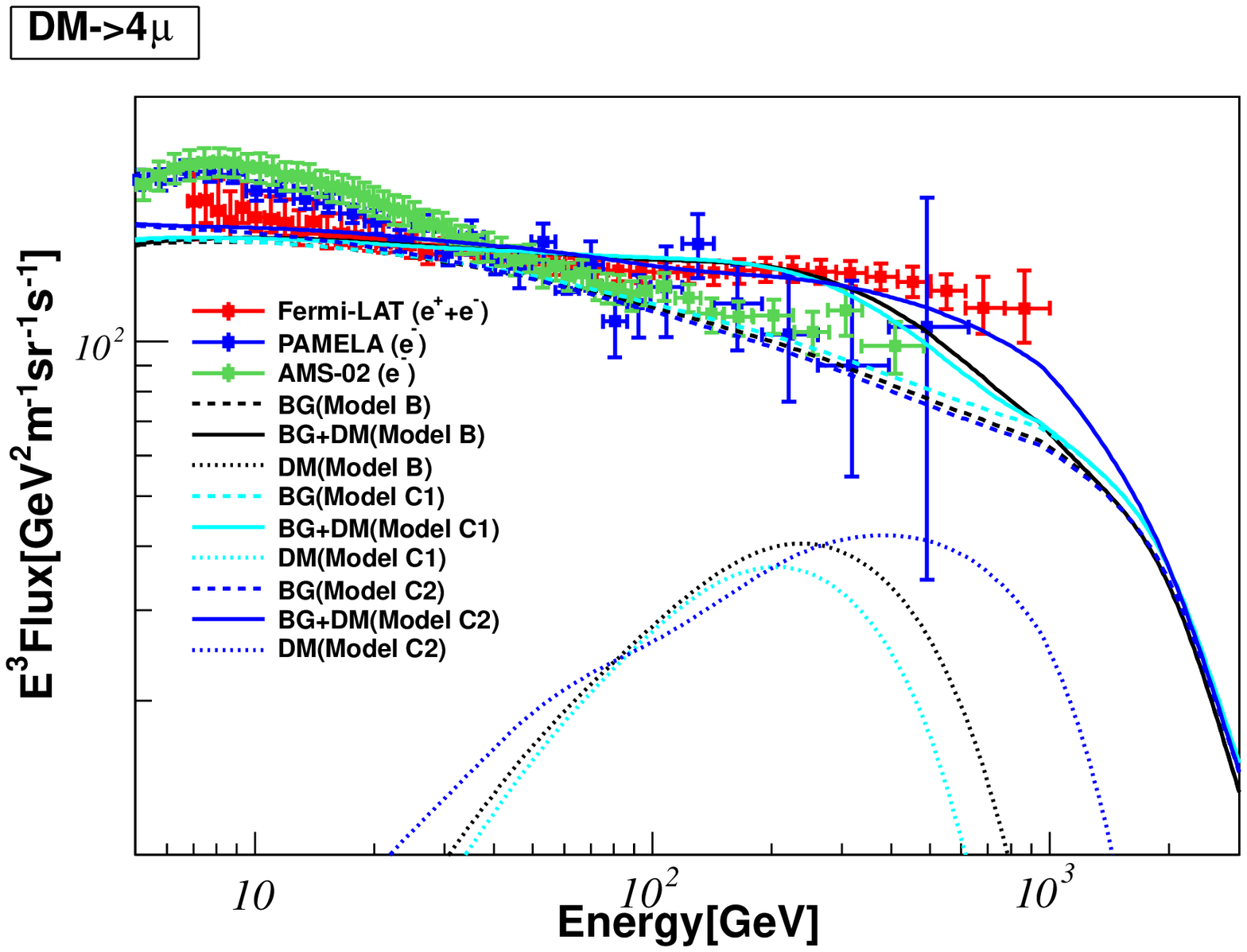}}
\caption{
(Left column) Comparison of  allowed regions in ($m_\chi,\langle \sigma v\rangle$)  plane at $99\%$ C.L. by
the data of AMS-02 on positron fraction~\cite{PhysRevLett.110.141102}
and Fermi-LAT on the total flux of electrons and positrons~\cite{Ackermann:2010ij}
for DM annihilating into $2\mu$ and $4\mu$ final states 
in Model B, C1 and  C2, corresponding to the variation of $Z_{h}$ and $D_{0}$.
(Middle column) predictions for the positron fraction with the best-fit parameters shown in 
\tab{tab:anni}  in the three models.
(Right column) predictions for the total flux of electrons and positrons with the best-fit parameters in the three models.
See text for explanation.
}
\label{fig:uncertaintiesElectronZD}
\end{center}
\end{figure}

The inconsistency between AMS-02 and Fermi-LAT data in 
$2\mu$ and $4\mu$ channels in Model A can be clearly seen  in \fig{fig:ann}
in which the allowed regions by AMS-02 and Fermi-LAT at $99\%$ C.L.
in  $(m_{\chi}, \langle \sigma v\rangle)$ plane  are plotted separately.
%
In the figure, neither the  cross section nor 
the DM particle mass  determined  by the AMS-02 data is 
consistent with Fermi-LAT data for $2\mu$ and $4\mu$ channels.  
For instance, for $2\mu$ channel, the Fermi-LAT data alone favour 
$m_\chi\approx 1.3$ TeV which is far away from the value $\sim 476$ GeV determined by AMS-02.
The best fitted cross section from Fermi-LAT is
$\langle \sigma v\rangle \approx 6.8\times 10^{-24}\text{ cm}^3\text{s}^{-1} $
which is also larger roughly  by a factor of two.
%
In \fig{fig:uncertaintiesContour99CL},   
detailed comparisons between the AMS-02 and Fermi-LAT favoured regions 
at higher confidence levels $99.999\%\ (99.99999\%)$ C.L. corresponding to $\Delta \chi^2=23.0\ (32.2)$ 
are shown for  model A.
One sees that even at $99.99999\%$ C.L. 
the two experimental results of AMS-02 and  Fermi-LAT
cannot be reconciled. 
The situation in model B is similar as  shown in \fig{fig:uncertaintiesElectronZD}.
In this model the favoured regions by AMS-02 and Fermi-LAT are also well separated.

%
It is necessary  to check  if such an observation is robust against 
the variations of the propagation parameters.  
We first consider the effect of 
changing the diffusion coefficient $D_{0}$ and the  diffusion halo height $Z_{h}$.
It is known that change in $D_{0}$ can result in 
changes in the ratio of secondaries to primaries. 
In \fig{fig:uncertaintiesElectronZD}, 
we show how the allowed regions change
in model C1 (C2), which corresponds to a typically  lower (upper) value of 
$D_{0}=5.45 \ (11.2)\times 10^{28} \text{ cm}^{2}\text{s}^{-1}$,
and $Z_{h}=3.2\ (8.6)$ kpc.
The corresponding spectra for positron fraction and the total flux of electrons
and positrons for best-fit parameters are also shown in  \fig{fig:uncertaintiesElectronZD}.
We find that
for larger diffusion constant $D_{0}=11.2\times 10^{28}\text{ cm}^{2}\text{s}^{-1}$, 
the disagreement between the results of AMS-02 and Fermi-LAT can be slightly reduced.
This is due to the fact that a larger $D_{0}$ reduces the background of the 
positron fraction, which leaves greater room for  heavier
DM particle. 
In $2\mu$ channel, the best-fit DM particle mass is $m_{\chi}=602$ GeV in Model C1.
But for Model C2, the best-fit value is $m_{\chi}=1.3$ TeV with
the $\chi^{2}/\text{d.o.f}$ value decreasing from $332.1/119$ (Model C1) to 
$189.6/119$ (Model C2), 
indicating a more consistent fit.
Note that in this case the value of  $Z_{h}=8.6$ kpc  is also quite  large.
%

\begin{figure}\begin{center}
{\includegraphics[width=0.45\textwidth]{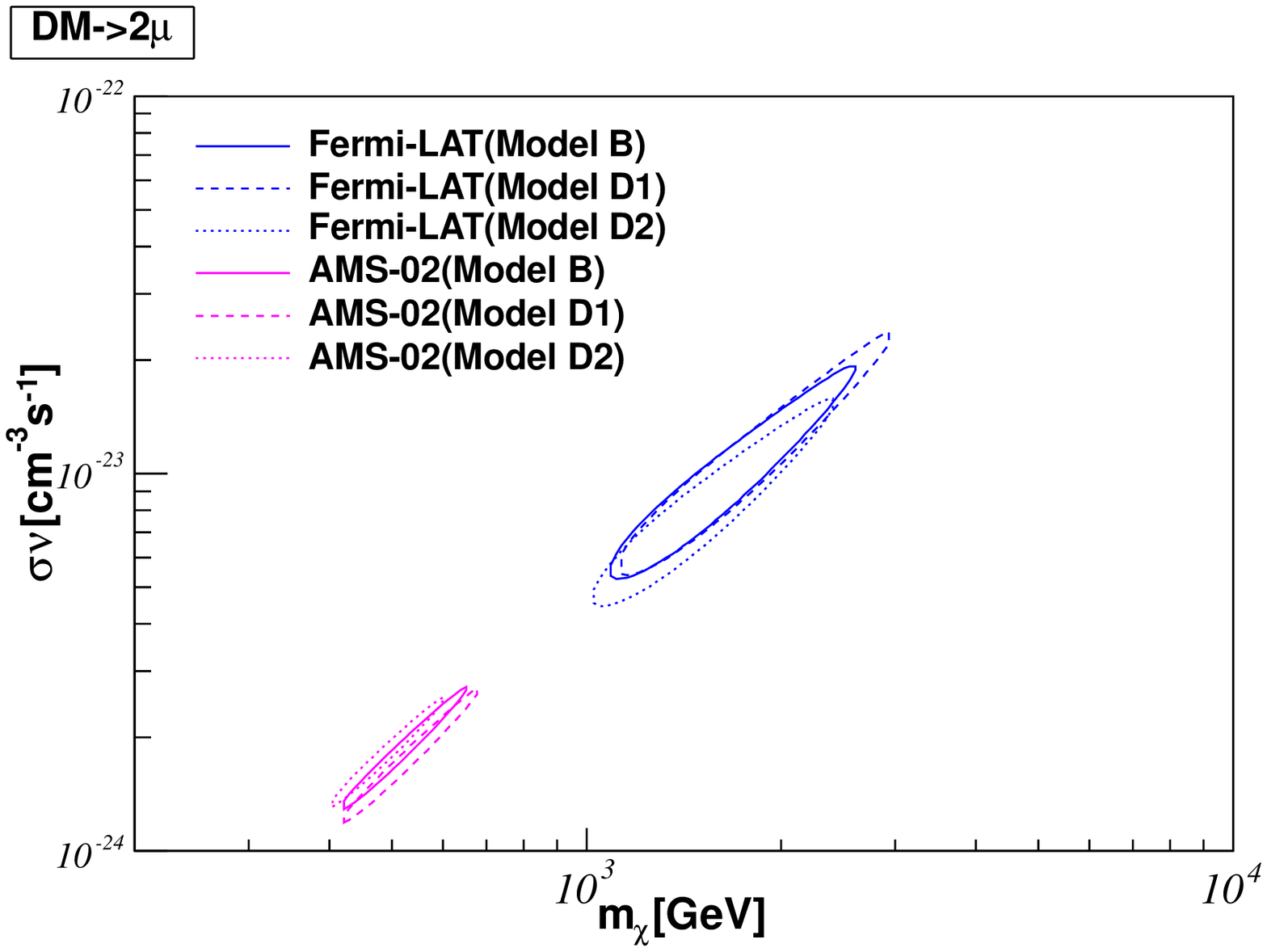}}
{\includegraphics[width=0.45\textwidth]{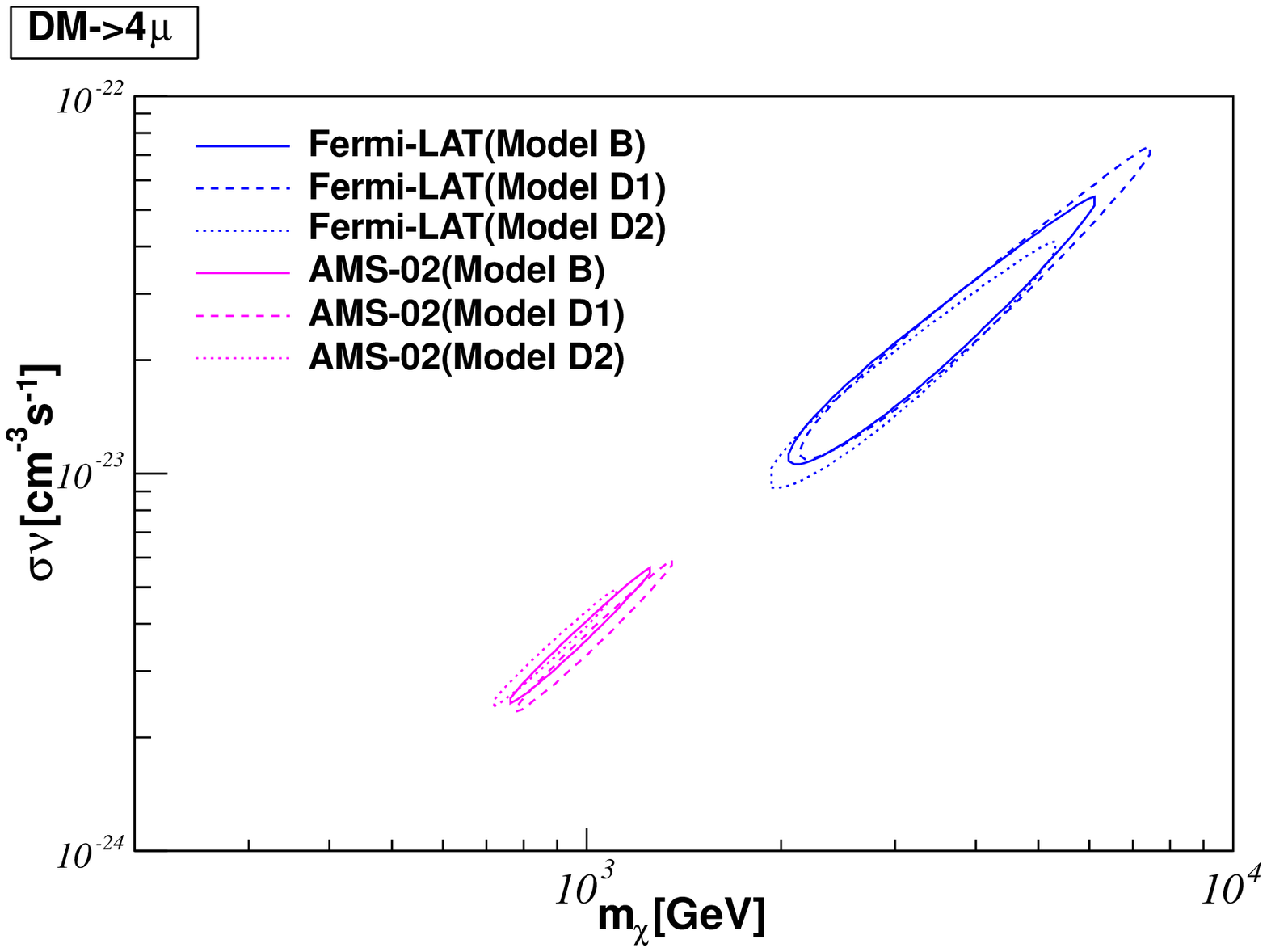}}
\\
{\includegraphics[width=0.45\textwidth]{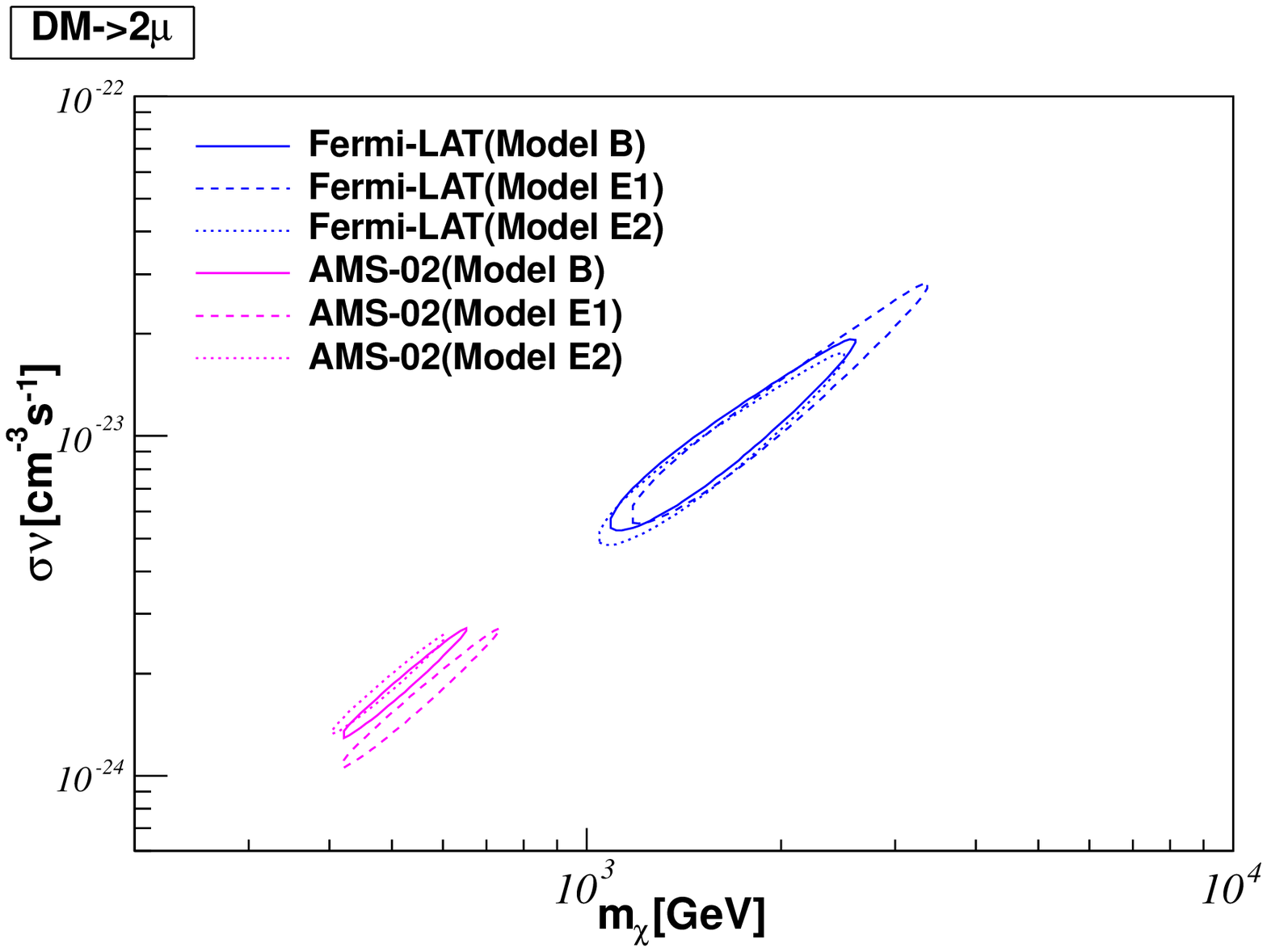}}
{\includegraphics[width=0.45\textwidth]{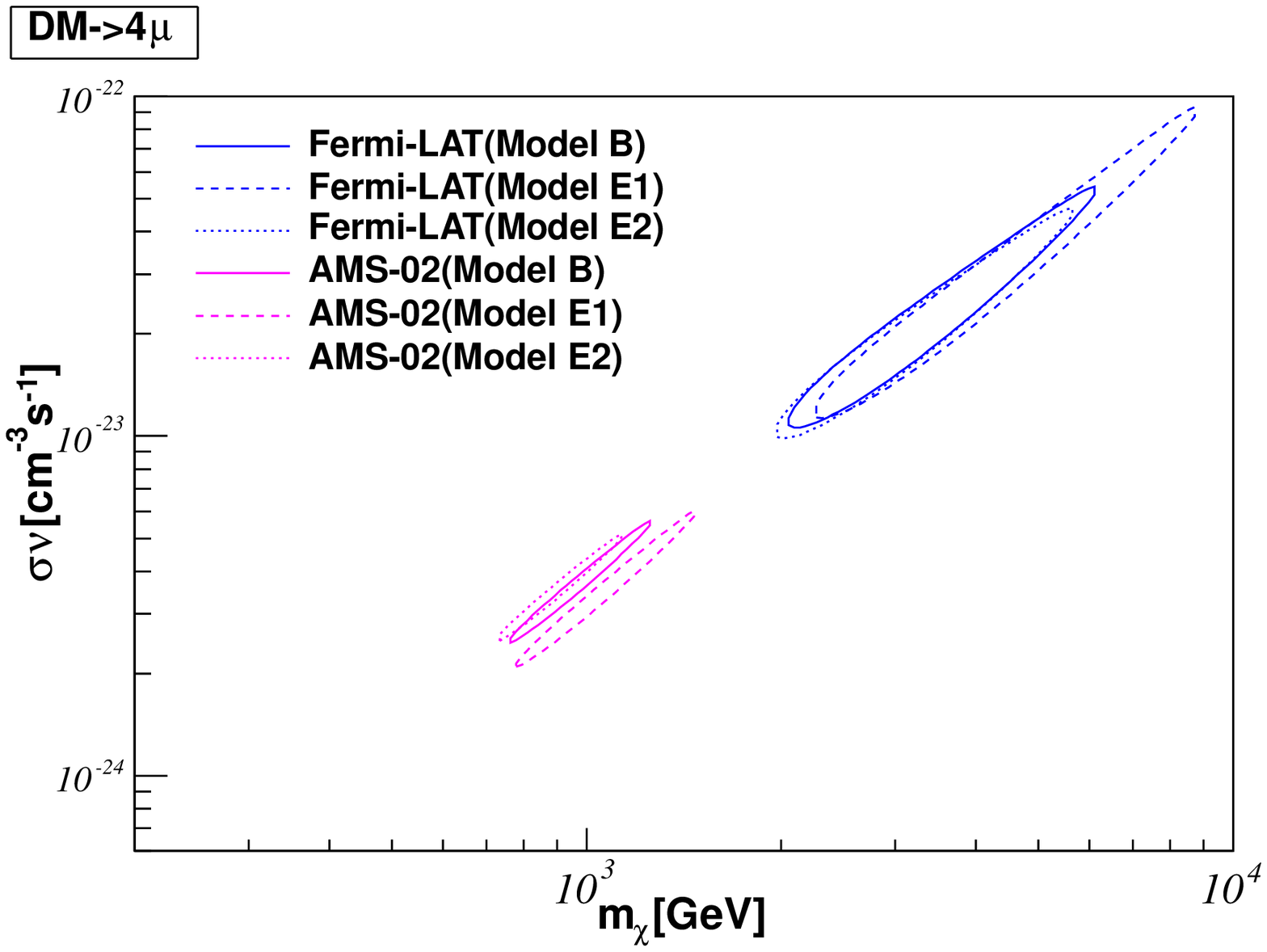}}
\caption{
Allowed regions in ($m_\chi,\langle \sigma v\rangle$)  plane at $99\%$ C.L. by
the data of AMS-02 on positron fraction~\cite{PhysRevLett.110.141102}
and Fermi-LAT on the total flux of electrons and positrons~\cite{Ackermann:2010ij}
for DM annihilating into $2\mu$ and $4\mu$ final states 
in D1 and  D2, corresponding to the variation of $\delta_{2}$ (upper row)
and in Model E1 and  E2, corresponding to the variation of $\gamma_{p2}$ (lower row).
The allowed regions in Model B are shown for a comparison.
}
\label{fig:uncertaintiesElectronDelta}
\end{center}
\end{figure}

We then consider the variation of the power index $\delta_{2}$  in the diffusion term. 
In Model D1 (D2) the value of $\delta_{2}$ is set to be $0.26\ (0.35)$. 
The results are shown in \fig{fig:uncertaintiesElectronDelta}. 
Since the value of $\delta_{2}$ is well constrained,
the changes in the allowed regions are less significant 
in comparison with Model B.
%
%
Finally consider the variation of the power index $\gamma_{p2}$ is considered. 
In model E1 (E2), the value of $\gamma_{p2}$ is set to  2.29 (2.47).
The fit results are also shown in \fig{fig:uncertaintiesElectronDelta}.
The change in $\gamma_{p2}$ has greater effect in positron fraction than
that in the total flux of electrons and positrons.  
This can be understood as the secondary positrons mainly arise from the interactions
between the primary proton and the interstellar  medium.
We find again that the variation of $\gamma_{p2}$ cannot relax the tension between 
the two experiments.

%




%
%


In a brief summary, we have found that in these models from A to E2, only the one with 
large diffusion coefficient $D_{0}$ and $Z_{h}$ can slightly reduce the tension between 
AMS-02 and Fermi-LAT.
Since the uncertainties in  $\kappa$ and $\delta$ have been considered in the fits,
the results indicate  that  such a  discrepancy between AMS-02 and Fermi-LAT 
are unlikely to  be removed by varying  the normalization  and  slope of  the 
astrophysical background.

In the case of  $2\tau$ channel,
the predicted fluxes are much smoother compared with 
that in the $2e$ and $2\mu$ channels. 
The agreement with the data is  improved. 
In Model A (B),
the goodness of fit is $\chi^{2}/\text{d.o.f}=1.8\  (1.7)$,
the favoured  DM particle mass is around $1.5\ (1.9)$ TeV with 
cross section 
$\langle \sigma v\rangle \approx 1.8 \ (2.2)\times 10^{-23}\text{ cm}^3\text{s}^{-1} $.
As shown in \fig{fig:uncertaintiesContour99CL}, 
the region of parameters favoured by the AMS-02 data is 
marginally consistent with that favoured by Fermi-LAT at $99\%$C.L..  
At higher confidence level such as $99.99999\%$, there is a visible overlap
between the two experiments.
In both Model A and B, 
the best-fit annihilation cross sections are very large,
which calls for a large boost factor of $~\mathcal{O}(600-700)$,
and  can be severely constrained by the nonobservation of 
gamma-rays from various sources in the sky.  
For instance, the constraint from the Galactic diffuse gamma-ray
can reach $\sim 1\times10^{-23}\text{ cm}^3\text{s}^{-1}$ at $m_{\chi}\approx 1.5$ TeV
for $2\tau$ channel with isothermal DM profile
\cite{Ackermann:2012rg
}.
Similar discussions only considering the AMS-02 data can be found in 
Refs.~\cite{DeSimone:2013fia,
Feng:2013zca
}.

The fit results for DM  particle annihilating into four-lepton channels
are shown in \tab{tab:anni}.
The injection spectra from the  four-lepton channels are 
in general  smoother than the corresponding two-lepton channels,
which results in better fits. 
Among all the channels, the $4\tau$ channel has the lowest 
$\chi^{2}/\text{d.o.f}\approx 1.7\ (1.6)$  in both Model A (B). 
The four-lepton channels in general  prefer larger DM particle masses.
For instance, the best-fitted DM mass is $\sim 1-1.5$ TeV for $4\mu$
and $\sim 3-4$ TeV for $4\tau$ final states. 
The required cross sections are also large, 
as it is shown in the right column of \fig{fig:ann}.
\subsection{Fits with decaying dark matter}
We proceed to perform  global fits for the DM decay scenario. 
Although the injection spectrum from a decaying DM particle with mass $m_\chi$
should be the same for the annihilating DM particle with mass $m_\chi/2$, 
the final electron/positron flux after propagation is slightly different due to
the different DM density distribution dependences, 
namely,
the source term is proportional to $\rho(\mathbf{r})$ in the case of DM decay,
but it is proportional to $\rho^2(\mathbf{r})$ in the case of DM annihilation.
For Einasto profile, the final electron/positron flux after propagation tends to be 
slightly steeper  in the case of DM decay. 
Such a small spectral difference between decay and annihilation  is 
unlikely to be distinguished by PAMELA and Fermi-LAT data, 
but can lead to significantly different results when fit includes 
precision AMS-02 data.

%
%
%

We first consider the case with charge symmetric decays, namely,  $\epsilon=0$.
The results of best-fit parameters are listed in \tab{tab:decay} for Model A and Model B.
We find that in general 
the fits in the case of  DM decay  have larger $\chi^2$ than that in DM annihilation.
%
For instance,  for the channels with electron final states,
the $\chi^2/\text{d.o.f}$ can reach $\sim 7.3 \ (6.9)$ in Model A (B), 
indicating  rather poor fits with electron final states. 
Thus we shall focus on $\mu$  and $\tau$ final states instead.

\begin{figure}[htb]
\includegraphics[width=0.49\textwidth]{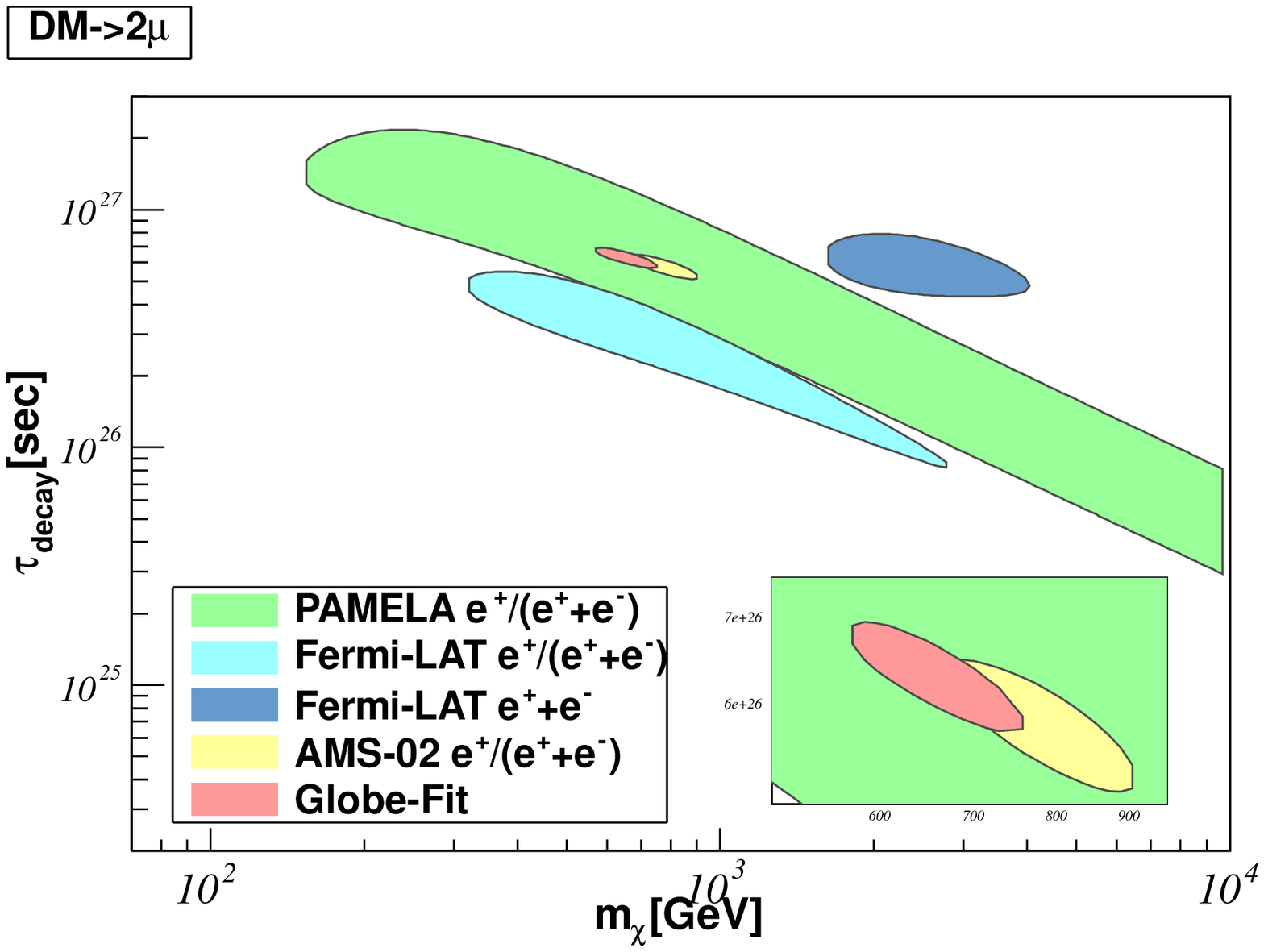}\includegraphics[width=0.49\textwidth]{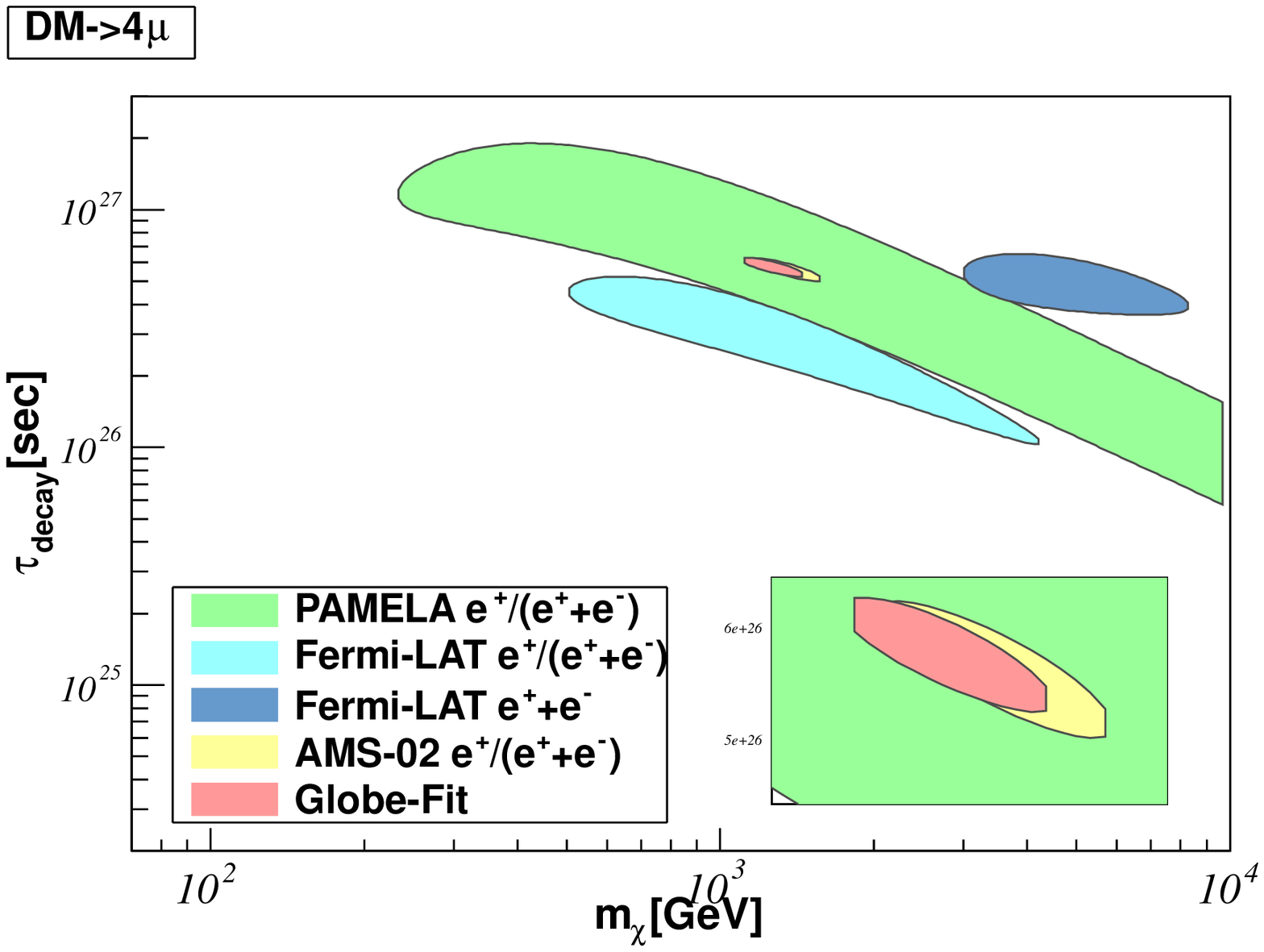}
\\
\includegraphics[width=0.49\textwidth]{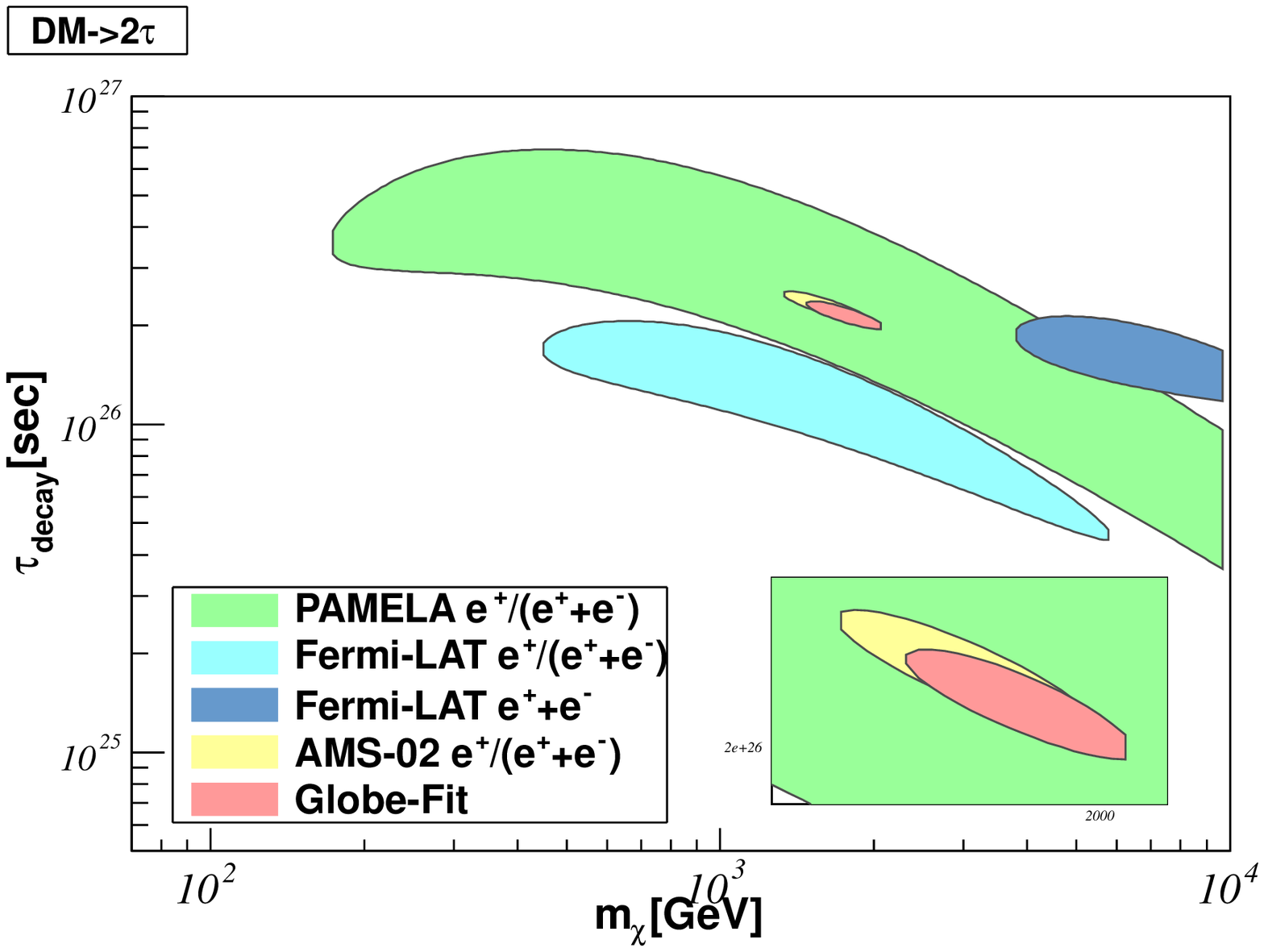}\includegraphics[width=0.49\textwidth]{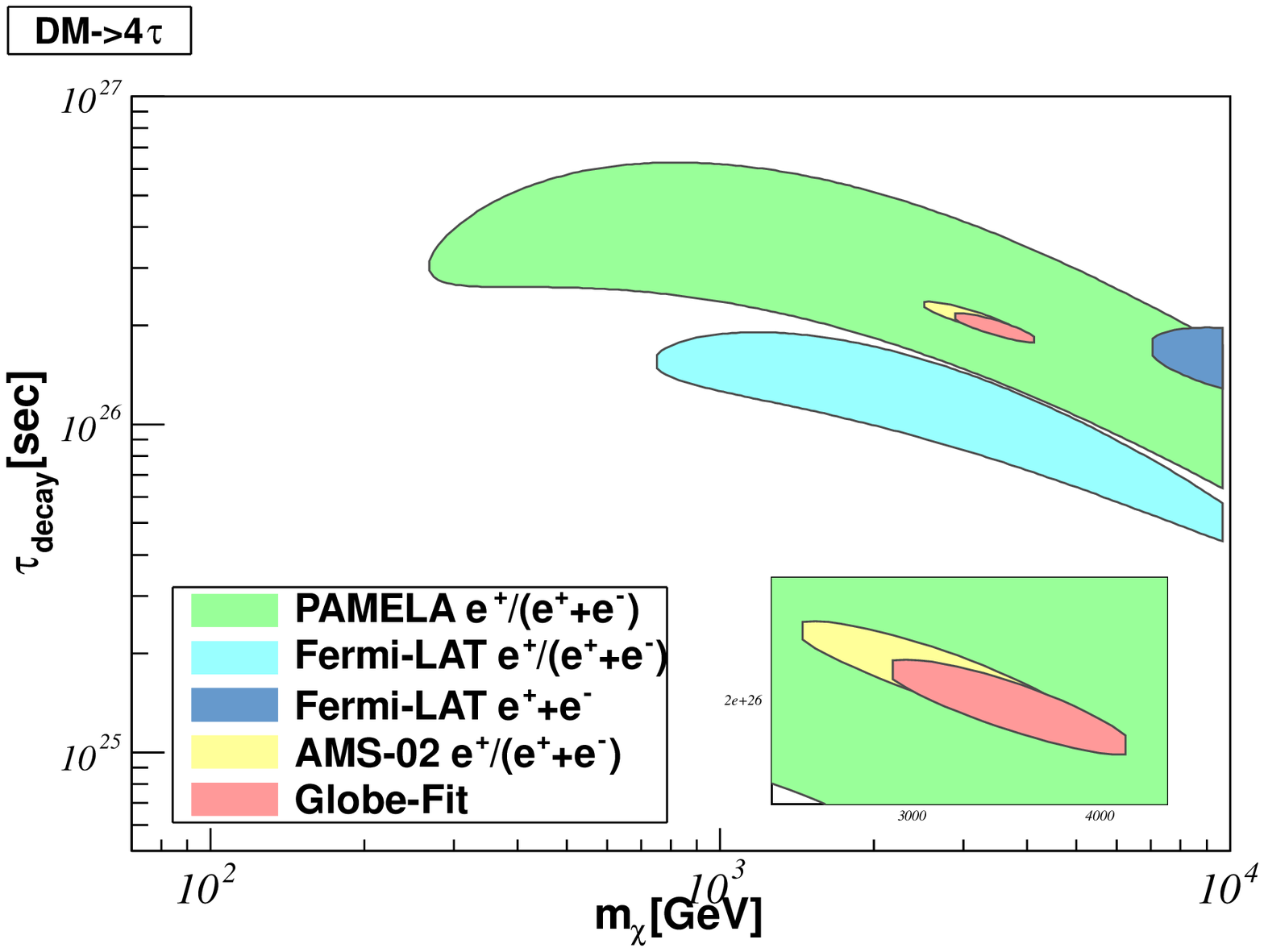}
\caption{ 
The same as \fig{fig:ann}, 
but for the allowed regions in ($m_\chi,\ \tau$)  plane at $99\%$ C.L. 
for DM decaying into  $2\mu$, $2\tau$, $4\mu$ and $4\tau$ final states 
from the Global fits.}
%
 \label{fig:decay}
\end{figure}

\begin{figure}
{\includegraphics[width=0.33\textwidth]{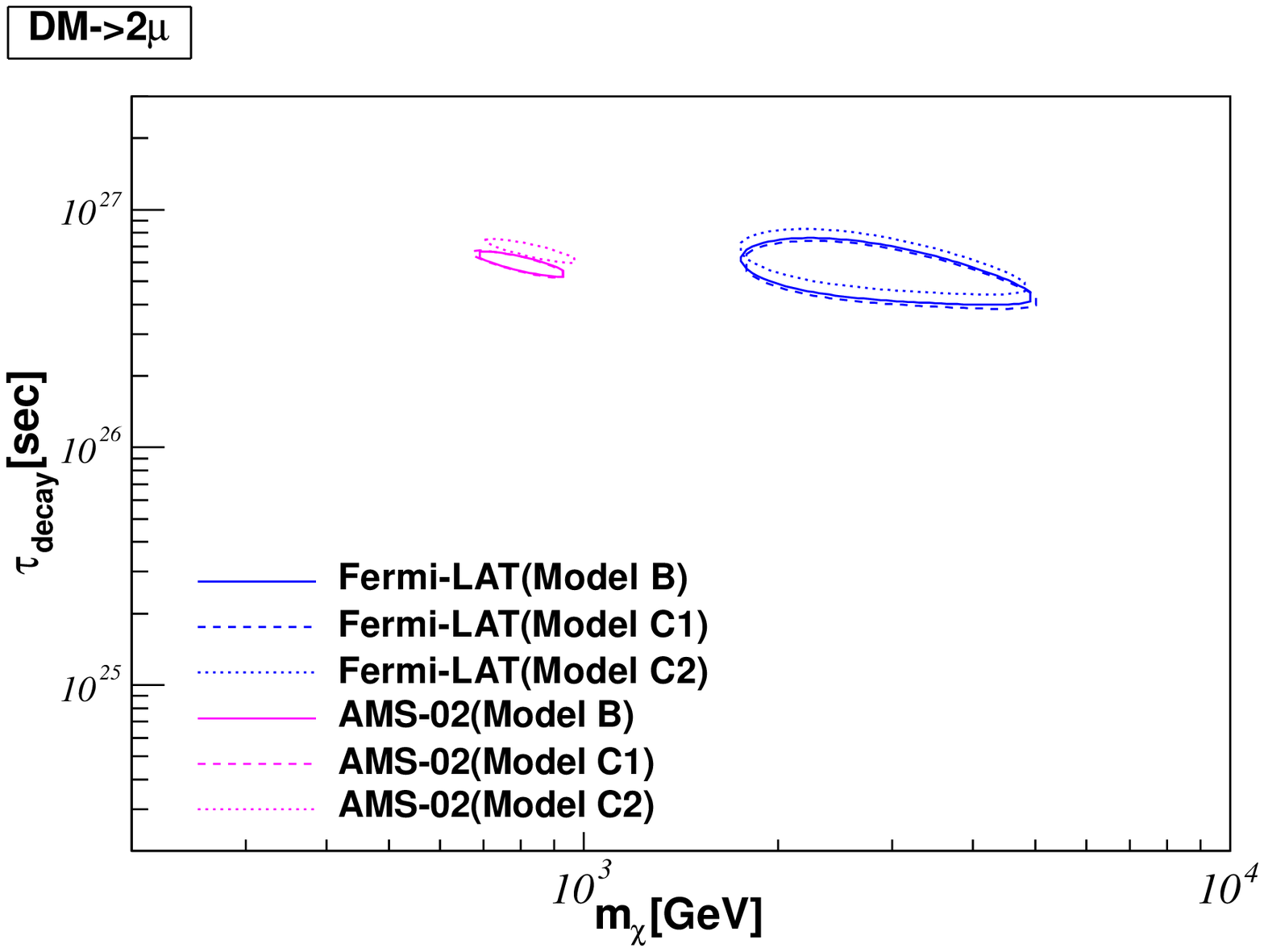}}{\includegraphics[width=0.33\textwidth]{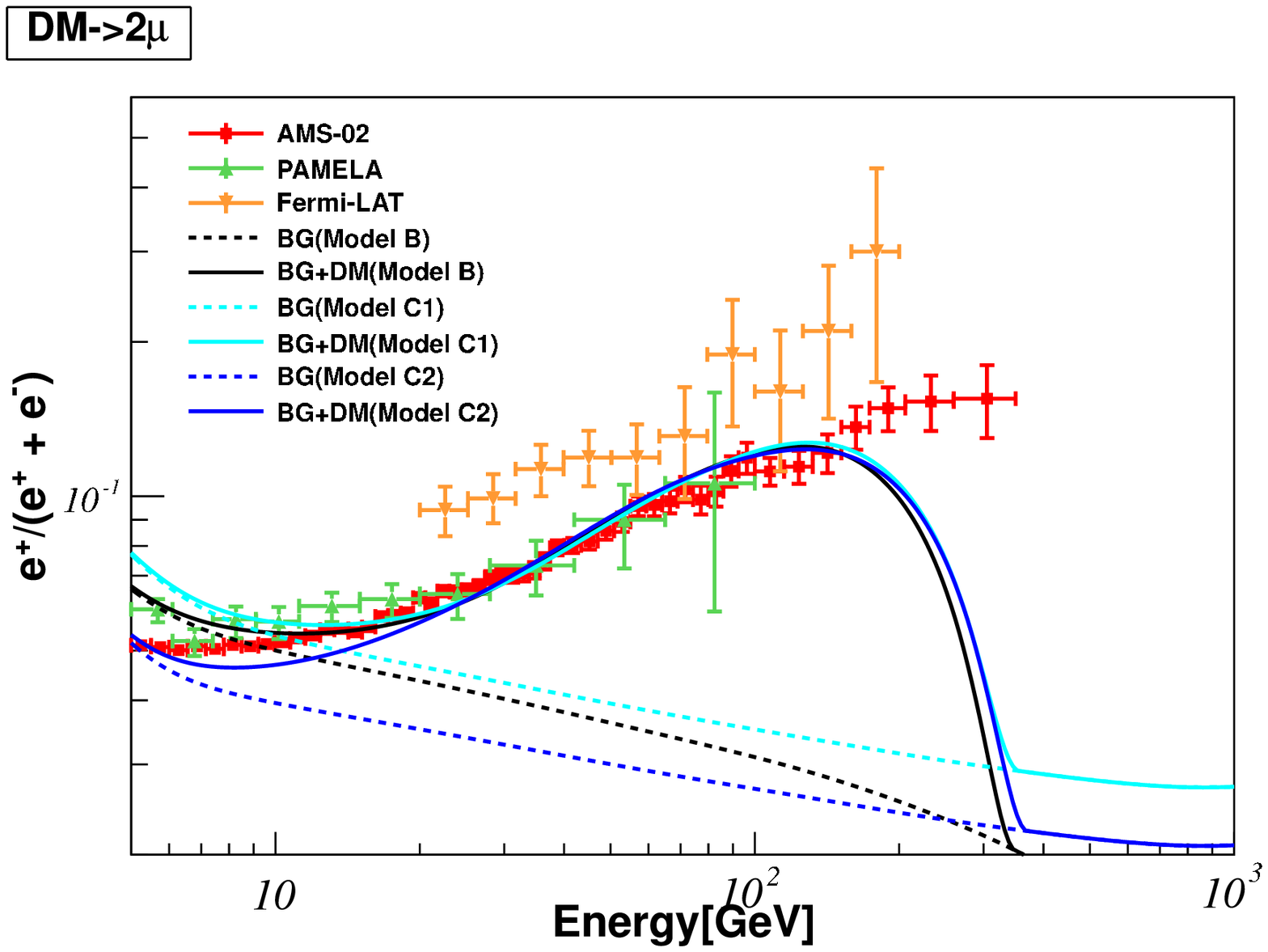}}{\includegraphics[width=0.33\textwidth]{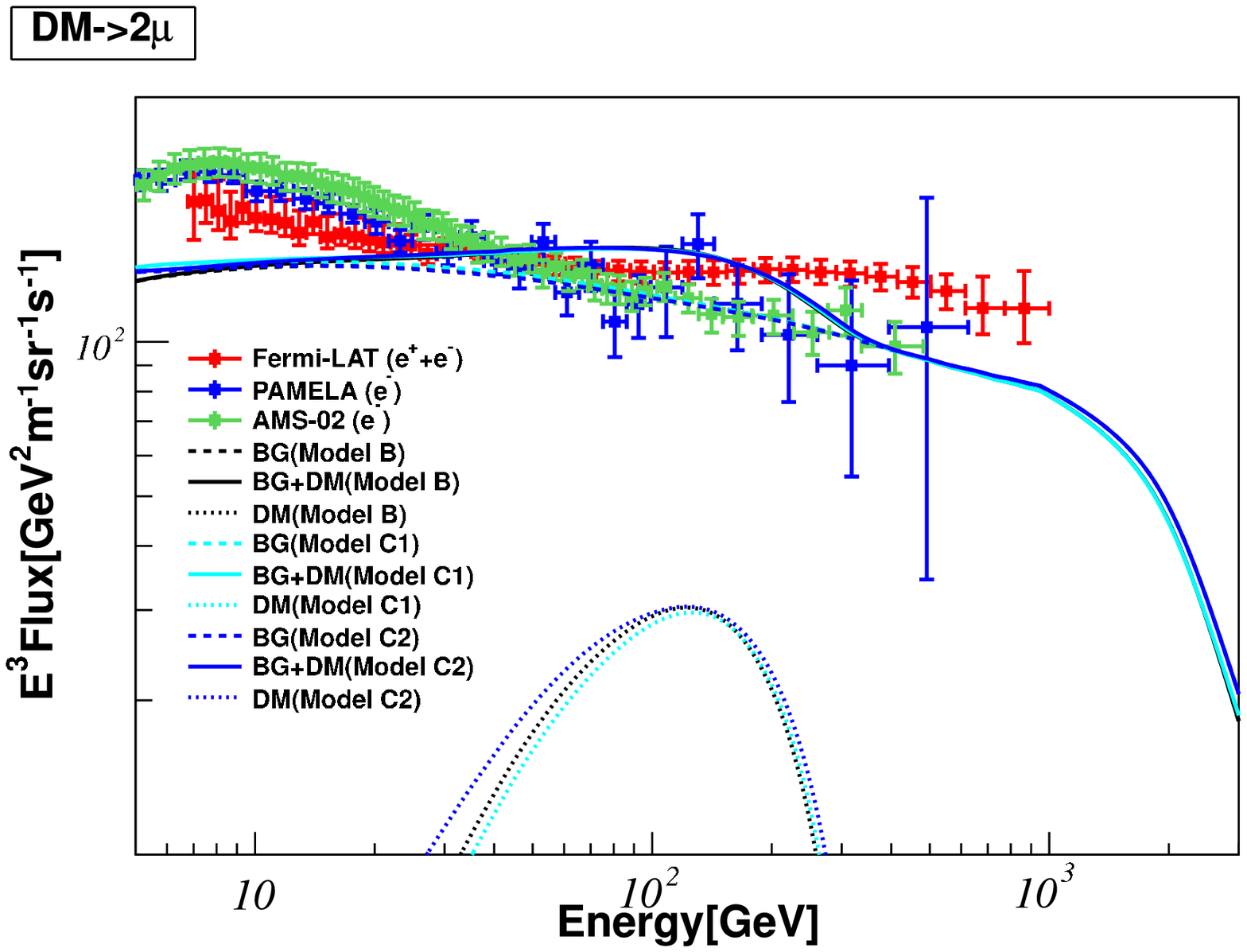}}
\\
{\includegraphics[width=0.33\textwidth]{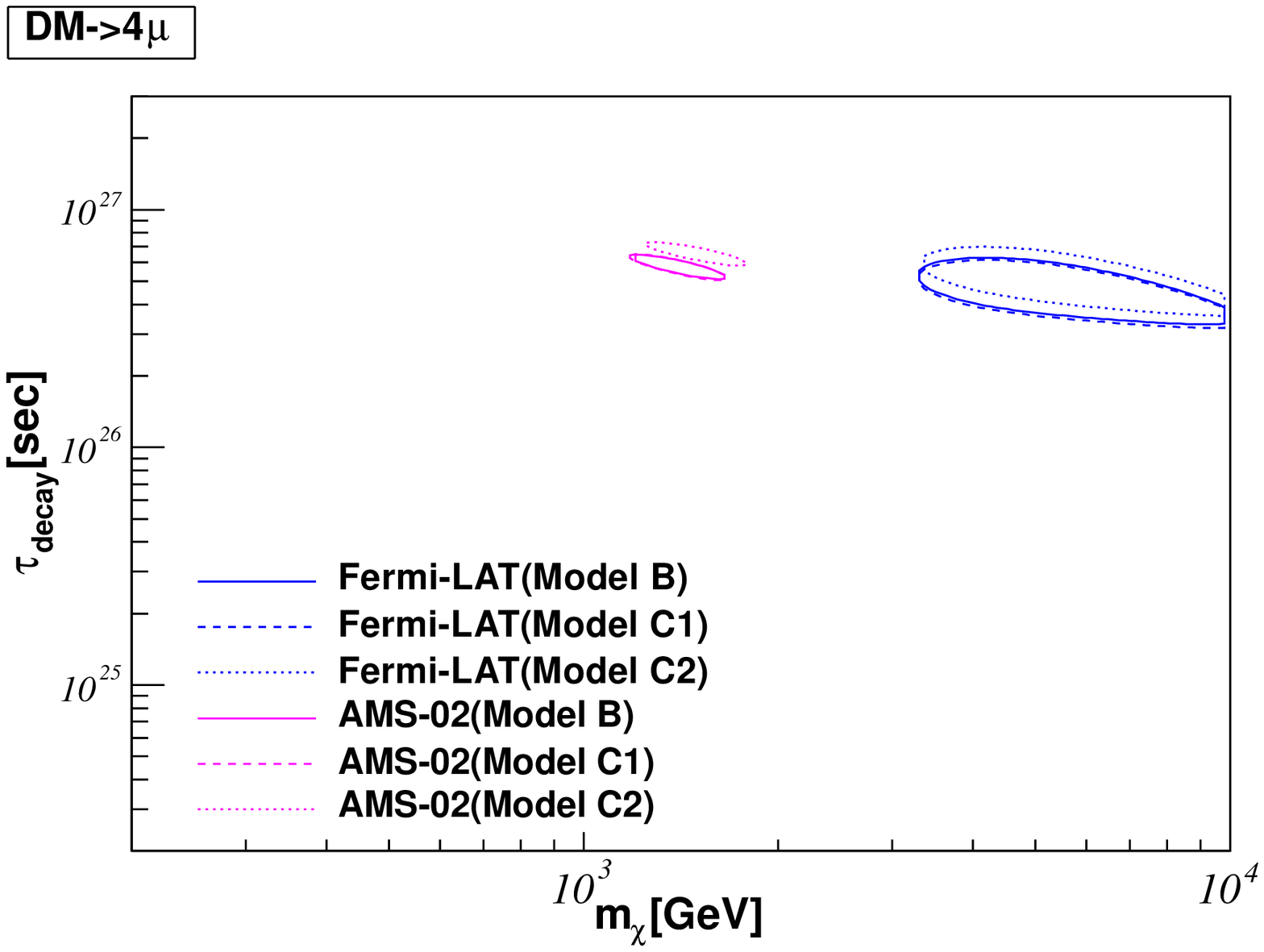}}{\includegraphics[width=0.33\textwidth]{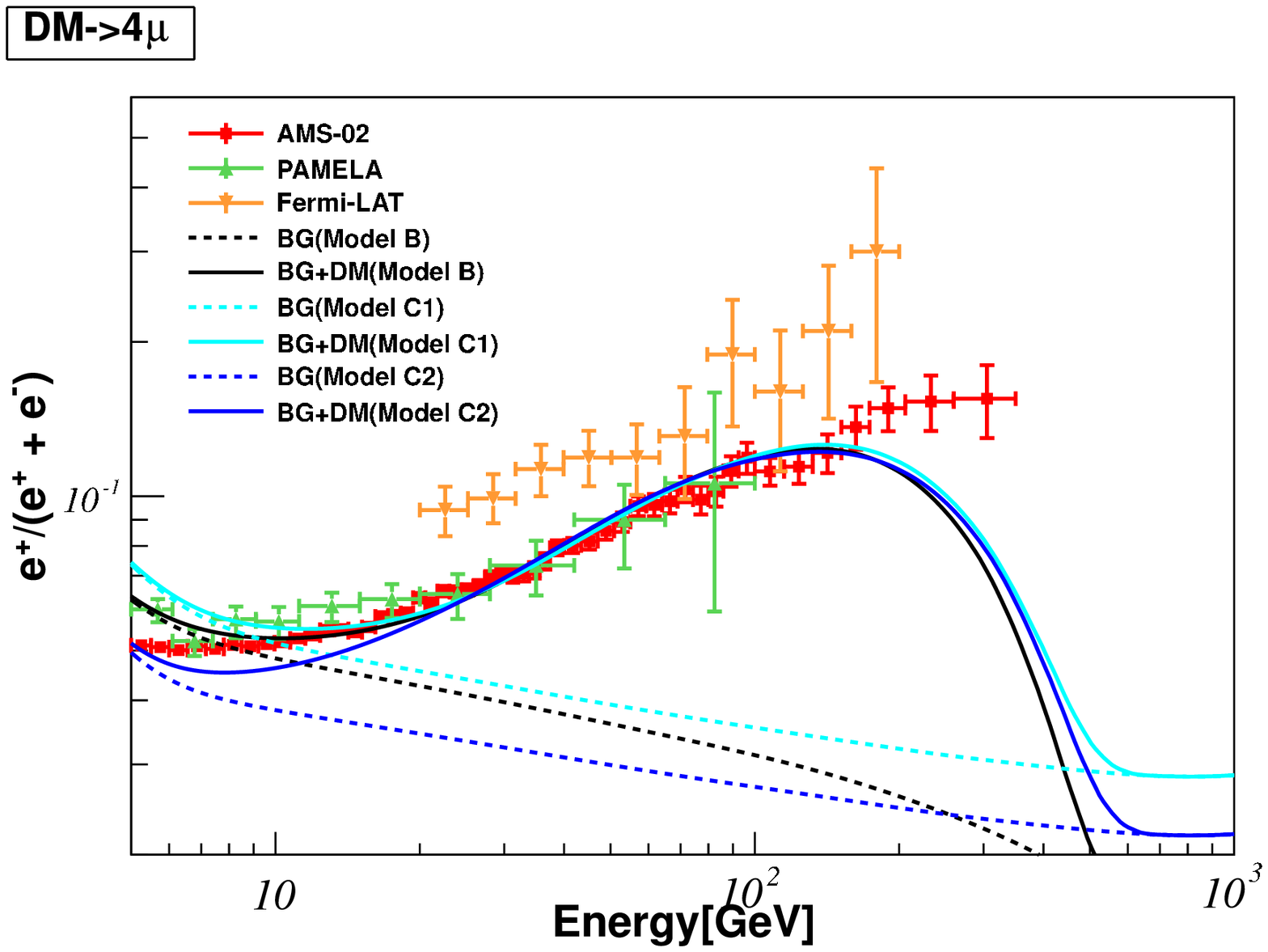}}{\includegraphics[width=0.33\textwidth]{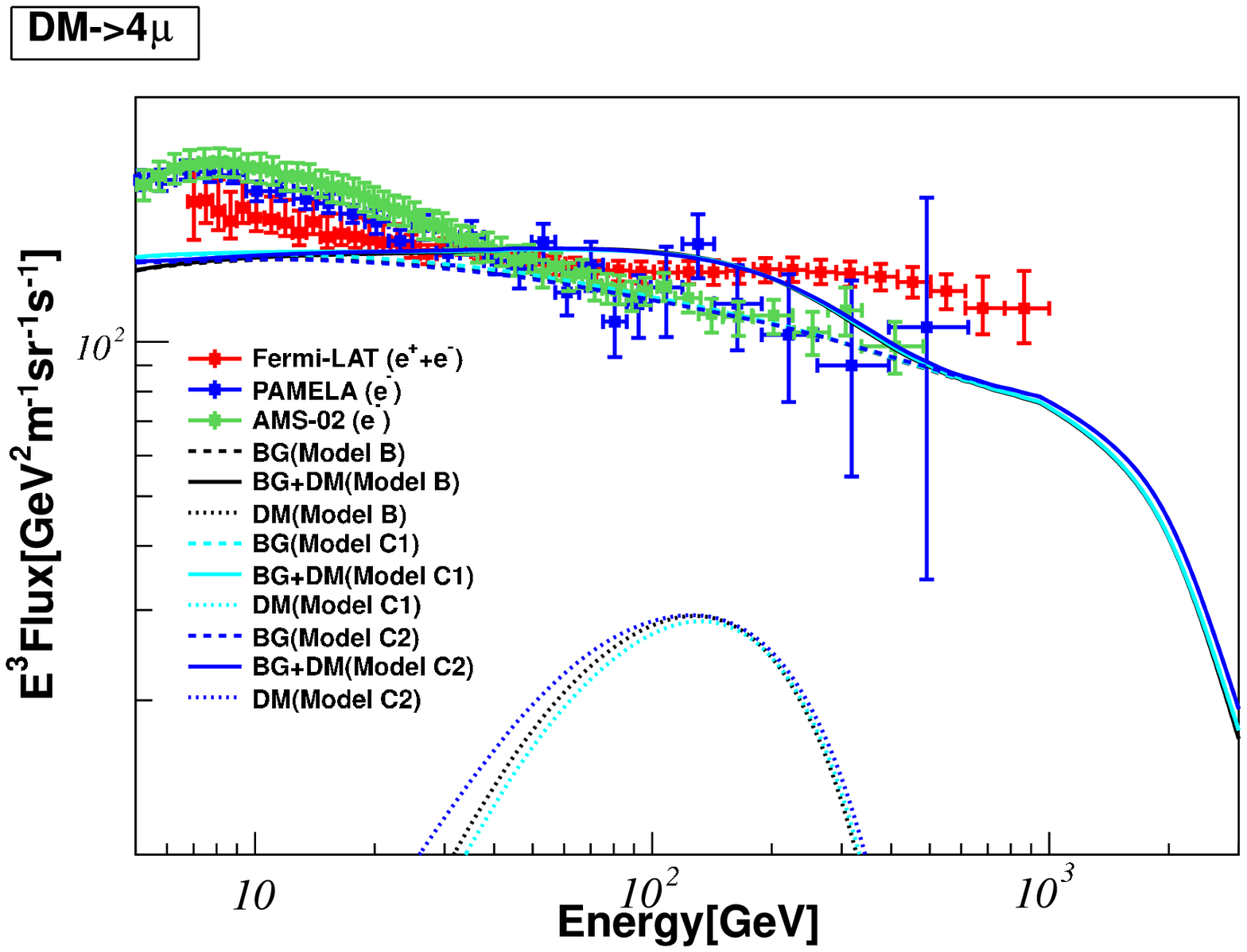}}
\caption{
The same as \fig{fig:uncertaintiesElectronZD}, but for the case of DM decay.
%
}
\label{fig:uncertaintiesElectronDecayZD}
\end{figure}

In \fig{fig:decay}, we show the allowed regions in the $(m_{\chi}, \tau)$ plane
at $99\%$ C.L. for  DM particle decaying into $2\mu$, $2\tau$, $4\mu$  and $4\tau$ final states
in Model A.
The corresponding allowed regions in $(\kappa, \delta)$ plane at 
the same confidence level
are shown in the right panel of \fig{fig:delta_anni}. 
We follow the same fitting strategy in obtaining 
the allowed region in $(m_{\chi}, \tau)$ plane by each single experiment 
as in the case of DM annihilation, namely, the values of $\kappa$
and $\delta$ from the global fit are taken as inputs. 
From the figure, one sees that there is no overlapping region between Fermi-LAT
and AMS-02 favoured region at $99\%$ C.L. in Model A for all the final states. 
In the case of Model B, very similar results are obtained. 
The contours of the allowed regions in $(m_{\chi},\langle \sigma v\rangle)$ plane
for Model C1 and C2 are shown in \fig{fig:uncertaintiesElectronDecayZD}.
We find that for the Model C2 with large diffusion coefficient, the 
tension between AMS-02 and Fermi-LAT is not reduced as that in the case 
of DM annihilation.  
In $2\mu$ channel, 
the best-fit DM particle mass is $m_{\chi}=705$ GeV in Model C1.
For Model C2, the best-fit value is $m_{\chi}=733$ GeV with
the $\chi^{2}/\text{d.o.f}$ value decreasing from $456.1/119$ (Model C1) to 
$461.3/119$ (Model C2).
Thus there is no improvement in the goodness-of-fit.

\begin{table}[thb]
\begin{center}
\begin{tabular}{cccccc}
  \hline\hline
mode & $m_\chi$(GeV) & $\tau (\times 10^{26}\text{s})$ & $\kappa$ & $\delta(\times 10^{-2})$ &$\chi^2_{\text{tot}}/$dof\\ 
\hline
$2e$	&334.0 	&21.1	&0.632 	&6.79	&892.87/119\\
      &332.1  &24.2       &0.673  &4.25      &836.39/119\\
\hline
$2\mu$	&654.8 	&6.27	&0.806 	&1.40	&510.77/119\\
	&691.1 	&6.39	&0.856 	&-1.24	&493.92/119\\
\hline
$2\tau$	&1762.4 	&2.15	&1.019 	&-4.41	&291.92/119\\
	&1860.1 	&2.19	&1.072 	&-6.79	&291.56/119\\
\hline
$4e$	&506.2 	&19.3	&0.737 	&3.54	&622.69/119\\
      &523.7  &19.9       &0.787  &0.81        &594.44/119\\
\hline
$4\mu$	&1258.6 	&5.76	&0.882 	&-0.78	&414.90/119\\
	&1328.4 	&5.85	&0.933 	&-3.32	&406.53/119\\
\hline
$4\tau$	&3455.5 	&1.97	&1.058 	&-5.34	&265.93/119\\
	&3647.0 	&2.01	&1.112 	&-7.69	&266.56/119\\
  \hline\hline
\end{tabular}
\end{center}
\caption{
Best-fit values of $m_{\chi}$, $\tau$, $\kappa$ and $\delta$,
as well as the $\chi^{2}/\text{d.o.f}$ for DM particles decaying into 
$2e$, $2\mu$, $2\tau$, $4e$, $4\mu$ and $4\tau$ final states, 
assuming no charge asymmetry.
For each final states, the values in the first (second) row corresponds to the results in Model A (B).
}
\label{tab:decay}
\end{table}


\begin{figure}
{\includegraphics[width=0.49\textwidth]{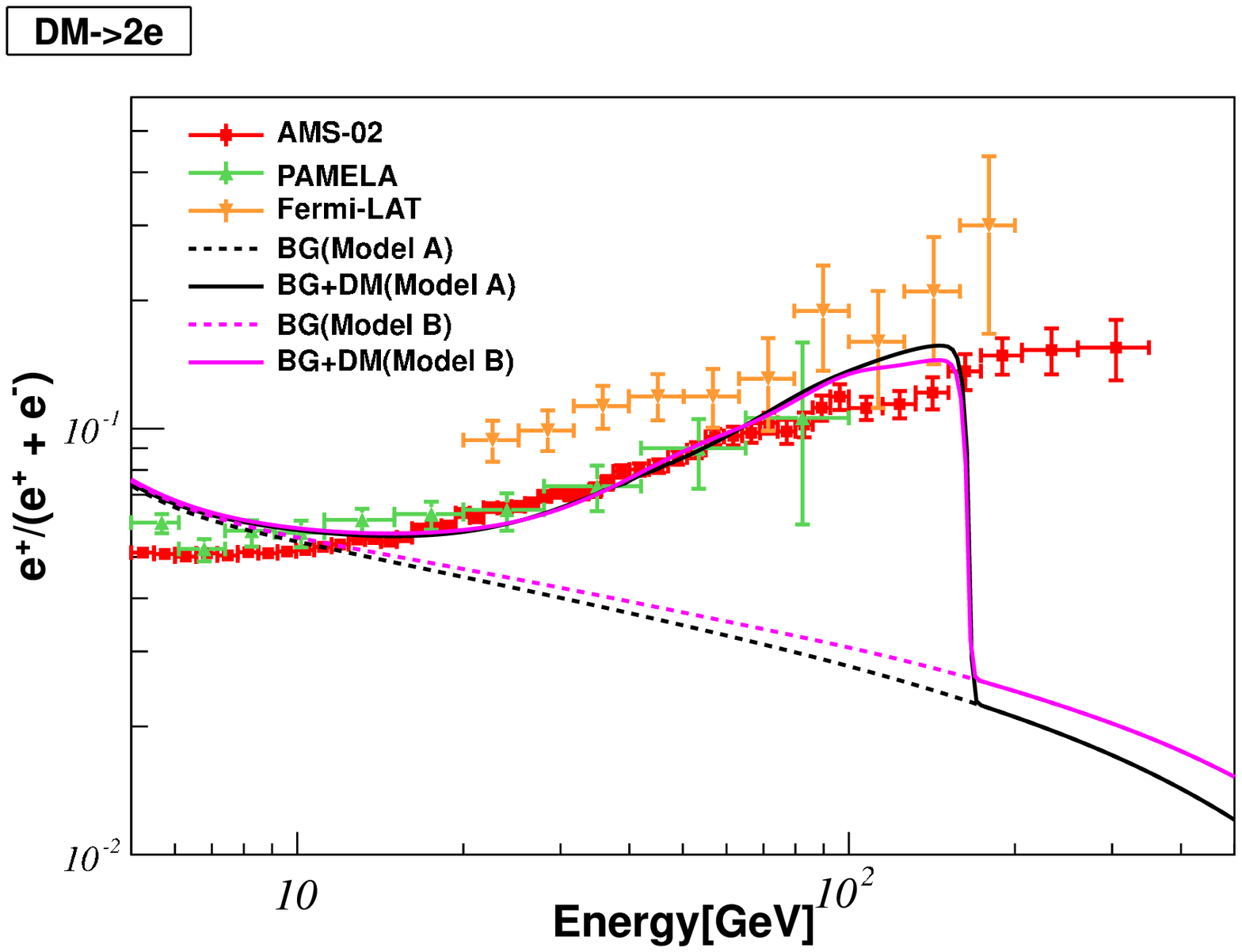}}{\includegraphics[width=0.49\textwidth]{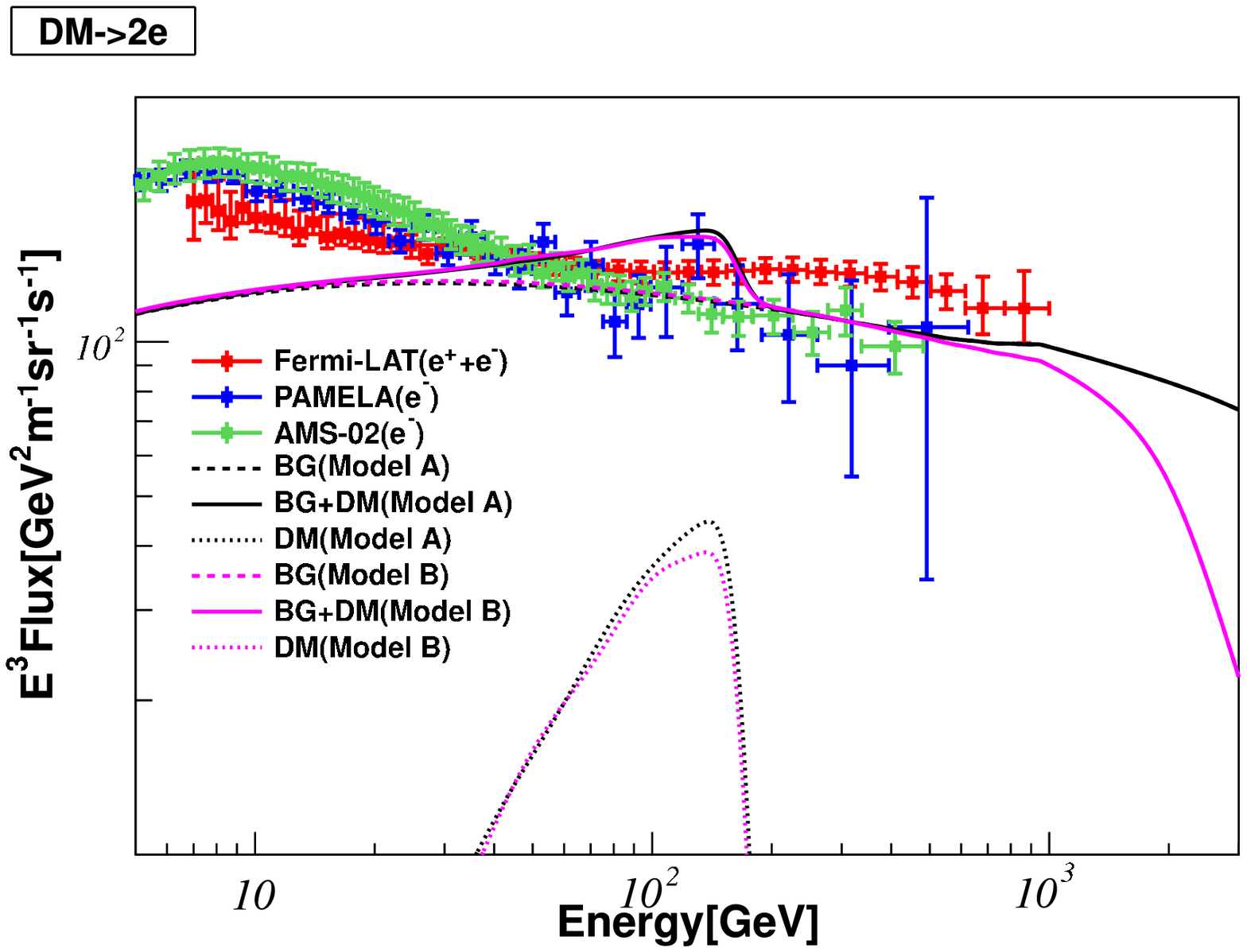}}
\\
{\includegraphics[width=0.49\textwidth]{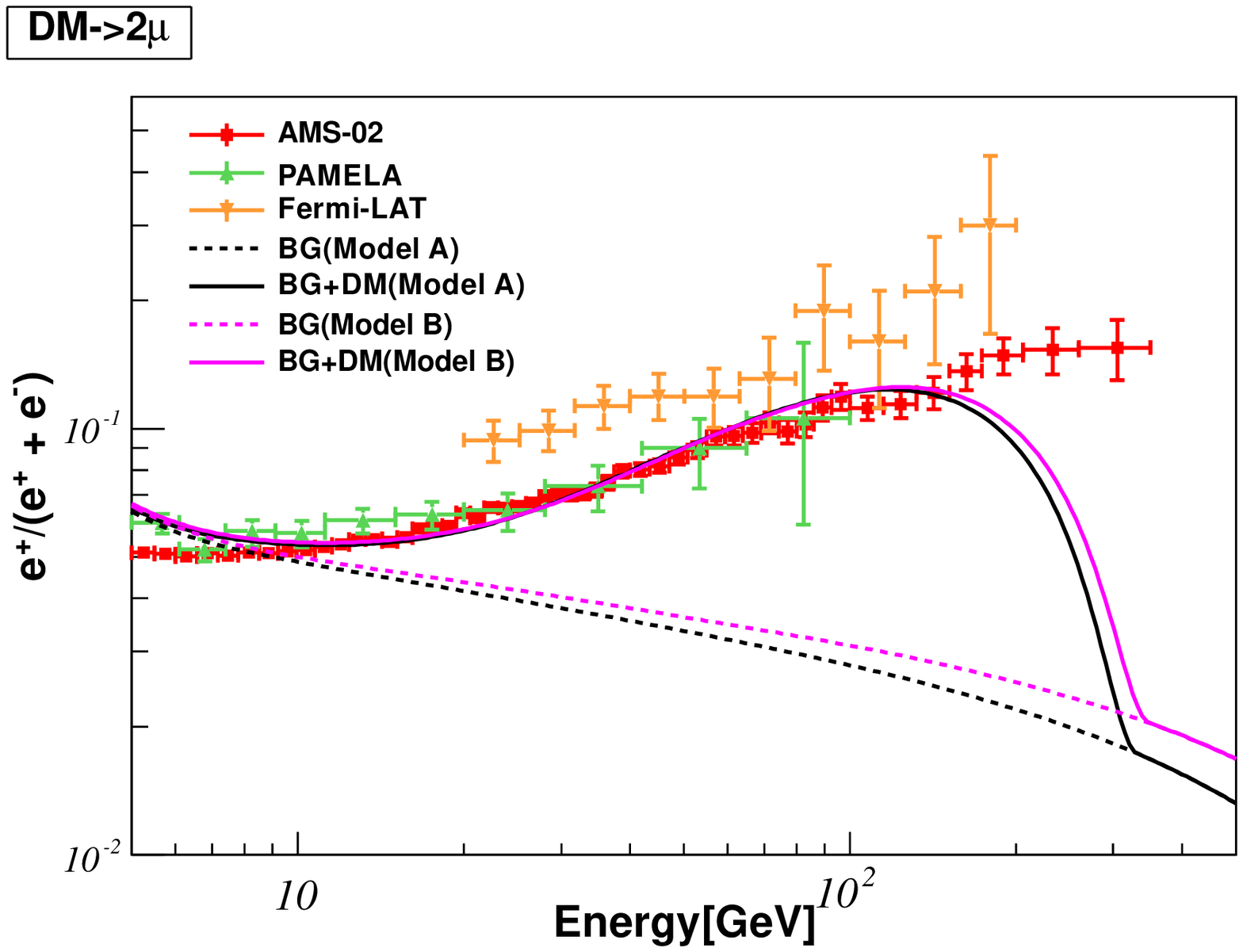}}{\includegraphics[width=0.49\textwidth]{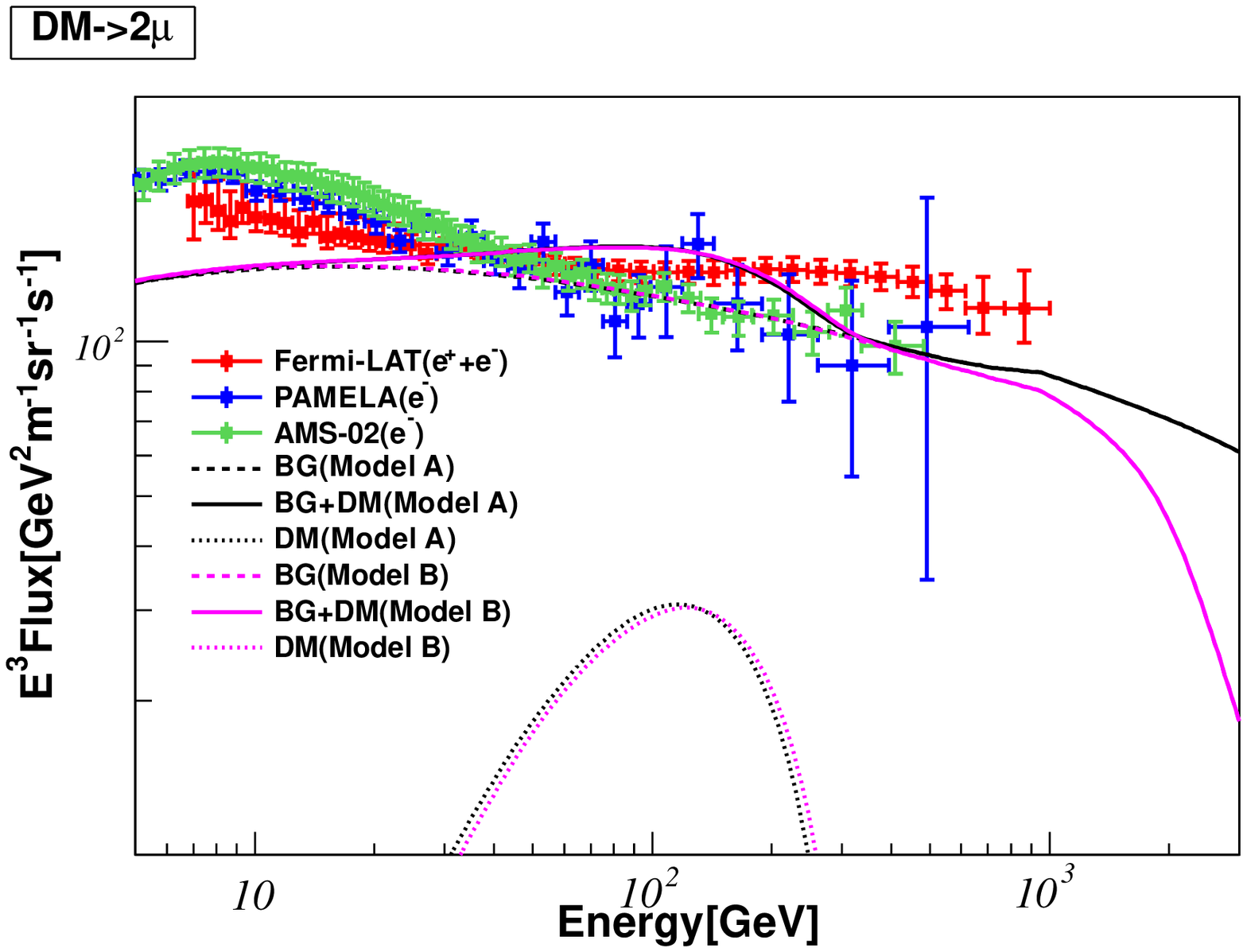}}
\\
{\includegraphics[width=0.49\textwidth]{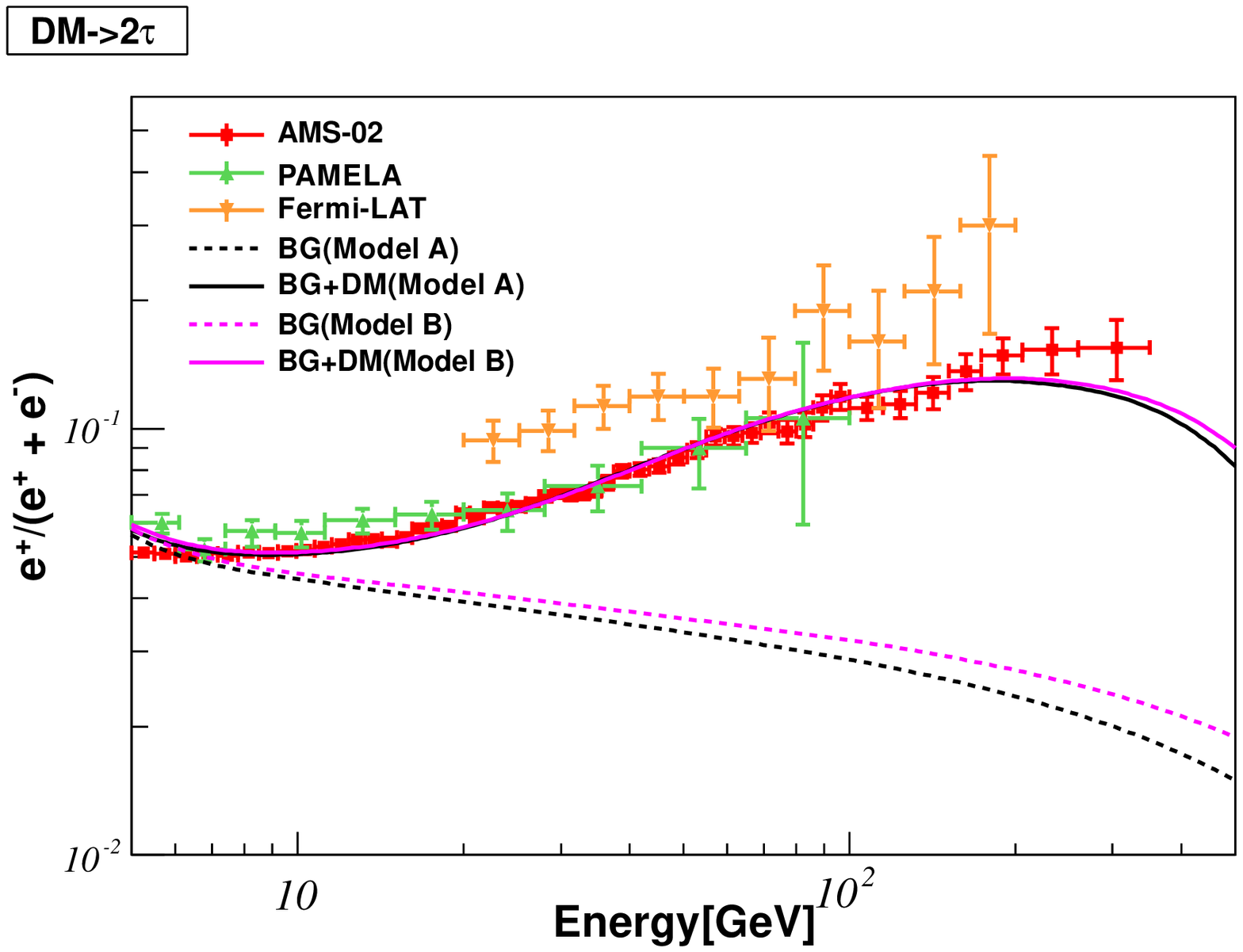}}{\includegraphics[width=0.49\textwidth]{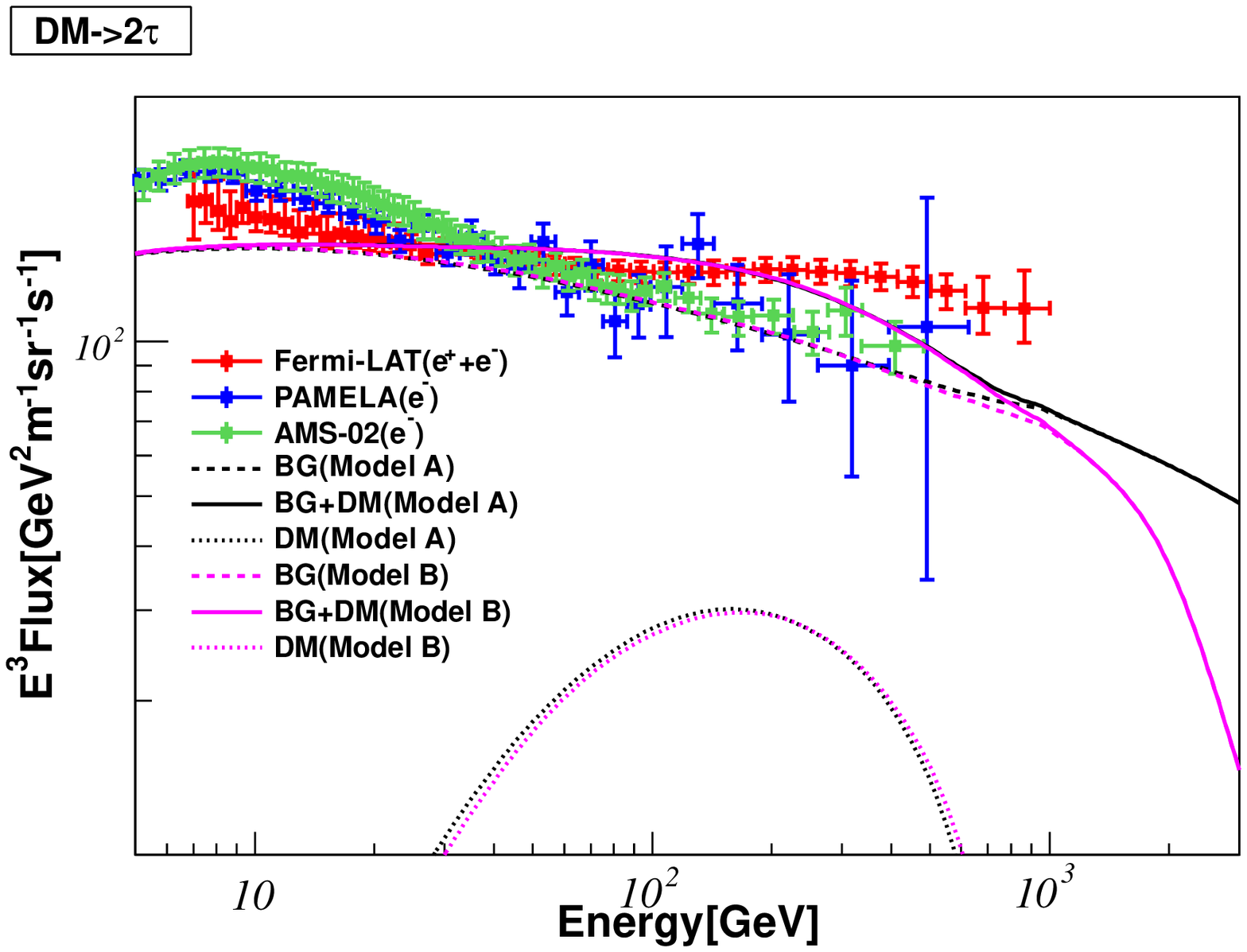}}
\caption{
The same as \fig{fig:uncertaintiesFlux2FinalAnni}, but for the case of DM decay.}
%
\label{fig:uncertaintiesFlux2FinalDecay}
\end{figure}

\begin{figure}
{\includegraphics[width=0.49\textwidth]{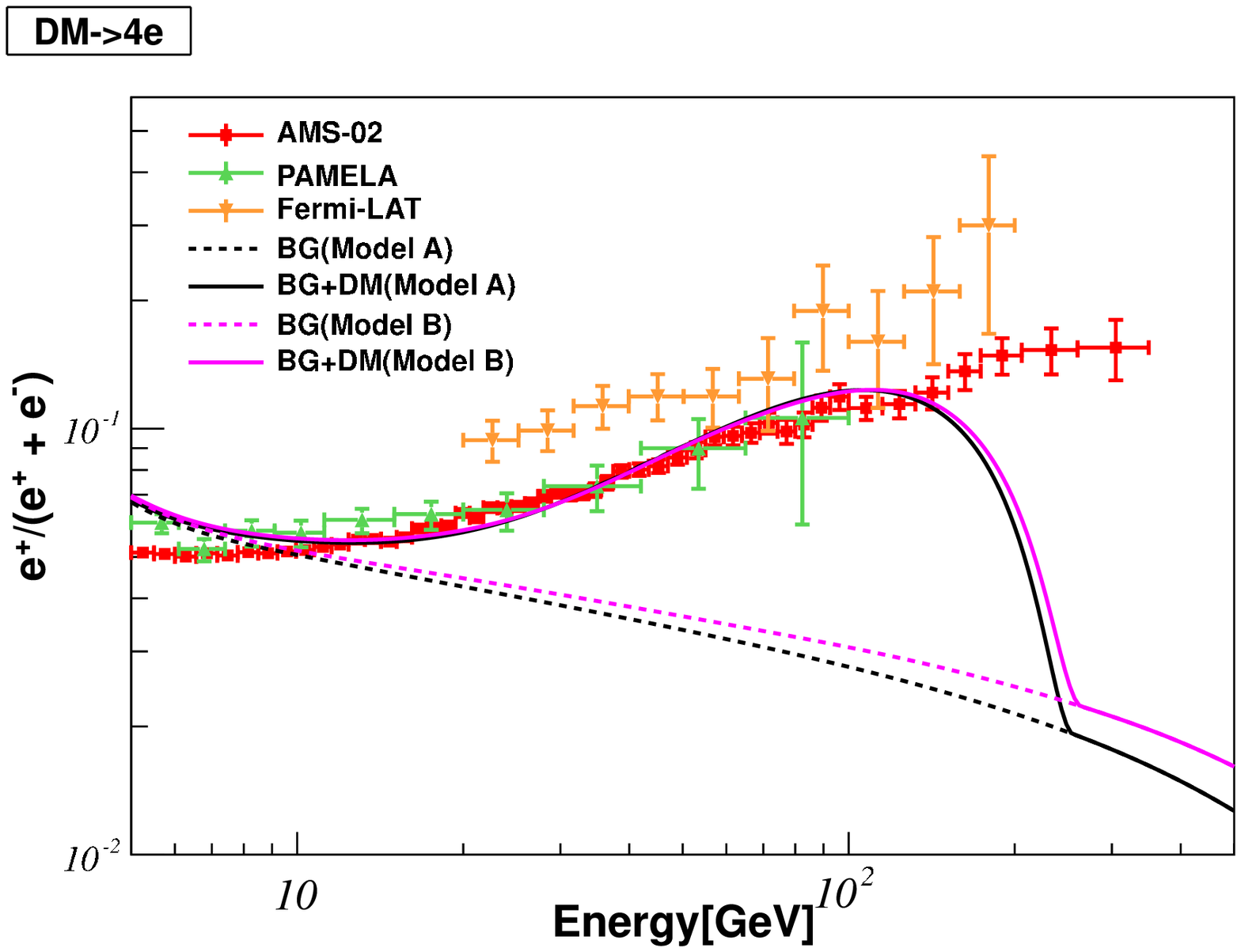}}{\includegraphics[width=0.49\textwidth]{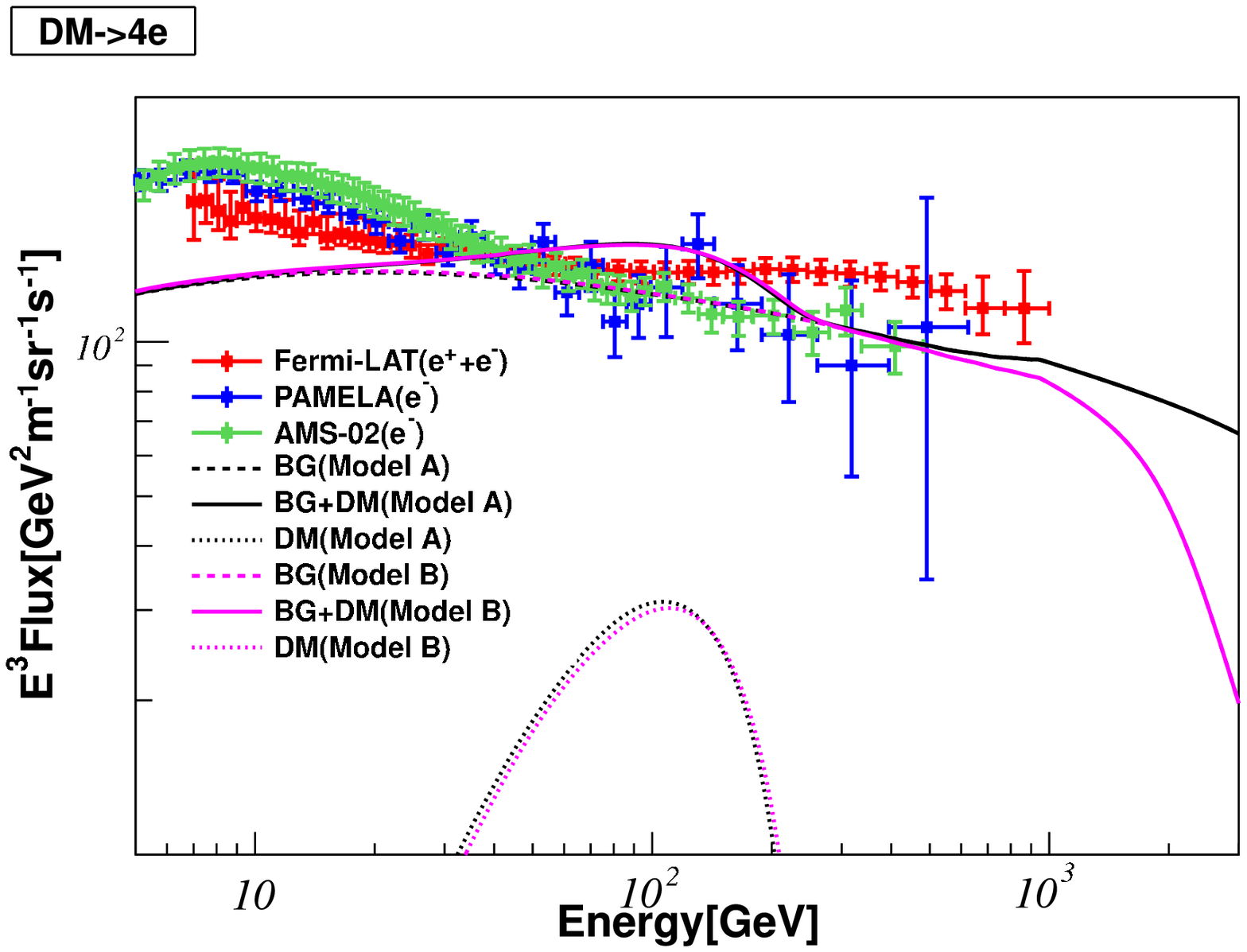}}
\\
{\includegraphics[width=0.49\textwidth]{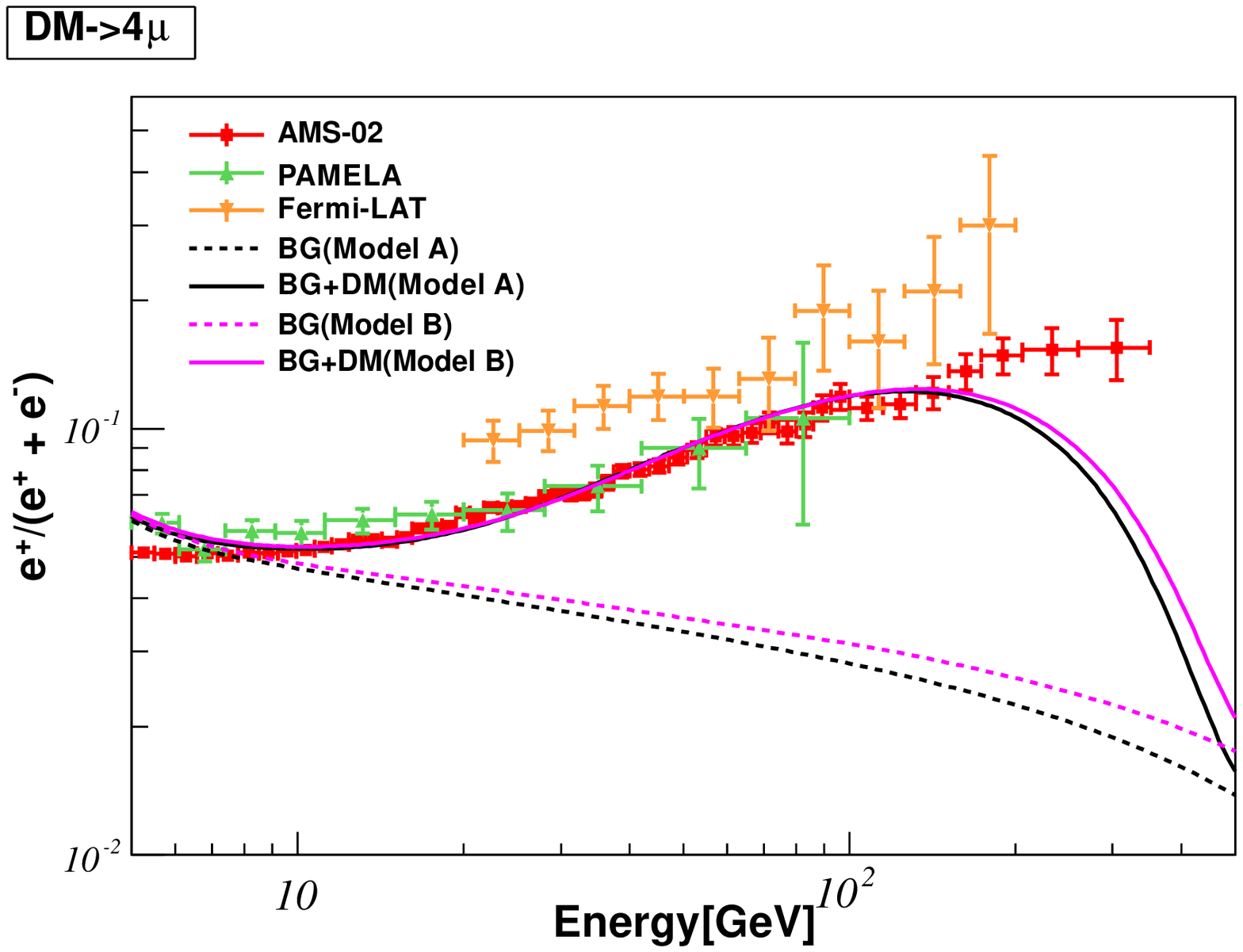}}{\includegraphics[width=0.49\textwidth]{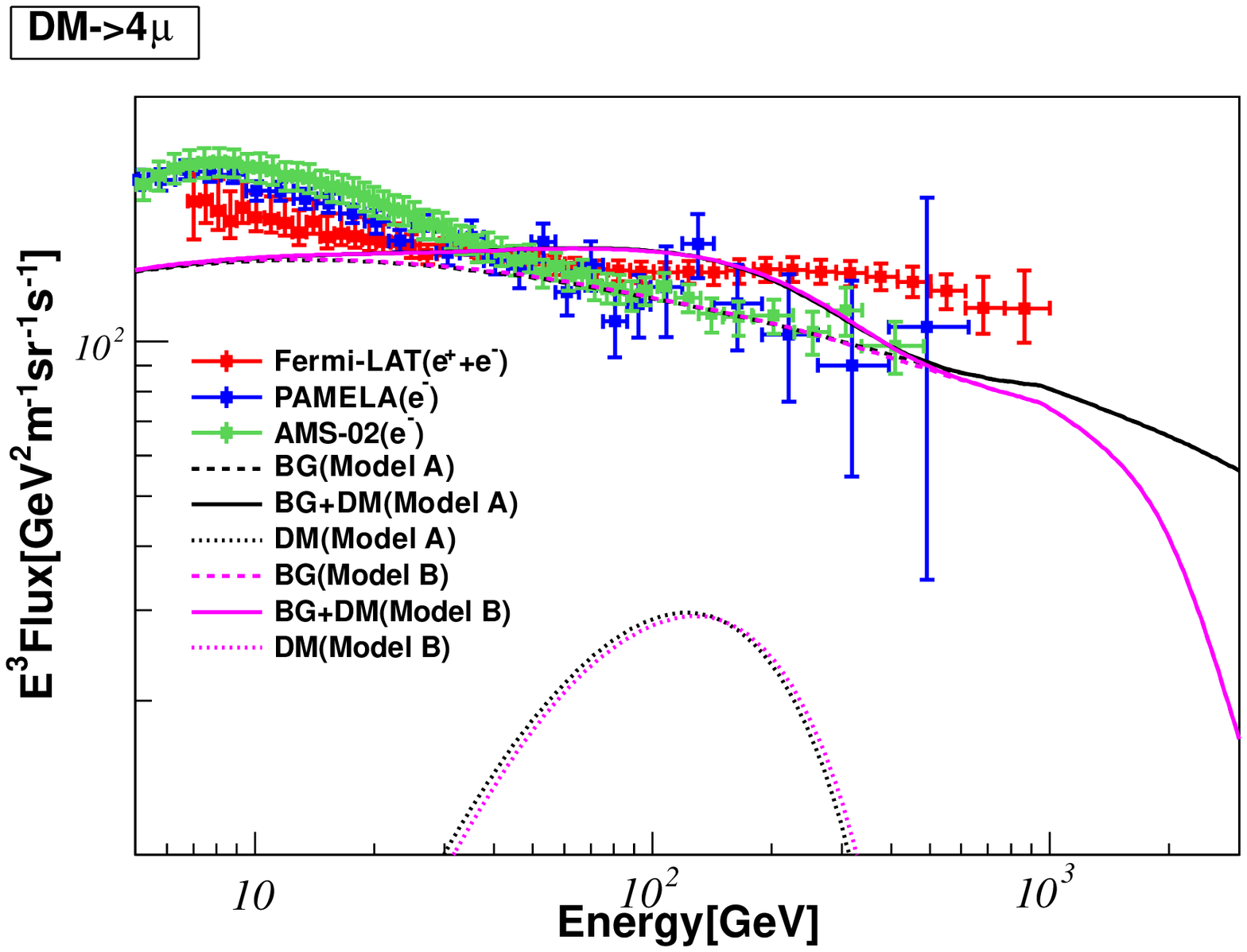}}
\\
{\includegraphics[width=0.49\textwidth]{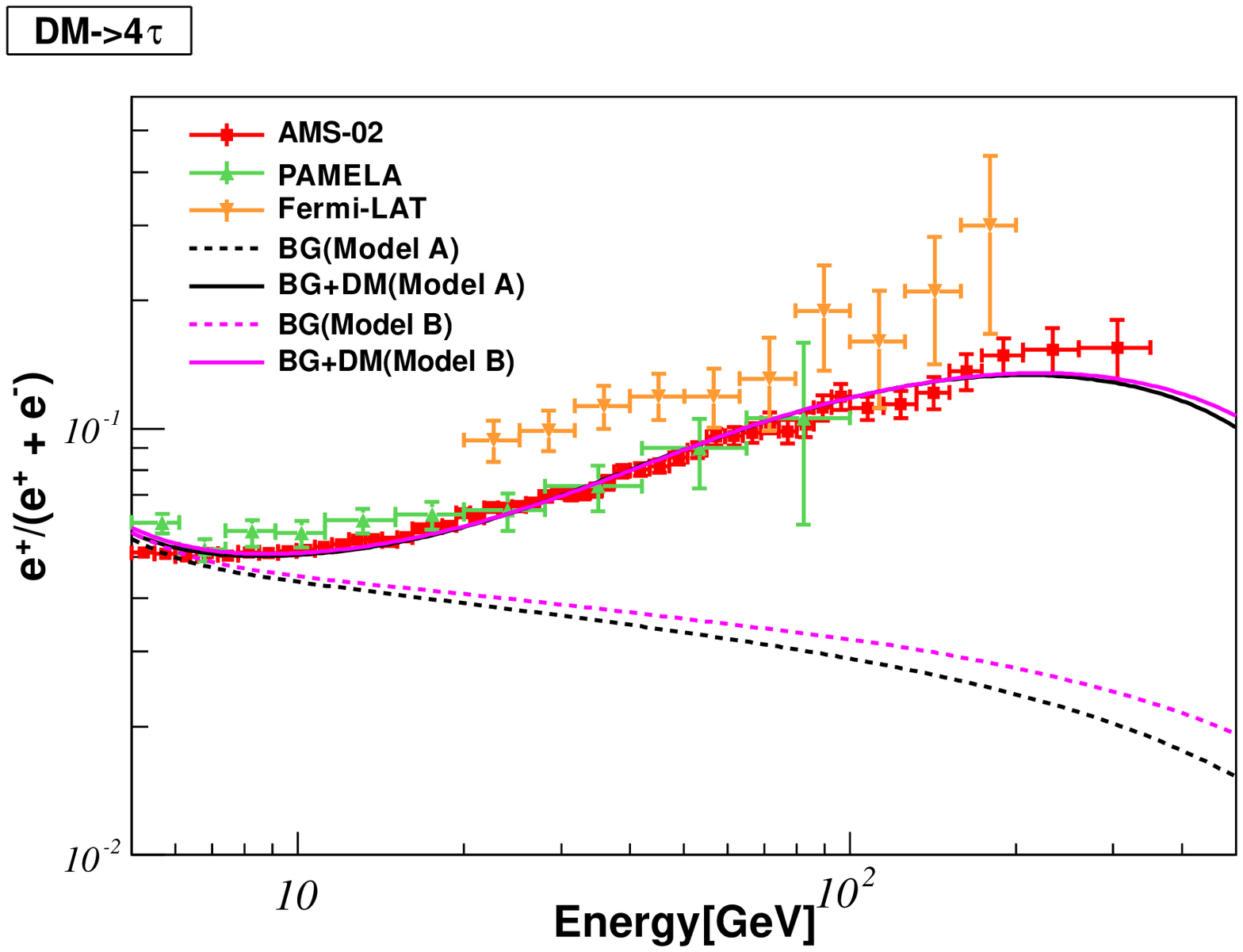}}{\includegraphics[width=0.49\textwidth]{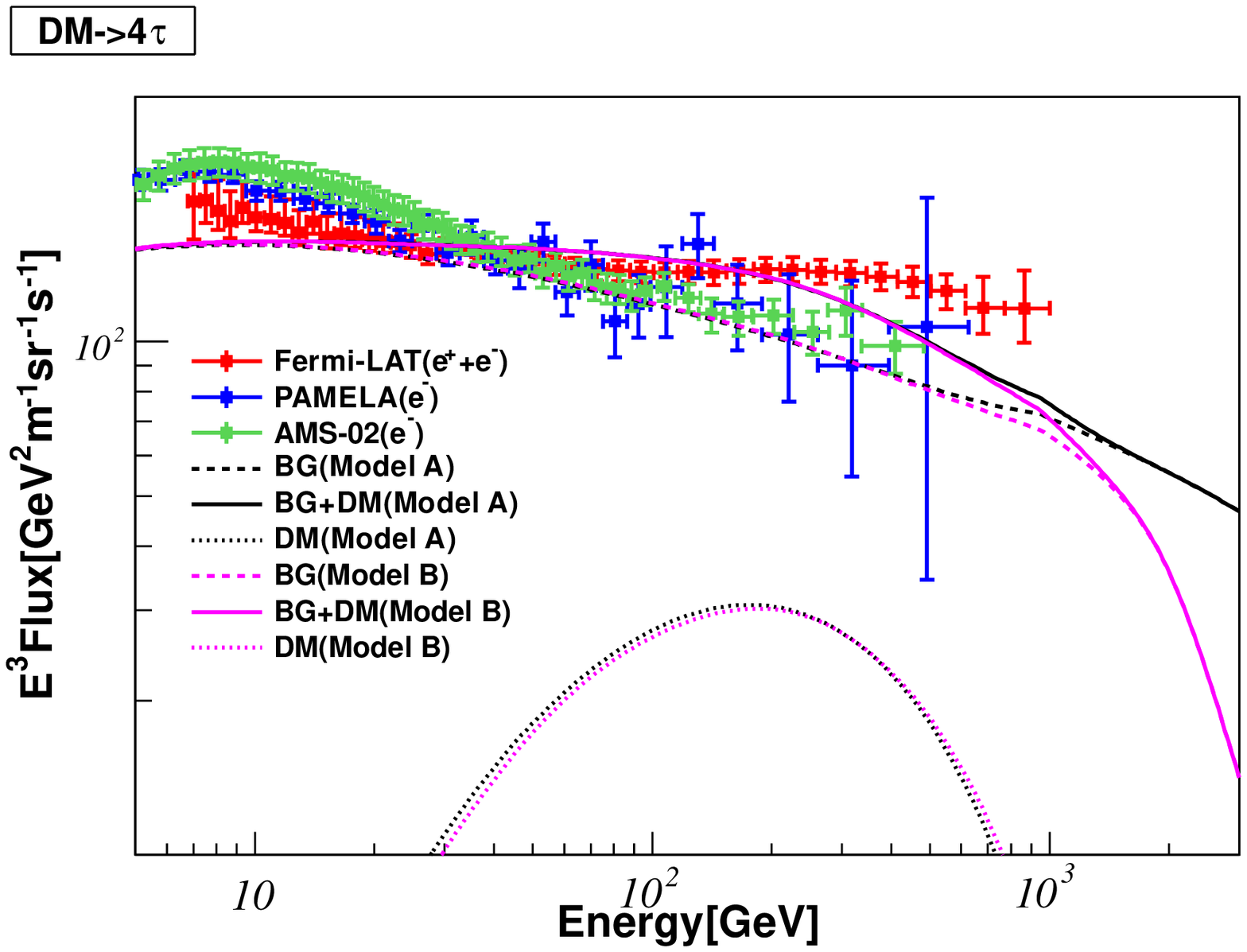}}
\caption{
The same as \fig{fig:uncertaintiesFlux4FinalAnni}, but for the case of DM decay.
}
\label{fig:uncertaintiesFlux4FinalDecay}
\end{figure}

In the case of charge asymmetric decay, i.e. $\epsilon\neq 0$, 
the predicted positron fraction can vary without 
changing the total flux of electrons and positrons.
For $\epsilon=1$, the positron fraction is increased by a factor of two
compared with the case where $\epsilon=0$.
On the other hand, $\epsilon <0$ will suppress the positron fraction. 
Introducing $\epsilon$ leads to more freedom to fit the data.
However, from \eq{eq:decay-source} and \eq{eq:Rapprox}, 
change in the factor $(1+\epsilon)$  can be compensated by
the changes in  the values of $\kappa$ and $\delta$.
Thus a precise determination of $\epsilon$ requires that 
the values of  $\kappa$ and $\delta$ should be precisely determined independently.

\begin{figure}[htb]
\begin{center}
\includegraphics[width=0.49\textwidth]{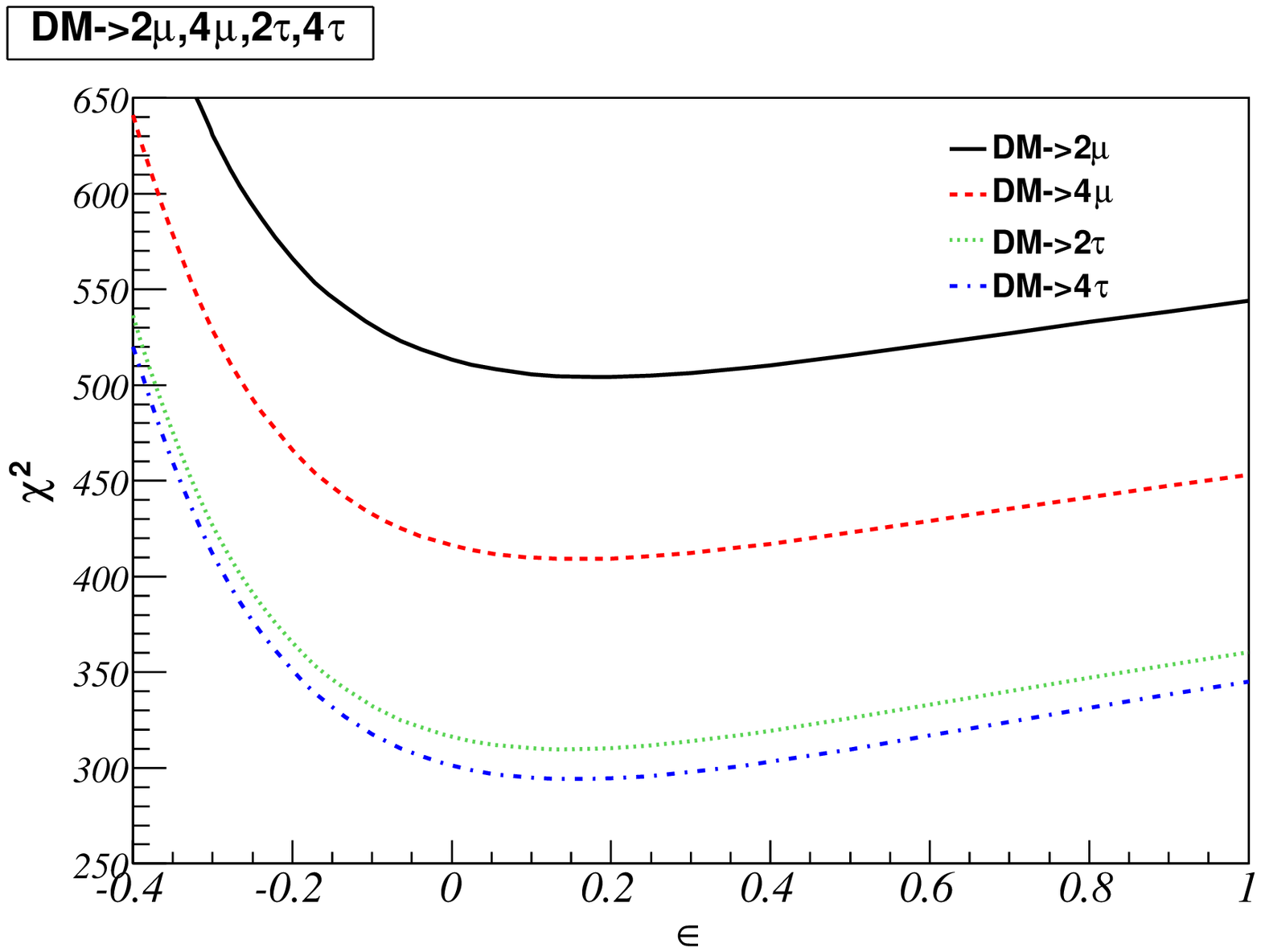} \includegraphics[width=0.49\textwidth]{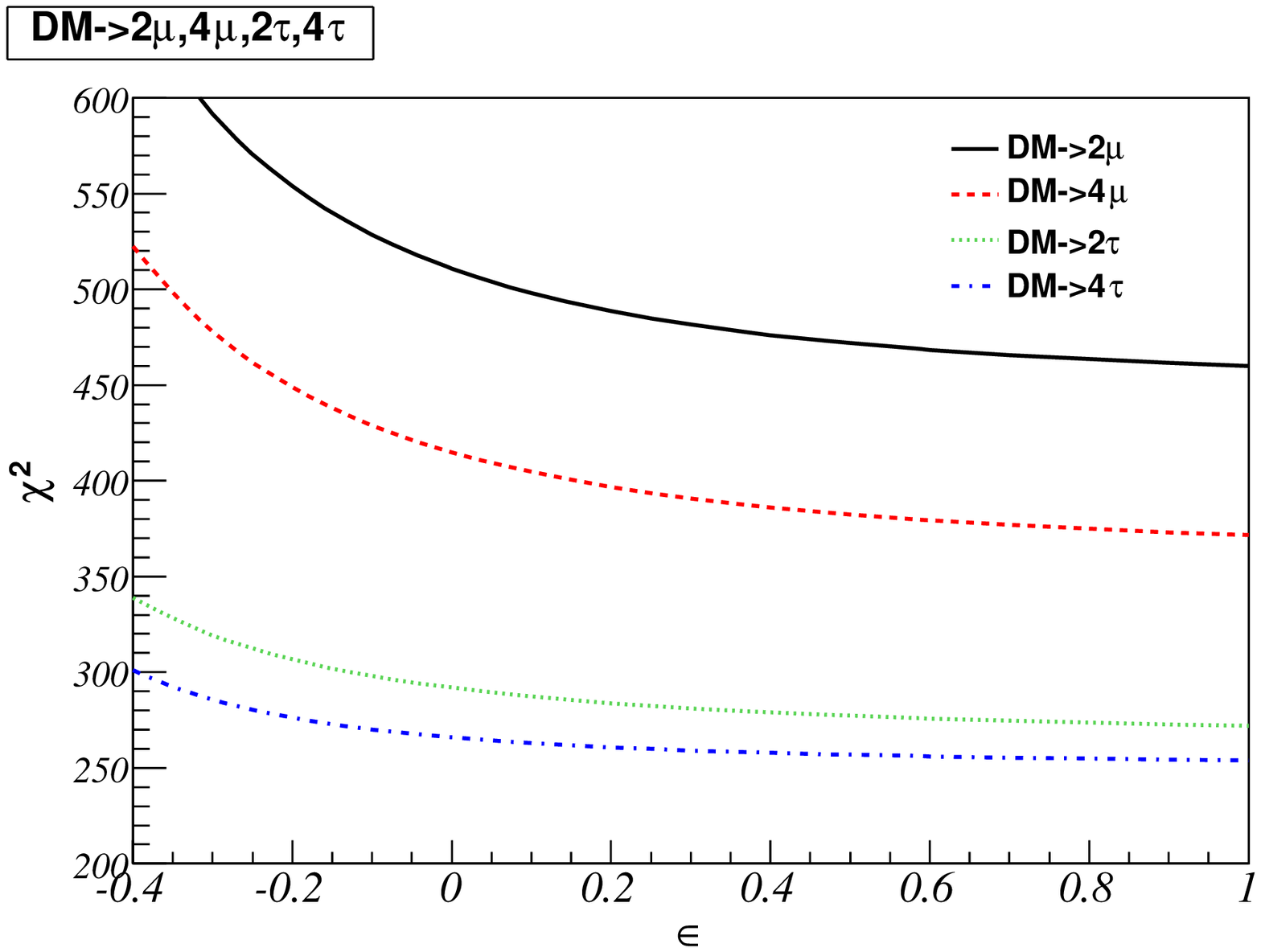}
\end{center}
\caption{
Values of $\chi^{2}$ as a function of charge asymmetric parameter $\epsilon$
from the global fits for the decay final states $2\mu$, $4\mu$, $2\tau$ and $4\tau$
for the case with fixed $\kappa=0.85$ and $\delta=0$ (left)
and with  $\kappa$ and $\delta$ as free parameters determined by the global fit (right).
 }
 \label{fig:decayAS}
\end{figure}

In the first step, 
we consider a simplified case where  $\kappa$ and $\delta$ are fixed at
some typical values $\kappa=0.85$ and $\delta=0$.
For fixed $\kappa$ and $\delta$, the values of $\epsilon$ can be well determined from
the global fit.
In the left panel of \fig{fig:decayAS} 
the values of $\chi^{2}$ as a function of $\epsilon $ are shown.
At $99\%$ C.L., 
only the $2\mu$ channel slightly prefers a nonzero $\epsilon$ in the range $0.02-0.41$.
The allowed values of $\epsilon$ for $2\tau$, $4\mu$ and $4\tau$ channels
are all compatible with zero. 

When  $\kappa$ and $\delta$ are treated as free parameters in the global fit,
the $\chi^{2}$ values decrease significantly. 
For instance, in $2\tau$ channel, the minimal value of $\chi^{2}$ is 
reduced from $294.5$ to $254.1$.
In all the four channels the best-fit values are $\epsilon \approx 1$. 
However, 
as shown in the right panel of \fig{fig:decayAS},
the corresponding $\chi^{2}$ curves are rather flat for $\epsilon>0$,
which indicates less accurate determinations of $\epsilon$,
especially for $2\tau$ and $4\tau$ channels.
At $99\%$ C.L., the allowed ranges for $\epsilon$ are
$\sim0.66-1.0$ and $\sim0.64-1.0$ for $2\mu$ and $4\mu$ channels, respectively.
For $2\tau$ and $4\tau$ channels, the allowed ranges are $\sim0.41-1.0$ 
and $\sim0.20-1.0$ respectively.
We thus conclude that the current data slightly favour 
the scenario of asymmetric DM decay. 
But the statistical significance is not very high.
For obtaining a robust conclusion, 
more experimental data are needed.

%


\section{Conclusions}\label{sec:conclusions}
The AMS-02 collaboration has released the first measurement of
the positron fraction with unprecedented accuracy.
Using the publicly available GALPROP code, 
we have performed  a global analysis  on  the latest data of 
PAMLEA, Fermi-LAT, and AMS-02, 
in terms of DM  annihilation and decay into 
$2e$, $2\mu$, $2\tau$, $4e$, $4\mu$ and $4\tau$ final states. 
A number of propagation models with different
halo heights $Z_{h}$, diffusion parameters $D_{0}$, $\delta_{1,2}$ and 
power indices of primary nucleon sources $\gamma_{p1,p2}$, etc. 
are considered. 
The normalization  and  slope of the background electron fluxes are 
also allowed to vary.
We have found that  
for $2\mu$ and $4\mu$ final states, 
the parameter regions determined by AMS-02 is significantly different from that
favoured by  Fermi-LAT data on the total flux of electrons and positrons. 
For the conventional background model (Model A),
the two allowed regions do not overlap even at $99.99999\%$ C.L. 
For other models, 
we find that the tension between  the two experiments can only 
be slightly reduced in the case of large $Z_{h}$ and $D_{0}$ in Model C2.
The consistency of fits are improved for  $2\tau$ and $4\tau$ final states, 
which favours TeV scale DM with large cross sections corresponding to 
the boost factor of $\mathcal{O}(1000)$.
However, such large annihilations can be in tension with the current measurements of 
cosmic gamma-rays.
In all the  considered leptonic channels,  
we find that the current data favour the scenario of DM annihilation over  DM decay. 
In the decay scenario, 
we have considered both charge symmetric and asymmetric decays.
The results are sensitive to the value of $\kappa$ and $\delta$.
For fixed typical values of  $\kappa$ and $\delta$, the charge asymmetric factor $\epsilon$
is well determined and compatible with zero at $99\%$ C.L.. 
When both $\kappa$ and $\delta$ are taken as free parameters, 
the global fits favour $\epsilon=1$, but the uncertainties in $\epsilon$
become significantly larger.
Thus, currently the charge asymmetric DM decay is only slightly favoured.

\textit{Note added:}
As we were  finalizing  the first version of the manuscript, 
a  preprint  with  similar global fitting analysis but different treatment of 
backgrounds and different focus came out \cite{Yuan:2013eja}.
The conclusions in this work 
are in agreement with theirs. 

\subsection*{Acknowledgments}
This work is supported in part by
the National Basic Research Program of China (973 Program) under Grants No. 2010CB833000;
the National Nature Science Foundation of China (NSFC) under Grants No. 10975170,
No. 10821504  
and No. 10905084;
and the Project of Knowledge Innovation Program (PKIP) of the Chinese Academy of Science.

\bibliography{amssmf,ivpbar}
\bibliographystyle{JHEP}

\end{document}